\documentclass{article}

\usepackage{arxiv}

\usepackage[utf8]{inputenc} 
\usepackage[T1]{fontenc}    
\usepackage{hyperref}       
\usepackage{url}            
\usepackage{booktabs}       
\usepackage{amsfonts}       
\usepackage{nicefrac}       
\usepackage{microtype}      
\usepackage{lipsum}		
\usepackage{graphicx}
\usepackage{natbib}
\usepackage{doi}

\usepackage[utf8]{inputenc} 
\usepackage[T1]{fontenc}    
\usepackage{hyperref}       
\usepackage{microtype}      
\usepackage{lipsum}		
\usepackage{natbib}
\usepackage{enumitem}
\usepackage{url}            
\usepackage{amsfonts}       
\usepackage{nicefrac}       
\usepackage{xcolor}         
\definecolor{peachfuzz}{RGB}{255, 190, 152}

\usepackage{orcidlink}

\usepackage{algorithm}
\usepackage{algpseudocode}
\usepackage{multirow}
\usepackage{colortbl}
\usepackage{comment}
\usepackage{amsmath}

\usepackage{cleveref}

\usepackage{subcaption}
\usepackage{xspace}

\usepackage{lscape}
\usepackage{pifont}
\newcommand{\cmark}{\ding{51}}%
\newcommand{\xmark}{\ding{55}}%

\newcommand{\spara}[1]{\smallskip\noindent{\bf #1}}

\newcommand{\ouralgo}{\textsc{HyDRA}\xspace}

\newcommand{\userprob}{\ensuremath{P(u|z_u)}\xspace}
\newcommand{\userprobzeta}{\ensuremath{P(u|z_u)^\zeta}\xspace}

\newcommand{\itemprob}{\ensuremath{P(i|y_i)}\xspace}
\newcommand{\itemprobxi}{\ensuremath{P(i|y_i)^\xi}\xspace}

\newcommand{\dataset}{\ensuremath{\mathcal{D}}\xspace}

\usepackage{xcolor}
\usepackage{xargs}
\usepackage[colorinlistoftodos,prependcaption,textsize=small]{todonotes}
\usepackage[framemethod=tikz]{mdframed}

\newcommandx{\unsure}[2][1=]{\todo[linecolor=red,backgroundcolor=red!25,bordercolor=red,caption={},#1]{#2}}
\newcommandx{\change}[2][1=]{\todo[linecolor=blue,backgroundcolor=blue!25,bordercolor=blue,caption={},#1]{#2}}
\newcommandx{\info}[2][1=]{\todo[linecolor=green,backgroundcolor=green!25,bordercolor=green,caption={},#1]{#2}}
\newcommandx{\improvement}[2][1=]{\todo[linecolor=purple,backgroundcolor=purple!25,bordercolor=purple,caption={},#1]{#2}}
\newcommandx{\discussion}[2][1=]{\todo[linecolor=blue,backgroundcolor=yellow!25,bordercolor=yellow,caption={},#1]{#2}}

\newcommandx{\thiswillnotshow}[2][1=]{\todo[disable,#1]{#2}}

\title{Flexible Generation of Preference Data for Recommendation Analysis}

\author{
\href{https://orcid.org/0000-0002-0961-4151}{\includegraphics[scale=0.06]{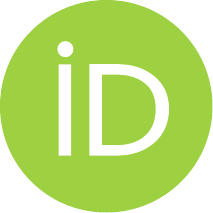}\hspace{1mm}Simone Mungari} \\
	University of Calabria\\
        ICAR-CNR\\
        Revelis s.r.l.\\
	\texttt{simone.mungari@unical.it} \\
    \And
    \href{https://orcid.org/0000-0002-4670-8157}{\includegraphics[scale=0.06]{fig/orcid.pdf}\hspace{1mm}Erica Coppolillo} \\
	University of Calabria\\
        ICAR-CNR\\
	\texttt{erica.coppolillo@unical.it} \\
   \AND
   \href{https://orcid.org/0000-0003-3978-9291}{\includegraphics[scale=0.06]{fig/orcid.pdf}\hspace{1mm}Ettore Ritacco} \\
	University of Udine\\
	\texttt{ettore.ritacco@uniud.it} \\
    \And 
    \href{https://orcid.org/0000-0001-9672-3833}{\includegraphics[scale=0.06]{fig/orcid.pdf}\hspace{1mm}Giuseppe Manco} \\
	ICAR-CNR\\
	\texttt{giuseppe.manco@icar.cnr.it}
   } 
\begin{document}

\maketitle



\begin{abstract}
Simulating a recommendation system in a controlled environment, to identify specific behaviors and user preferences, requires highly flexible synthetic data generation models capable of mimicking the patterns and trends of real datasets. In this context, we propose \ouralgo, a novel preferences data generation model driven by three main factors: user-item interaction level, item popularity, and user engagement level.
The key innovations of the proposed process include the ability to generate user communities characterized by similar item adoptions, reflecting real-world social influences and trends. Additionally, \ouralgo\ considers item popularity and user engagement as mixtures of different probability distributions, allowing for a more realistic simulation of diverse scenarios. This approach enhances the model's capacity to simulate a wide range of real-world cases, capturing the complexity and variability found in actual user behavior.
We demonstrate the effectiveness of \ouralgo\ through extensive experiments on well-known benchmark datasets. The results highlight its capability to replicate real-world data patterns, offering valuable insights for developing and testing recommendation systems in a controlled and realistic manner. The code used to perform the experiments is publicly available:~\url{https://github.com/SimoneMungari/HYDRA}.
  
\end{abstract}





\maketitle

\section{Introduction}

Ensuring clean and reliable interaction data is a major concern in the context of recommendation~\citep{Heinrich2019DataQI, Shalom2015DataQM}, social network analysis~\citep{Reda2023AssessingTQ, 10.1007/978-3-319-10518-5_23}, and machine learning in broad sense~\citep{10.1145/3394486.3406477}. 
As algorithms become more powerful and sophisticated, the demand for reliable benchmarking studies to evaluate and compare their capabilities across various perspectives and scenarios is growing~\citep{10.1145/3383313.3412489}. However, the availability of benchmark open-source datasets is limited since large industrial companies generally do not release their vast amounts of proprietary data to the public. As a result, the need for dependable datasets is more urgent and valuable than ever.

Traditionally, evaluation takes place using a variety of publicly available real-life datasets. Notable examples are Movielens~\citep{movielens} and Netflix~\citep{bennett2007netflix}, Lastfm~\citep{Bertin-Mahieux2011} and Yahoo! Music~\citep{DBLP:journals/jmlr/DrorKKW12},
Epinions~\citep{epinions}, and Amazon~\citep{hou2024bridginglanguageitemsretrieval}. However, these real-life benchmark datasets exhibits several limitations. First, they tend to focus on specific domains, limiting the generalizability of findings across diverse applications. Additionally, many of these datasets lack the scale needed to assess the performance of algorithms in large, real-world settings, which can lead to inaccurate performance assessments. 
Lastly, intrinsic biases present in these datasets, such as filter bubbles or echo chambers resulting from feedback loops in operational environments~\citep{10.1145/3306618.3314288,10.1002/aaai.12055,GAO20211166}, can distort the evaluation of recommender systems, leading to misleading conclusions about their effectiveness.

Standardized datasets can help mitigate these issues by offering a broader, unbiased, and scalable framework for testing and improving recommendation algorithms. To obtain them, crowd-sourcing can be considered as a viable option~\citep{Lee2014ImpactOR, recsys_sales_diversity, sales_diversity_impact, Zhu2018TheIO}. However, recruiting real users is generally expensive and time-consuming. Besides, user's behavior data poses several privacy concerns and challenges that prevent actual recruitinig strategies and in general the disclosure and availability of preferences.
Indeed, a more practical and comprehensive solution to fill the aforementioned gaps and address the scarcity and inadequacy of real data consists of generating data synthetically. 

In this regard, some methods try to replicate the characteristics of real-world data by learning their distribution \citep{dagzang, DBLP:conf/icml/ArjovskyCB17}, by performing augmentation~\citep{novel-items} or compression~\citep{data-condensation}. These methods learn directly from the ground truth, thus providing a reliable replication of real-world data. Nevertheless, such approaches are bound to the underlying originary scenarios, thus lacking flexibility. On the other hand, generative probabilistic frameworks represents viable alternatives for producing synthetic datasets \citep{DBLP:conf/ai/SmithCP17, barabasi1999emergence, kronecker,BPR,cf-augmentation}. 
However, current synthetic generation methods of preference data face significant limitations, including narrow customization and unrealistic distributions. These methods struggle to accurately reflect nuanced user preferences and behaviors. Additionally, the data often lacks variability, leading to overly simplistic patterns that fail to capture the complexity of real-world user interactions and network structures.

In this paper, we present \ouralgo (Hyper-Parametric Data generation for Recommendation Analysis), a flexible generative probabilistic framework for generating preference data.  
The theoretical foundation of our method relies on three main components: modeling \textit{user beliefs, leanings, and attitude to engage and interact}; assessing \textit{item exposure and potential popularity}; and evaluating the likelihood that \textit{users' and items' inherent features match}. From this theoretical model, we parameterize the framework by introducing multiple control factors to leverage a fully controllable generation process. Our method notably allows for: (\textit{i}) regulating user-item interactions; (\textit{ii}) freely tweaking the underlying data distributions; and (\textit{iii}) faithfully replicating the characteristics of pre-existing datasets.
We perform an extensive evaluation of \ouralgo, assessing its efficacy under the aforesaid perspectives.

The rest of the paper is structured as follows. Section~\ref{sec:related_work} provides an overview of state-of-the-art synthetic generation procedures. In Section~\ref{sec:modeling}, we formally define our methodology and discuss its theoretical foundation. Section~\ref{sec:generation} describes the instantiation of \ouralgo based on insights from the modeling discussion. We present the experimental results in Section~\ref{sec:experiments}, demonstrating the flexibility of \ouralgo. Finally, Section~\ref{sec:conclusions} offers pointers for future developments.

\section{Related Work}
\label{sec:related_work}
Given the scarcity of high-quality real-world datasets for recommender systems, there has been considerable effort in the literature to generate synthetic data. In this broad context, we specifically concentrate on producing realistic user-item interactions. This task poses several challenges, including the necessity to replicate the topological characteristics of real-world scenarios \citep{erdHos1960evolution, barabasi1999emergence, kronecker,dagzang}. The present literature covers a wide range of different approaches, which we outline below.


\spara{Data Augmentation and Condensation.} Data augmentation consists of expanding an existing dataset while preserving its structural properties. This valuable task  has been extensively studied in the literature. 
\citet{novel-items} propose a framework based on Variational Autoencoders (VAEs) for generating novel items that users will probably interact with. \citet{Belletti2019ScalableRR} propose an expanding approach based on an adaption of Kronecker Graphs. More recent approaches have also explored the adoption of Large Language Models in this regard~\citep{llms-data-augmentation}.

In an opposite perspective, condensation refers to the task of compressing the original data while still maintaining their properties. \citet{data-condensation} propose a novel framework for condensing the original dataset while addressing the long-tail problem with a reasonable choice of false negative items.  \citet{graph-condensation} propose a strategy aimed at compacting a graph preserving its features for classification tasks. The process is performed by a gradient matching loss optimization and a strategy to condense node features and structural information simultaneously.

\spara{Semi-Synthetic Generation.} 
Differently from data augmentation methods, these approaches involve models learning directly from a ground-truth dataset to generate a fully synthetic one. For instance,~\citet{wasserstein-gan} introduce a Wasserstein-GAN architecture~\citep{DBLP:conf/icml/ArjovskyCB17} for this purpose. The resulting synthetic dataset exhibits similar patterns and distributions of users, items, and ratings compared to the real dataset used in the experimental evaluation. 

A specific sub-field of this research area focuses on generating synthetic data to protect sensitive real-world information~\citep{privacy-preserving}. These methods are typically applied in fields such as finance and healthcare, and in general in any context where data privacy is crucial \citep{privacy-preserving, DBLP:journals/corr/abs-2311-03488}. For instance, in \citep{DBLP:journals/corr/abs-2008-03797}, the framework alters a subset of real data values to produce a new semi-synthetic dataset. Similarly, in \citep{DBLP:conf/ai/SmithCP17}, the model starts by generating a dense user-item matrix using a probabilistic matrix factorization approach based on Gaussian distributions, further masking some user preferences.

\spara{Probabilistic Models.} The generation of synthetic data is often performed by adopting probabilistic approaches. Different kinds of distributions have been explored for user-item content sampling, such as the typical Gaussian~\citep{DBLP:conf/aaai/WangWSSL20}, the Bernoullian~\citep{DBLP:conf/aaai/ZhangCSH21}, the Dirichlet and Chi-square distributions~\citep{evaluation-attribute-aware}.
Among the others, ~\citet{non-negative-mf} propose a Bayesian Generative Framework and modeling procedure based on Gibbs Sampling for binary count data with side information. 

\spara{Simulation-based approaches.} Driven by the objective of integrating generation and recommendation, other methods aim at simulating realistic interactions for inactive users \citep{cf-augmentation, usersim, Wang2021LearningTR}. In \citep{DBLP:conf/icwsm/RibeiroV023}, the intent is to generate a binary preference matrix with five different topics spanning from the Far Left to the Far Right of the political spectrum. Their main limitation is the lack of a study of the generated user and item distributions.
\citet{chaney} extend the model proposed in~\citep{Schmit2017HumanIW} for allowing multiple interactions for the same user. The histories are generated by sampling from a (noisy) utility matrix which represents the actual preferences of users.

Table \ref{tab:competitors} reports the portion of methods presented in this section that generate synthetic datasets from scratch, thus presenting similarities with our proposal. The table summarizes the differences with our methodology.

\begin{table}[!ht]
\centering
\resizebox{\columnwidth}{!}{%
\begin{tabular}{@{}ccccccc@{}}
\toprule
Model &
  No training &
  Realistic & 
  \multicolumn{3}{c}{Flexibile} &
  Source Code \\
  & & & Distributions & Interactions & Topics \\ 
  \midrule
\ouralgo (ours) &
  \cmark &
  \cmark &
  \cmark &
  \cmark &
  \cmark &
  \cmark \\
  Tso et al. \citep{evaluation-attribute-aware} &
   \cmark &
   \xmark &
   \xmark &
   \xmark & 
   \cmark &
   \xmark \\
\citet{DBLP:conf/ai/SmithCP17} &
   \cmark &
   \cmark &
   \xmark &
   \cmark & 
   \xmark &
   \xmark \\
   \citet{chaney} &
   \xmark &
   \xmark &
   \xmark & 
   \xmark &
   \xmark &
   \xmark \\
   
   \citet{DBLP:conf/aaai/ZhangCSH21} &
   \cmark & 
   \xmark &
   \xmark &
   \cmark & 
   \cmark &
   \xmark \\
   \citet{DBLP:conf/icwsm/RibeiroV023} &
   \cmark &
   \cmark &
   \xmark &
   \xmark &
   \cmark &
   \cmark \\
   
   \bottomrule
\end{tabular}%
}
\caption{Comparison between the proposed framework and the state-of-the-art models which generate synthetic data from scratch. 
The column ``No training" indicates whether the method allows to generating interactions without a training phase;
``Realistic" refers to the conformity of the generated data to the typical properties of real-world data distributions (e.g. long-tailed); ``Source Code" denotes the availability of the code for the experiments; finally, the column ``Flexible" refers to the adaptability of the approach in terms of (\textit{i}) fitting a specific distribution, (\textit{ii}) ad-hoc tuning the synthetic user-item interactions, and (\textit{iii})  introducing a given number of topics, further regulating the users interest accordingly.}
\label{tab:competitors}
\end{table}

\section{Data Modeling}
\label{sec:modeling}
The foundational intuition behind our proposal is to jointly model the observations of users and items in a preference dataset along with intrinsic information within the data, represented as data statistics. While the latter are essentially hidden information embedded within the list of user-item pairs, they could be distorted by a probabilistic model if not explicitly considered.
From this insight, considering the number of items adopted by each user and the number of users for each item as key statistics, it follows that \ouralgo is governed by three main components: User-Item Matching, User Engagement Level, and Item Popularity.

\spara{User-Item Matching.} This factor summarizes the level of affinity between user preferences and item characteristics. Typically, users select items that meet and satisfy their tastes and needs. We believe this phenomenon is not absolute; rather, it depends on other two factors. First, a user does not have visibility of the entire set of items, making it unrealistic for a choice to be driven solely by preferences. Secondly, a user-item pair will be observable in the preference matrix only if both the user and the item have achieved a certain level of exposure: user engagement level and item popularity, respectively.

\spara{User Engagement Level.} The probability of observing a user is a consequence of their affiliation with interaction platform. Thus, the generative model must define a distribution capable of simulating the frequency of user occurrences in the dataset.

\spara{Item Popularity.} Previous studies in the field of recommendation systems~\citep{Zhu2018,CE17} have shown that user behavior is significantly influenced by the phenomenon of popularity bias. Users tend to be influenced by global trends, and an item is often chosen because of its popularity, regardless of its actual affinity with various users. This reinforces the principle that an item's value increases as it is adopted by more individuals~\citep{LiuGZL19}.

\vspace{0.5em}
\noindent Let $U=\{1,\ldots,n\}$ and $I=\{1,\ldots, m\}$ be the set of users and items, respectively. Let $\dataset$ be the dataset of all user preferences, expressed as pairs $(u,i)$, where $u\in U$ and $i\in I$. From $\dataset$, we can derive the vectors $\mathbf{z} \in \mathbb{N}^n$ and $\mathbf{y} \in \mathbb{N}^m$, such that $z_u = |\{i: (u,i)\in \dataset\}|$ represents the level of engagement of user $u$ and $y_i = |\{u: (u,i)\in \dataset\}|$ represents the popularity of item $i$.
Formally, given a set $\nu$ of parameters governing the preferences, \ouralgo defines the likelihood of the observed tuple $(\dataset, \mathbf{z}, \mathbf{y}, \nu)$ by utilizing an implicit feedback approach. This means we define a random variable $r \in \{0, 1\}$ to indicate whether a pair $(u, i)$ exists in the dataset. 
Consequently, we have:
\begin{equation}
\begin{split}
    P(\dataset, \mathbf{z}, \mathbf{y}, \nu | \Theta)
    &= 
     P(\nu) \, P_\nu(\dataset|\mathbf{z}, \mathbf{y})\,  P(\mathbf{z}, \mathbf{y}|\Theta)\, ,
\end{split}
\label{eq:dataset_probability}
\end{equation}
where
\begin{equation}
\begin{split}
    P_\nu(\dataset|\mathbf{z}, \mathbf{y})  &= \prod_{u,i \in \dataset} P_\nu(r=1, u, i|z_u,y_i) \\
    &\quad \prod_{u,i \notin \dataset} \left[1 - P_\nu(r=1, u, i|z_u,y_i)\right]
\end{split}
\label{eq:prob_data}
\end{equation}
and $P(\mathbf{z}, \mathbf{y}|\Theta) = P(\mathbf{z}|\Theta) \, P(\mathbf{y}|\Theta)$ represents the probability of observing the frequencies $\mathbf{z}$ and $\mathbf{y}$. 
As mentioned, the probabilities $P_\nu(r, u, i | z_u, y_i)$, $P(\mathbf{z}|\Theta)$ and $P(\mathbf{y}|\Theta)$
represent the User-Item Matching, User Engagement Level, and Item Popularity components, respectively.
In particular, $P_\nu(r, u, i | z_u, y_i)$ can be factored as
\begin{equation}
    P_\nu(r=1, u, i|z_u, y_i)
    = P_\nu(r=1 | u, i) \, P(u | z_u)\, P(i | y_i) \, ,
\label{eq:implicit_feedback}
\end{equation}
assuming independence between the pairs $(u, z_u)$ and $(i, y_i)$.

\color{black}
In addition, we model the distributions governing User Engagement Level and Item Popularity as a mixture of components:
\begin{equation}\label{eq:eng_pop}
    P(\mathbf{z}|\Theta) =  \sum_{k=1}^K \pi_k \prod_{u \in U} P_{\theta_k}(z_u) \, ; \; \; \;
    P(\mathbf{y}|\Theta) =  \sum_{h=1}^H\psi_h\prod_{i \in I} P_{\vartheta_h}(y_i).
\end{equation}
Here, $\{1, \ldots, K\}$ and $\{1, \ldots, H\}$ represent partitions over two probabilistic spaces containing the chosen distributions for the User Engagement Level and the Item Popularity, respectively, whose parameters are $\theta_k$ and $\vartheta_h$, with $k \in \{1, \ldots, K\}$ and $h \in \{1, \ldots, H\}$.
Within the equations, the probabilities $\pi_k$ and $\psi_h$ indicate mixing distributions. 
Combining Equations~\ref{eq:prob_data},  \ref{eq:implicit_feedback} and~\ref{eq:eng_pop}, we can rewrite Equation~\ref{eq:dataset_probability} as:
\begin{equation}
\begin{split}
    P(\dataset, \mathbf{z}, \mathbf{y}, \nu | \Theta) 
    & = P(\nu)\, P_\nu(\dataset|\mathbf{z}, \mathbf{y})
    \\ &\quad
    \sum_{k=1}^K \sum_{h=1}^H\pi_k \psi_h \prod_{u \in U} P_{\theta_k}(z_u) \, \prod_{i \in I} P_{\vartheta_h}(y_i) 
\end{split}
\label{eq:hydra}
\end{equation}

\subsection{User-Item Matching}
Drawing inspiration from \citep{chaney} and \citep{DBLP:conf/ai/SmithCP17}, we instantiate $\nu = \{\nu_u, \nu_i\}$ as the set of the model parameters  $\nu_u = \{\rho_u, \mu^{\rho}_u\}_{u\in U}$ and $\nu_i = \{\alpha_i, \mu^{\alpha}_i\}_{i\in I}$, relative to users and items, that govern the probability of observing preferences. First, we define \emph{User-Item Matching} by exploiting latent factors: 
\begin{equation}
    P_\nu(r=1 | u, i) = \lambda \cdot \rho_u^\top \alpha_i \, ,
    \label{eq:user-item v1}
\end{equation}

where $\rho_u$ and $\alpha_i$ are latent vectors of size $f$, representing instantiations over a domain of features that characterize user preferences and item properties. The $\lambda$ component is a regularization term based on the likelihood of observing an interaction in the specific domain of interest, regardless of the users/items involved, thus modeling the intrinsic expected density of \dataset. 

Both $\rho_u$ and $\alpha_i$ are modeled considering a hierarchical process based on Dirichlet distributions. The first step draws the prior probabilities $\mu_u^\rho$ and $\mu_i^\alpha$ for each user and item over the latent space, while the second step draws their posterior distributions.
For users, we define:
\begin{equation}
\begin{split}
    \rho_u \sim \textrm{Dirichlet}(10 \, \mu_u^\rho) \quad \mbox{and} \quad
    \mu_u^\rho \sim \textrm{Dirichlet}({\bf 1}_f)\, ,
\end{split}
\label{eq:rho}
\end{equation}
where ${\bf 1}_f$ denotes a vector composed of $f$ ones. The ${\bf 1}_f$ parameter of the Dirichlet distribution that models $\mu_u^\rho$  allows the prior probability of users to evenly distribute in the latent space. This implies that a user may uniformly assume any kind of role, ranging from a \textit{generalist}, who is interested in all features within the latent space, to a \textit{specialist}, who focuses their attention on a specific topic.
However, once a point is selected, the user is likely to exhibit weighted preferences for a subset of features. This behavior is modeled by the Dirichlet distribution with parameters $10 \, \mu_u^\rho$, where $\mu_u^\rho$ represents the center of the user's preferences. The scaling factor of $10$ encourages users to explore a broader range of latent factors, counteracting the tendency of the Dirichlet distribution to prioritize vertices of the latent space when its parameters are in the interval $[0, 1]$, such as in the case of $\mu_u^\rho$.

A notable difference in our approach compared to~\citep{chaney} is in the sampling of $\mu_u^\rho$ that is performed for each user. This means that all generated users are evenly mapped in the latent space, thus potentially exhibiting a wide range of behaviors, from generalist to specialist. In contrast, single sampling leads to user stereotyping, where users cluster around a shared behavioral profile that acts as an unpredictable center of mass. Figure~\ref{fig:multi vs single dir} in  Appendix~\ref{sec:appendix_user_item} illustrates this difference.

Similarly to $\rho_u$, for each item $i$ we define $\alpha_i$ as:
\begin{equation}
    \alpha_i \sim \textrm{Dirichlet}(0.1 \, \mu_i^\alpha)
    \quad \mbox{and} \quad
    \mu_i^\alpha \sim \textrm{Dirichlet}({\bf 100}_f) \, ,
\label{eq:alpha}
\end{equation}
where ${\bf 100}_f$ denotes a vector composed of $f$ entries, each equal to 100. In this case, we constrain $\mu_i^\alpha$ to be at the center of the space.
In our opinion, if users have the freedom to explore the entire latent space, items should be confined to a limited subset of topics. Therefore, we model $\alpha_i$ according to a Dirichlet distribution with parameters $\mu_i^\alpha$ scaled down by a factor of $0.1$. This approach emphasizes the vertex prioritization phenomenon of the Dirichlet distribution, making it highly unlikely for items to encompass all latent factors. Again, Figure~\ref{fig:user item dir} in Appendix~\ref{sec:appendix_user_item} illustrates this aspect.


In summary, the probability of observing a preference is conditioned by a global prior component, defined as:
\begin{equation}
    P(\nu) = \prod_{u \in U} P(\nu_u) \, \prod_{i \in I} P(\nu_i) \, ,
\label{eq:prior_dir}
\end{equation}
where:
\begin{align}
P(\nu_u) &= \textrm{Dirichlet}(\rho_u; 10 \, \mu_u^{\rho}) \, \textrm{Dirichlet}(\mu_u^{\rho}; \mathbf{1}_f) \label{eq:dir_u}\\
P(\nu_i) &= \textrm{Dirichlet}(\alpha_i; 0.1 \, \mu_i^{\alpha}) \, \textrm{Dirichlet}(\mu_i^{\alpha}; \mathbf{100}_f) \, . \label{eq:dir_i}
\end{align}




\spara{Generating partitions.}\label{sec:generating-communities}
The proposed protocol can be easily adapted to generate topical user \textit{communities} and item \textit{categories}.
Specifically, we can partition $U = \{U_1, U_2, \ldots, U_c \}$, and $I = \{I_1, I_2, \ldots, I_g\}$ to represent populations of users and items that exhibit strong internal homophily and external heterogeneity across the latent space. This ensures that \ouralgo can simulate realistic topical connections between groups of similar users who share preferences for a subset of features and groups of items characterized by those features.
To disable a feature in a user community $c$ (or item category $g$), it is sufficient to set, for each user in $c$ (or each item in $g$), the related component of $\mu_u^\rho$ (or $\mu_i^\alpha$, respectively) to an arbitrarily small constant $\epsilon$. Users within the community will be evenly distributed around the remaining latent factors, while items will maintain their uneven distribution across these factors.

In terms of likelihood, the proposed partitions can be modeled by a suitable adaptation of Equation~\ref{eq:prior_dir}, to include a mixture of priors over the $\mu_u^\rho$ and $\mu_i^\alpha$ components.

\subsection{User Engagement and Item Popularity}
An additional element of flexibility can be achieved by appropriately modeling the factors of User Engagement Level and Item Popularity, which in Equation~\ref{eq:hydra} act as regularization terms for User-Item Matching. Empirical evidence shows that  engagement and popularity  typically adhere to Long-Tail distributions~\citep{Newman_2005,Clauset_2009,10.1145/3650043}. As aforesaid, our approach models these factors as mixtures of such distributions: $P_{\theta_k}$ for users and $P_{\vartheta_h}$ for items, with $k \in \{1, \ldots, K\}$ and $h \in \{1, \ldots, H\}$.
The distributions we focused on in our modeling are show in Table~\ref{tab:distributions}.

\begin{table}[ht!]
\resizebox{0.99\linewidth}{!}{
\begin{tabular}{@{}cccc@{}}
\toprule
Distribution & Parameters $\theta/\vartheta$ & $f(x$) & $C$ \\ \midrule
Power-Law & $\alpha$ & $x^{-\alpha}$ & $(\alpha - 1)x_{\min}^{\alpha - 1}$ \vspace{0.3em}\\
\arrayrulecolor{gray}\hline
\multirow{2}{*}{\begin{tabular}[c]{@{}c@{}}Power-Law with \\ Exponential cut-off\end{tabular}} & $\alpha, \lambda$ & $x^{-\alpha} e^{-\lambda x}$ & $\frac{\lambda^{1 - \alpha}}{\Gamma(1 - \alpha, \lambda x_{\min})}$ \vspace{1.3em} \\
\arrayrulecolor{gray}\hline
Exponential & $\lambda$ & $e^{-\lambda x}$ & $\lambda e^{\lambda x_{min}}$  \vspace{0.3em}\\
\arrayrulecolor{gray}\hline
Stretched Exponential & $\lambda, \beta$ & $x^{\beta - 1} e^{-\lambda x^{\beta}}$ & $\beta \lambda e^{\lambda x_{\min}^{\beta}}$ \vspace{0.3em}\\
\arrayrulecolor{gray}\hline
Log-Normal & $\mu, \sigma$ & $\frac{1}{x} \exp \left[ - \frac{(\ln x - \mu)^2}{2\sigma^2} \right]$ & $\sqrt{\frac{2}{\pi \sigma^2}} \left[ \operatorname{erfc} \left( \frac{\ln x_{\min} - \mu}{\sqrt{2} \sigma} \right) \right]^{-1}$  \vspace{0.3em}\\ \bottomrule
\end{tabular}
}
\caption{List of Long-Tail distributions employed. The complete form is $P(x) = C \, f(x)$, such that $\int_{x_{min}}^{\infty} C \, f(x) dx = 1$.}
\label{tab:distributions}
\end{table}

In practice, the modeling is based on the assumption that observations $\mathbf{z}$ and $\mathbf{y}$ adhere to long-tailed distributions, in a two step process devised as follows: 
\begin{equation}
\begin{aligned}[c]
 k \sim \mathrm{Discrete}(\pi)\\
z_u  \sim P_{\theta_k}, \; u \in U
\end{aligned}
\qquad
\begin{aligned}[c]
h  \sim \mathrm{Discrete}(\psi)\\
y_i  \sim P_{\vartheta_h}, \; i \in I
\end{aligned}
\end{equation}
Here, $P_{\theta_k}(z_u)$ represents the probability of observing the frequency $z_u$ relative to user $u$. 
The resulting mixture of distributions enables adaptation to different probability shapes, thus resulting a more faithful representation of the characteristics of real-world observations. 




The final component in modeling the User Engagement is the probability of observing user $u$ (respectively, item $i$ for Item Popularity) within a preference, which is proportional to its frequency\footnote{Here, $\mathit{Beta}'(\mu, \sigma)$ represents an alternative parametrization of the $\mathit{Beta}(\alpha, \beta)$ distribution based on the specification of mean and variance, as suggested in~\citep{chaney}. 
}:

\begin{equation}
\begin{array}{ccc}
\userprob = \mathit{Bernoulli}(\varrho_u) && \itemprob = \mathit{Bernoulli}(\varphi_i)\\
\varrho_u \sim \mathit{Beta}'(z_u / m, \sigma_u)
 && 
\varphi_i \sim \mathit{Beta}'(y_i / n, \sigma_i)
\label{eq:user-item_relevance}
\end{array}
\end{equation}
\color{black}




Notably, the engagement level $z_u$ of a user is crucial for both the component $\prod_u P_{\theta_k}(z_u)$ and the component $P(u|z_u)$. The first term represents the global likelihood of observing the degree distribution of users. By modeling it as a long-tail distribution, this probability tends to favor realistic situations where many users with low occurrence in a few highly frequent users. The second term, however, encourages the selection of frequent users among those generated when creating user-item pairs.
A similar reasoning can be done focusing on the items.




\subsection{Inference and Estimation}
The above formalization relies on a parameter set $\Theta$ that can be readily estimated from real-world data.
More formally, given $\hat{\mathcal{D}}= \{\dataset, \mathbf{z}, \mathbf{y}, \nu\}$, we aim at estimating the optimal parameter set $\Theta$. To achieve this, we employ a variational approach by defining $Q(h,k|\hat{\mathcal{D}}, \Theta)$ as a proposal probability distribution. Then, the following inequality holds:
\begin{equation}
\begin{split}
\log P(\hat{\mathcal{D}}| \Theta) &\geq \log P_\Theta(\nu) +  \log P_\nu(\dataset| \mathbf{z}, \mathbf{y}) \\
&\quad+ \sum_{k=1}^K \sum_{h=1}^H Q(h,k|\hat{\mathcal{D}}, \Theta)\Bigg\{\log (\pi_k \, \psi_h) \\
&\quad\qquad + \sum_{u \in U} \log P_{\theta_k}(z_u) + \sum_{i \in I} \log P_{\vartheta_h}(y_i)\Bigg\}\\
&\quad- \sum_{k=1}^K \sum_{h=1}^H Q(h,k|\hat{\mathcal{D}}, \Theta) \log Q(h,k|\hat{\mathcal{D}}, \Theta) \, .
\end{split}
\end{equation}
Within the inequality, the lower bound is characterized by the log-likelihood of the preferences $\mathcal{D}$, the expected log-likelihood of the frequencies $\mathbf{z}$ and $\mathbf{y}$, and the cross-entropy of the proposal distribution $Q$. Specifically, the latter serve as a regularization component, indicating that in the absence of additional information, all probability distributions are considered equally likely. 
It can be shown that the optimal $Q$, given the other components, can be expressed as:
\begin{equation}
    Q(h,k|\hat{\mathcal{D}}, \Theta) = \frac{\pi_k \, \psi_h \prod_u P_{\theta_k}(z_u)  \prod_i P_{\vartheta_h}(y_i)} {\sum_{k=1}^K \sum_{h=1}^H \pi_k \, \psi_h \prod_u P_{\theta_k}(z_u)  \prod_i P_{\vartheta_h}(y_i)} \, .
\label{eq:e-step}
\end{equation}

Using the optimal $Q$, we can then formulate a loss component over two subsets $S \subseteq \dataset$ and $\bar{S} \subseteq U \times I \setminus \dataset$, related to the other model parameters: 
\begin{equation}
\begin{split}
    &\ell(\Theta |\hat{S}) =
    -\log P_\Theta(\nu) \\
    &\qquad\quad\;\;\, -\sum_{u, i \in S}\log \left(  P_\nu(r=1 | u,i)  \, P(u|z_u) \, P(i|y_i) \right) \\
    &\qquad\quad\;\;\, -\sum_{u, i \in \bar{S}}\log \left(1 -  P_\nu(r=1 | u,i)  \, P(u|z_u) \, P(i|y_i) \right)\\
    &\qquad\quad\;\;\, -\sum_{u\in U}\log \left(P_\Theta(\varrho_u)\right) - \sum_{i\in I}\log \left(P_\Theta(\varphi_i)\right)\\
    &- \mathbb{E}_{{h,k} \sim Q} \Bigg[\log (\pi_k \, \psi_h) +\sum_{u \in U} \log P_{\theta_k}(z_u) + \sum_{i \in I} \log P_{\vartheta_h}(y_i)\Bigg]\, .
\end{split}
\label{eq:m-step}
\end{equation}
where $\hat{S} = \{S, \bar{S}, \mathbf{z}, \mathbf{y}, \nu\}$ and $\lambda_\Theta = \exp\upsilon$.

\begin{algorithm}
\caption{Parameter Estimation}\label{alg:estimalgorithm}
\small
\begin{algorithmic}[1]
\State \textbf{Input:} $\dataset, \nu$
\State \textbf{Output:} $\Theta$
\State Initialize $\Theta^{(0)}$ randomly
\State Compute $\mathbf{z}$ and $\mathbf{y}$ from \dataset
\State $\hat{\mathcal{D}} \gets \{\dataset, \mathbf{z}, \mathbf{y}, \nu\}$
\State $t \gets 0$
\While{$\Theta$ has not converged}
    \State Compute $Q(h,k|\hat{\mathcal{D}}, \Theta^{(t)})$ \Comment{\textcolor{blue}{Expectation step, Eq.~\ref{eq:e-step}}}
    \State Sample a batch $S\subseteq \dataset$ of preferences
    \State Sample a batch $\bar{S} \subseteq U \times I \setminus \dataset$ of negative pairs
    \State Update $\Theta^{(t+1)}$ by gradient descent \Comment{\textcolor{blue}{Maximization step}} \\
    $ \qquad \qquad
    \nabla_{\Theta} \ell\left(\Theta^{(t)}|S, \bar{S}, \mathbf{z}, \mathbf{y}, \nu \right) 
    $ \Comment{\textcolor{blue}{Eq.~\ref{eq:m-step}}}
    \State $t \gets t + 1$
\EndWhile
\end{algorithmic}
\end{algorithm}
An Expectation-Maximization scheme for parameter estimation can thus be devised by combining the alternating computation of $Q$  with gradient descent, as outlined in Algorithm~\ref{alg:estimalgorithm}.

\section{Synthetic Data Generation}
\label{sec:generation}
The probabilistic modeling framework outlined so far enables the definition of a generative stochastic process for preferences. The general structure of the process is detailed as follows:

\mdfsetup{skipabove=5pt,skipbelow=5pt}
\begin{mdframed}[backgroundcolor=gray!10,linecolor=gray!60!,roundcorner=0pt,linewidth=1pt,
rightline=false,
leftline=false] 
\begingroup
\fontsize{8.5pt}{10.5pt}\selectfont
\textbf{Sampling Process}
\begin{enumerate}[leftmargin=0.25cm]
    \item[1:] Draw user distribution choice $k \sim \textrm{Discrete}(\pi)$
    \item[2:] Draw item distribution choice $h \sim \textrm{Discrete}(\psi)$
    \item[3:] For $u \in U$:
    \begin{enumerate}
        \item[3.1:] Draw $\rho_u \sim \textrm{Dirichlet}(10 \, \mu_u^\rho), \;  \mbox{where } \mu_u^\rho \sim \textrm{Dirichlet}(\mathbf{1}_f)$
        \item[3.2:] Draw $\varrho_u \sim \mathit{Beta}'(z_u/m,\sigma_u), \;  \mbox{where } z_u \sim P_{\theta_k}$
    \end{enumerate}
    \item[4:] For $i \in I$:
    \begin{enumerate}
        \item[4.1:] Draw $\alpha_i \sim \textrm{Dirichlet}(0.1 \, \mu_i^\alpha), \,\ \mbox{where } \mu_i^\alpha \sim \textrm{Dirichlet}(\mathbf{100}_f)$
        \item[4.2:] Draw $\varphi_i \sim \mathit{Beta}'(y_i/n, \sigma_i), \; \mbox{where } y_i \sim P_{\vartheta_h}$
    \end{enumerate}
    \item[5:] For $u \in U$ and $i \in I$:
    \begin{enumerate}
        \item[5.1:] If $\textrm{Bernoulli} \big(\lambda\cdot \rho_u^\top \alpha_i \cdot \varrho_u \cdot \varphi_i ) \big)$ then:
        \begin{enumerate}
            \item[5.1.1] Generate $(u,i)$
        \end{enumerate}
    \end{enumerate}
\end{enumerate}
\endgroup
\color{black}
\label{box:theoretical_model}
\end{mdframed}

Taking inspiration from Equation 3, the first step of the algorithm is to select the distributions $ k \in \{1, \ldots, K\} $ and $ h \in \{1, \ldots, H\} $ that govern User Engagement Level and Item Popularity, respectively, through two Discrete distributions (lines 1-2). Subsequently, the process characterizes users and items (lines 3-4), by (a) choosing the latent features $(\mu_u^\rho, \rho_u)$ and $(\mu_i^\alpha, \alpha_i)$ ; (b) sampling Engagement Level and the Popularity from the chosen distribution $k$ and $h$; devising their probability of occurrence \userprob and \itemprob,  as indicated in Equation~\ref{eq:user-item_relevance}.
Finally, in line 5, leveraging Equations~\ref{eq:user-item v1} and \ref{eq:hydra}, a Bernoulli trial is conducted for each user $u \in U$ and each item $i \in I$. The User-Item Matching, i.e., the probability of generating the user-item pair $(u, i)$, is given by the combination of the sampled components.  If the trial succeeds, the pair $(u, i)$ then is generated.


\spara{\ouralgo.}
Based on the general schema described so far, we tweak the probabilistic generative model to adapt the synthetic generation of datasets of implicit preferences for recommendation systems. The resulting algorithm, \ouralgo offers extensive control options, enabling the simulation of various user behaviors and item characteristics.
The pseudo-code of \ouralgo is shown in Algorithm~\ref{alg:hydra}. In the following, we discuss its control options.

\begin{algorithm}[ht]
\caption{\ouralgo}\label{alg:hydra}
\begin{algorithmic}[1]
\footnotesize
\State \textbf{Input:}
user set $U$ and item set $I$; user communities $U_1,\ldots, U_c$ and item categories $I_1, \ldots, I_g$ with associated features; 
Distribution priors $\pi$, $\psi$;
number of latent features $f$ and disabling factor $\varepsilon$; 
noise weight $\delta$ and prior $\chi$,  density control weights $\lambda$, $\zeta$, $\xi$, minimum frequency $\tau$; variability of the \textrm{Beta}' distributions $\beta$, $\sigma_u$, and $\sigma_i$.
\State \textbf{Output:} $\dataset$.
\State Draw user distribution $k \sim \textrm{Categorical}(\pi)$
\State Draw item distribution $h \sim \textrm{Categorical}(\psi)$
\For{$s \in \{1, \ldots, c\}$ and $u \in U_s$}
    \State $\hat{f} = \texttt{FilterPrior}(\textbf{1}_f, s,  \varepsilon)$ 
    \Comment{\textcolor{blue}{Generate communities}}
    \State Sample user hyper-parameters $\mu^\rho_u \sim \textrm{Dirichlet}(\hat{f})$
    \State Sample user latent factors $\rho_u \sim \textrm{Dirichlet}(10 \, \mu^\rho_u)$
    \State Draw the user interactions $z_u \sim P_{\theta_k}$
    \State Define the \emph{User Engagement Level} $\userprob$ as $\varrho_u \sim \mathit{Beta}'(z_u / m, \sigma_u)$
\EndFor
\For{$s \in \{1, \ldots, g\}$ and $i \in I_s$}
    \State $\hat{f} = \texttt{FilterPrior}(\textbf{100}_f, s, \varepsilon)$ 
    \Comment{\textcolor{blue}{Generate categories}}
    \State Sample item hyper-parameters $\mu_i^\alpha \sim \textrm{Dirichlet}(\hat{f})$
    \State Sample item latent factors $\alpha_i \sim \textrm{Dirichlet}(0.1 \, \mu^\alpha_i)$
    \State Draw the item interactions $y_i \sim P_{\vartheta_h}$
    \State Define the \emph{Item Popularity} $\itemprob$ as $\varphi_i \sim \mathit{Beta}'(y_i / n, \sigma_i)$
\EndFor
\State $\dataset \gets \emptyset$
\State Sample $\omega \sim \textrm{Beta}'(\chi, \beta)$ \Comment{\textcolor{blue}{Noise factor}}
\For{$u \in U$}
    \State ${\dataset}_u \gets \emptyset$
    \While{$|{\dataset}_u| < \tau$} \Comment{\textcolor{blue}{Imposing a minimum frequency}}
        \For{$i \in I$}
            \State Define the \emph{User-Item Matching} $T_{u,i} \gets \rho_u^\top \, \alpha_i$ 
            \State $T_{u,i} \gets \delta \, T_{u,i} + (1-\delta) \, \omega$ \Comment{\textcolor{blue}{Adding noise}}
            \State $T_{u,i} \gets \lambda \cdot T_{u,i} \cdot \varrho_u^\zeta \cdot \varphi_i^\xi$
            \Comment{\textcolor{blue}{Densification/Sparsification}}
            \State Sample $\eta_{u,i} \sim \mathit{Beta}'(T_{ui}, \beta)$
            \Comment{\textcolor{blue}{Adding variability}}
            \If{$\textrm{Bernoulli}(\eta_{u,i})$}
                \State ${\dataset}_u \gets {\dataset}_u \cup \{(u, i)\}$
            \EndIf
        \EndFor
    \EndWhile
    \State $\dataset \gets \dataset \cup {\dataset}_u$
\EndFor
\end{algorithmic}
\label{alg:generation}
\end{algorithm}


\spara{Generation of communities and categories.} Manipulating the priors of the Dirichlet distributions allows for the generation of user communities and item categories. Given a partition of $U$ and a partition of $I$, lines 3 and 4 of the Sampling Process can be extended by filtering the hyper-parameters $\mathbf{1}_f$ and $\mathbf{100}_f$. By utilizing a custom function, referred to as \texttt{FilterPrior} (lines 6 and 13 in \ouralgo), which maps elements of the partition to distinct subsets of latent features, it becomes possible to generate well-defined communities and categories. This is achieved by specializing them on the features whose prior is not set to $\varepsilon$, an arbitrarily small constant. This approach ensures that the generated data reflects the desired structural properties and relationships within the latent space.

\spara{Noise injection.} The User-Item Matching component represents the degree of appreciation a user has for an item. However, real-world navigation through a catalog of items often involves some degree of randomness. To simulate this phenomenon, \ouralgo introduces noise into the matching process by incorporating a noise term, $\omega$, into Equation~\ref{eq:user-item v1} using a linear combination parameterized by $0 \leq \delta \leq 1$ (line 26). The value of $\omega$ is sampled (line 20) from a \textrm{Beta} distribution, parametrized to guarantee mean (the hyper-parameter $\chi$) and variance $\beta$. 

\spara{User minimum frequency.} \ouralgo ensures that every user in $U$ is involved in at least a minimum number of observations, denoted as $\tau$. Line 23 of the algorithm mandates that for each user $u$, the generation of pairs $(u, i)$ with $i \in I$ is repeated until the user's preference list, $\mathcal{D}_u$, has a size greater than or equal to $\tau$. This guarantees sufficient representation of each user in $\mathcal{D}$.

\spara{Density manipulation.} To adjust the density of the preference matrix generated from the user-item observations (line 27), \ouralgo introduces a coefficient $\lambda$ and two exponents $\zeta$ and $\xi$. Specifically, as already mentioned, $\lambda$ scales the User-Item Matching, either amplifying ($\lambda > 1$) or diminishing ($\lambda < 1$) the overall probability of occurrence. In particular, higher values of $\lambda$ increase the likelihood of generating more successes in the Bernoulli trial (line 29). The exponents $\zeta$ and $\xi$ adjust the distributions of users and items, respectively. Higher values amplify the importance of user engagement and item popularity over the feature matching. By contrast,  values approaching zero result in a dominance of the bernoulli trial governed by the feature matching. 

\spara{Increased variability.} A further element of randomness is introduced in line 28, where, rather than directly relying on a Bernoulli trial, the process incorporates an element of variance $\beta$ within the sampling process.

A summary notation table for both the modeling and the resulting sampling process can be found in the Appendix.

\section{Experimental Evaluation}
\label{sec:experiments}

To evaluate the effectiveness of \ouralgo in generating preference data, we address two key research questions:
\begin{itemize}
    \item[\textbf{RQ1.}] How flexible is \ouralgo in generating  user-item \textit{interactions}? Does it provide control on the resulting \textit{data distributions}?
    \item[\textbf{RQ2.}] Are the interactions generated by \ouralgo realistic? Can the framework \textit{replicate} the properties of real benchmark datasets?
\end{itemize}

\noindent To address \textbf{RQ1}, we conducted experiments to demonstrate the flexibility of the proposed hyper-parametrization in controlling the distributions. The experimental code is publicly available\footnote{\url{https://github.com/SimoneMungari/HYDRA}}.

\spara{Interactions among user communities and item categories.} 
To conduct this set of experiments, we generate a synthetic dataset featuring two user communities and two item categories. For simplicity, we divide both users and items into two equal-sized groups, denoted as communities $U_1$ and $U_2$ and categories $I_1$ and $I_2$, respectively. The communities and categories are correlated: users in $U_1$ primarily prefer items in category $I_1$, while users in $U_2$ exhibit a preference for items in $I_2$.

\begin{figure}[ht!]
    \centering
    \begin{subfigure}[b]{0.49\linewidth}
    \centering
    \includegraphics[width=\textwidth]{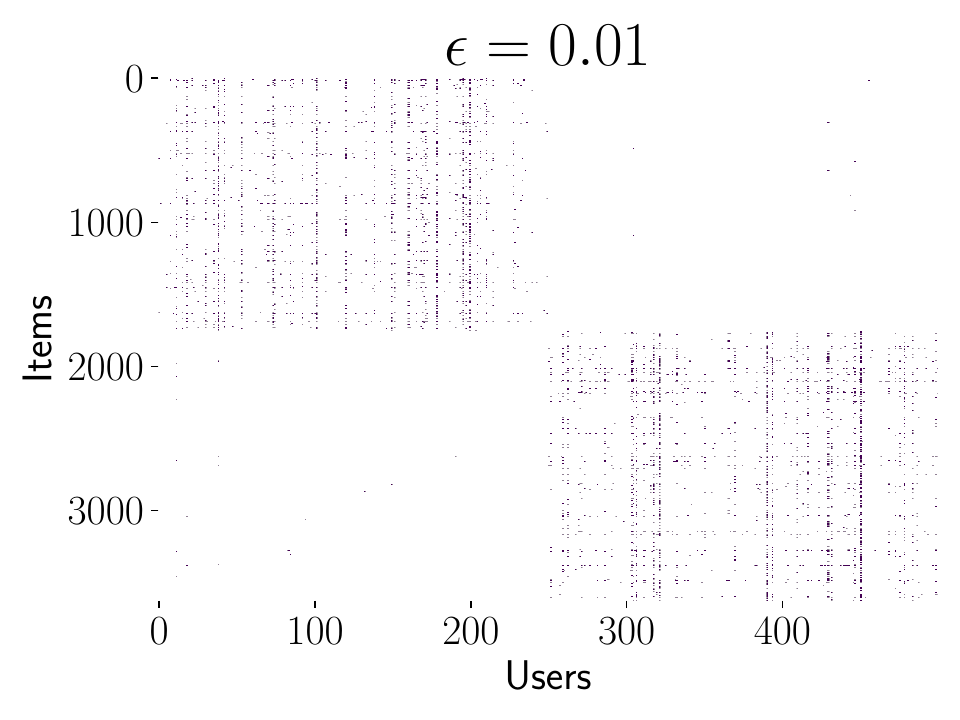}
    \caption{}
    \label{fig:interactions_matrix_0.01}
    \end{subfigure}
    \begin{subfigure}[b]{0.49\linewidth}
    \centering
    \includegraphics[width=\textwidth]{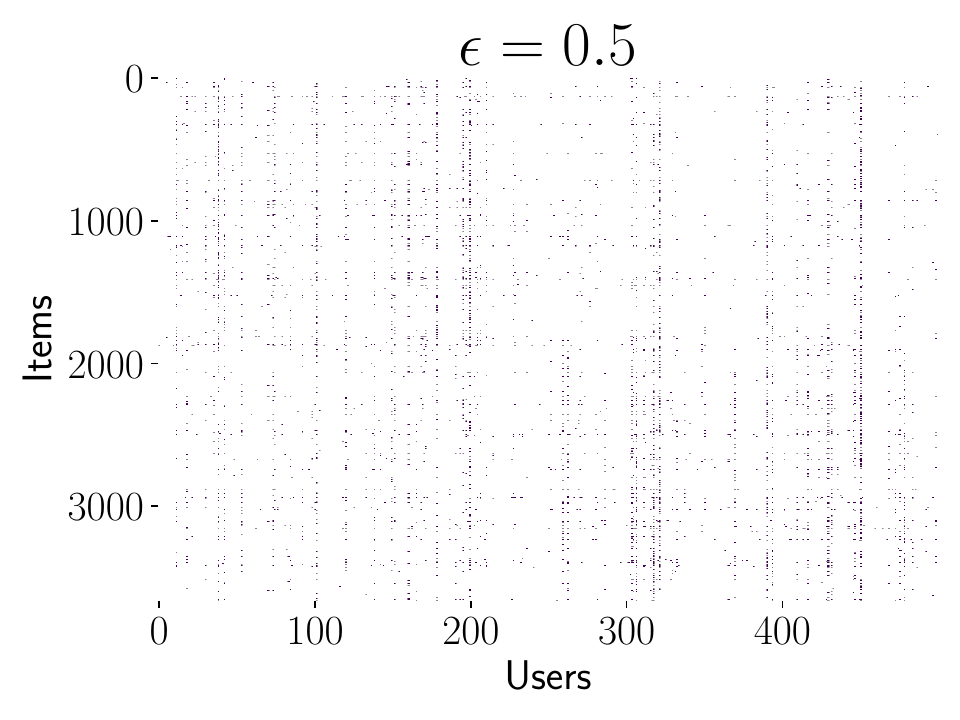}
    \caption{}
    \label{fig:interactions_matrix_0.5}
    \end{subfigure}
    \caption{Visualization of the user-item interaction matrix by varying the $\varepsilon$ parameter. The X-axis reports the users, while the Y-axis represents the items. A dot in the position ($u, i$) indicates the user $u$ interacted with item $i$.}
    \label{fig:interactions_matrix}
\end{figure}

\begin{figure}[ht!]
    \centering
    \includegraphics[width=.3\columnwidth]{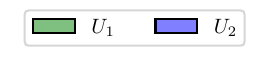}\\
    \includegraphics[width=.48\columnwidth]{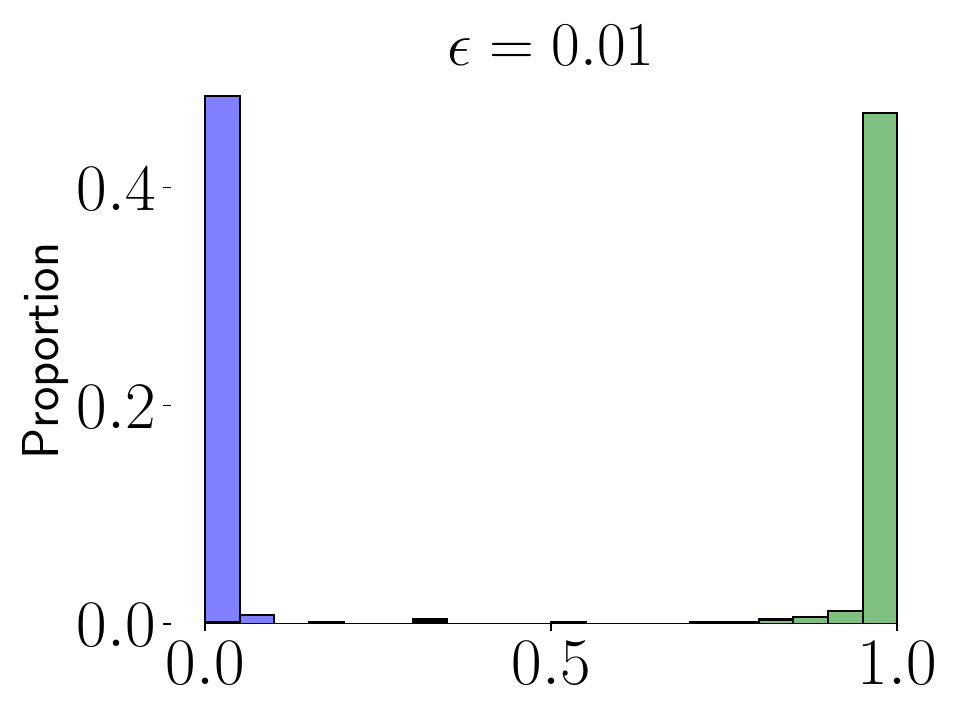}
    \includegraphics[width=.48\columnwidth]{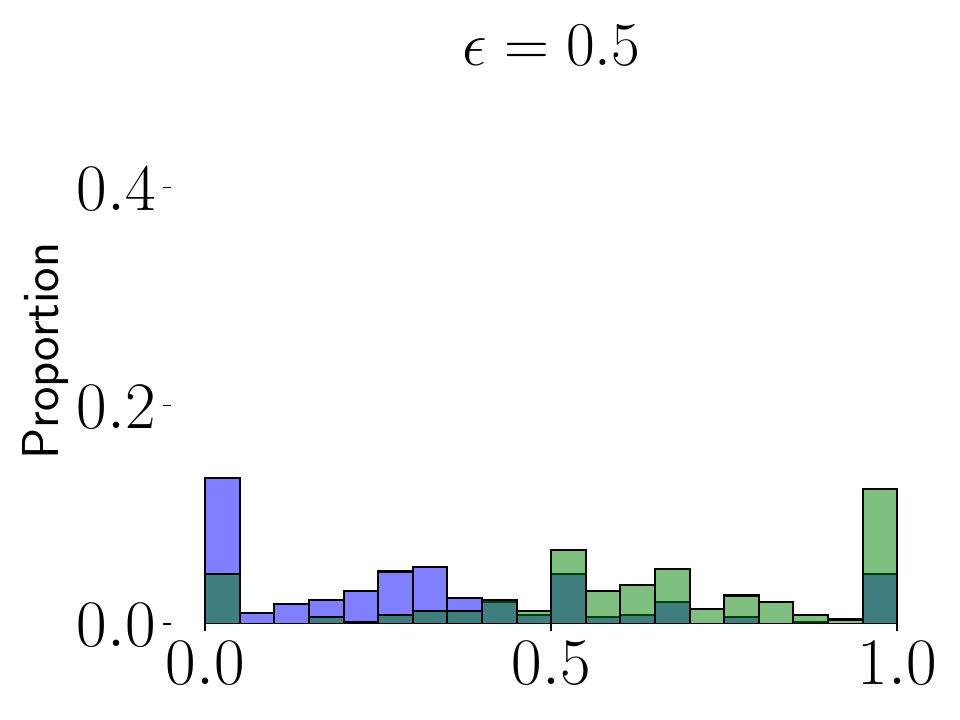}
    \caption{Histograms of user interactions with a specific topic of interest. The X-axis represents the percentage of items in $I_1$ within the users history. The Y-axis shows the proportion of users having that percentage.} \label{fig:category_percentage_distributions}
\end{figure}

Figure~\ref{fig:interactions_matrix} illustrates the impact of the $\varepsilon$ parameter on the visibility of communities and categories within the interaction matrix. When $\varepsilon$ is close to 0 (Figure~\ref{fig:interactions_matrix_0.01}), the matrix is dominated by two strongly connected components. As $\varepsilon$ increases, the matrix becomes more dispersed, eventually reaching a state of complete homogeneity, as seen in Figure~\ref{fig:interactions_matrix_0.5}. Additionally, Figure~\ref{fig:category_percentage_distributions} presents histograms of user interactions within category $I_1$ for each community, demonstrating a transition from fully clustered to entirely overlapping interactions.

\begin{figure*}[!ht]
    \centering
    \begin{tabular}{ccccc}
    \includegraphics[width=.17\textwidth]{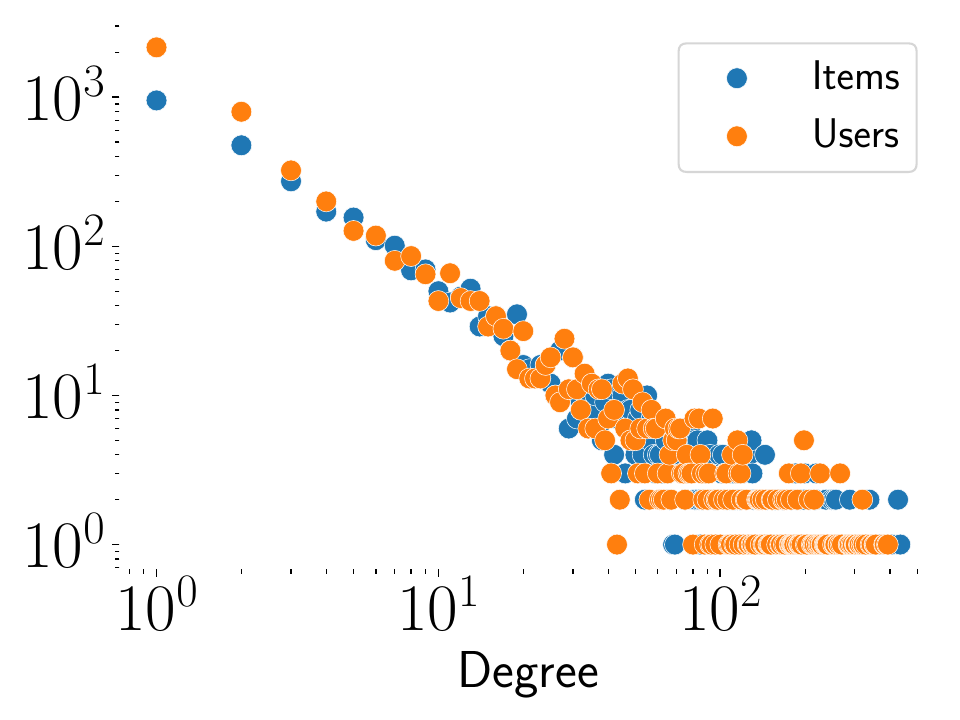}
    & 
    \includegraphics[width=.17\textwidth]{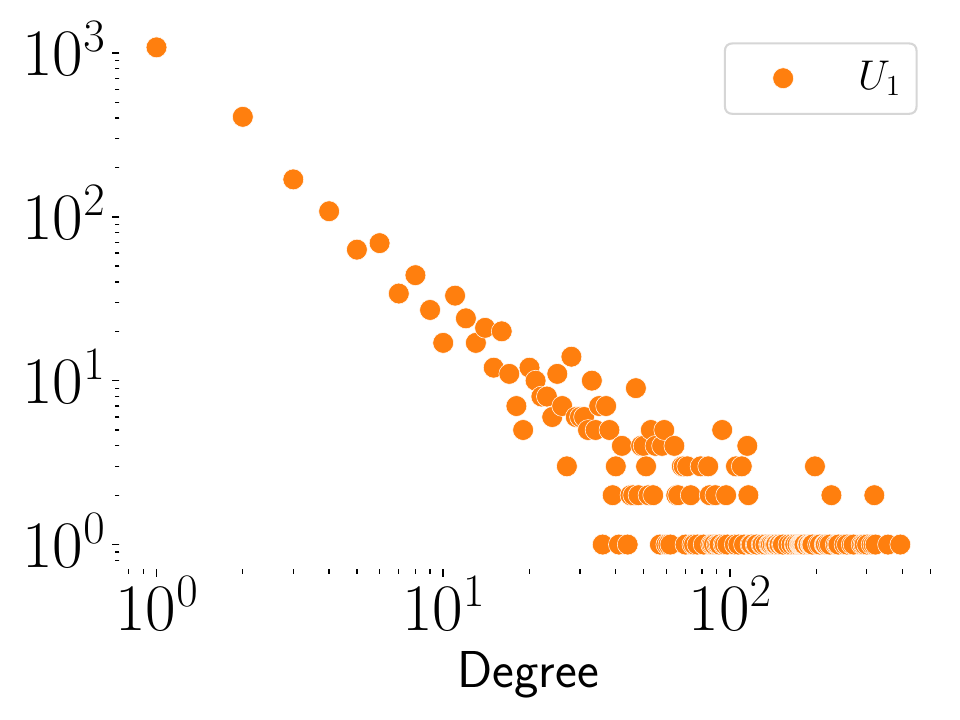}
    & 
    \includegraphics[width=.17\textwidth]{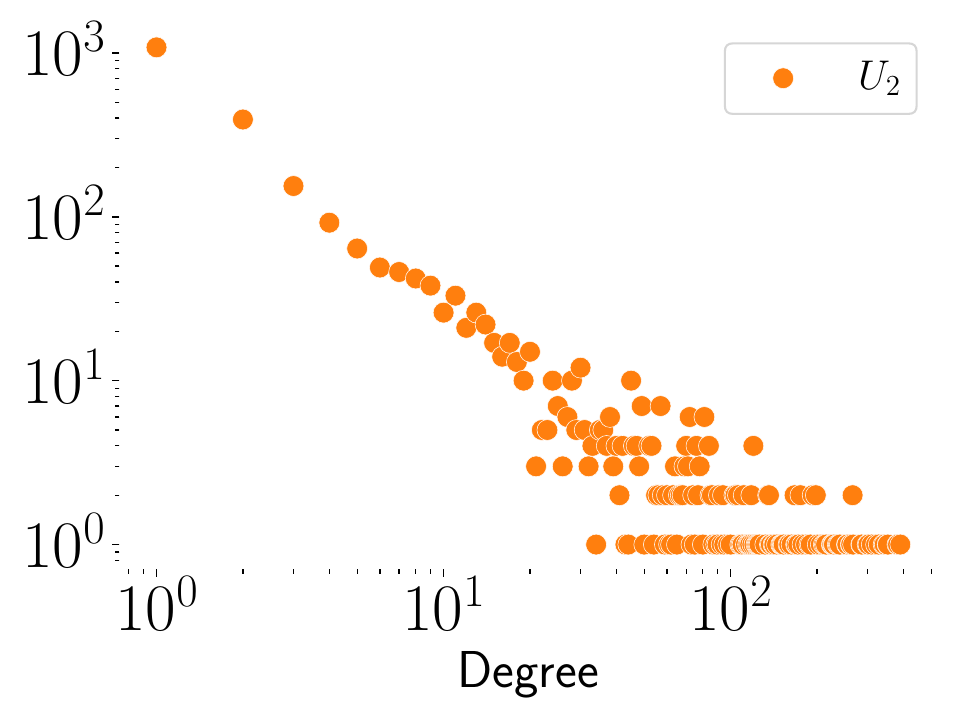}
    &
    \includegraphics[width=.17\textwidth]{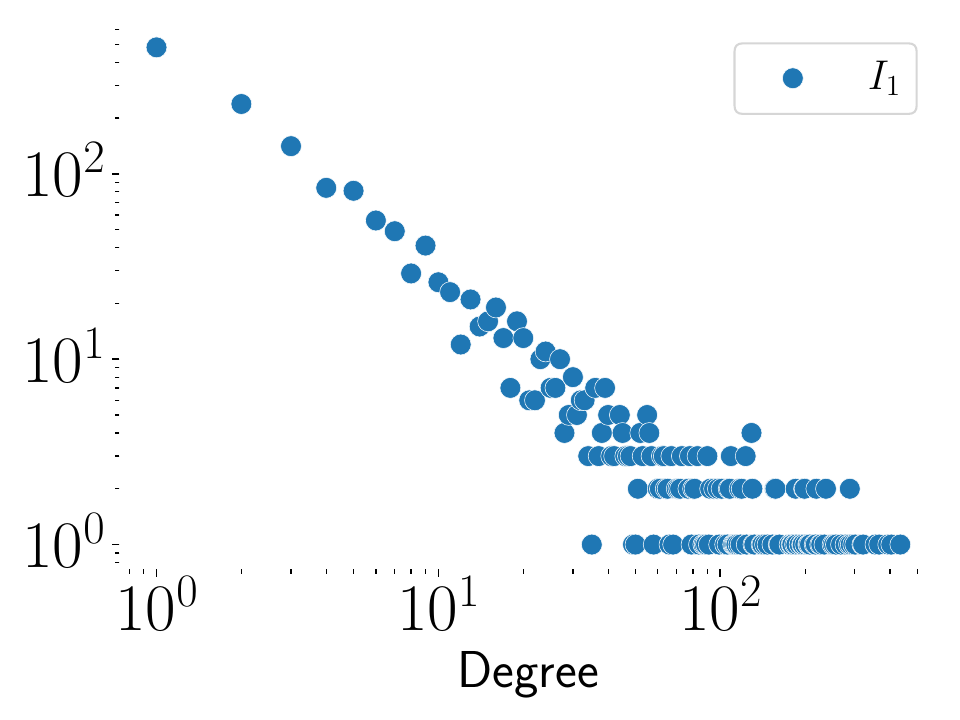}
    & 
    \includegraphics[width=.17\textwidth]{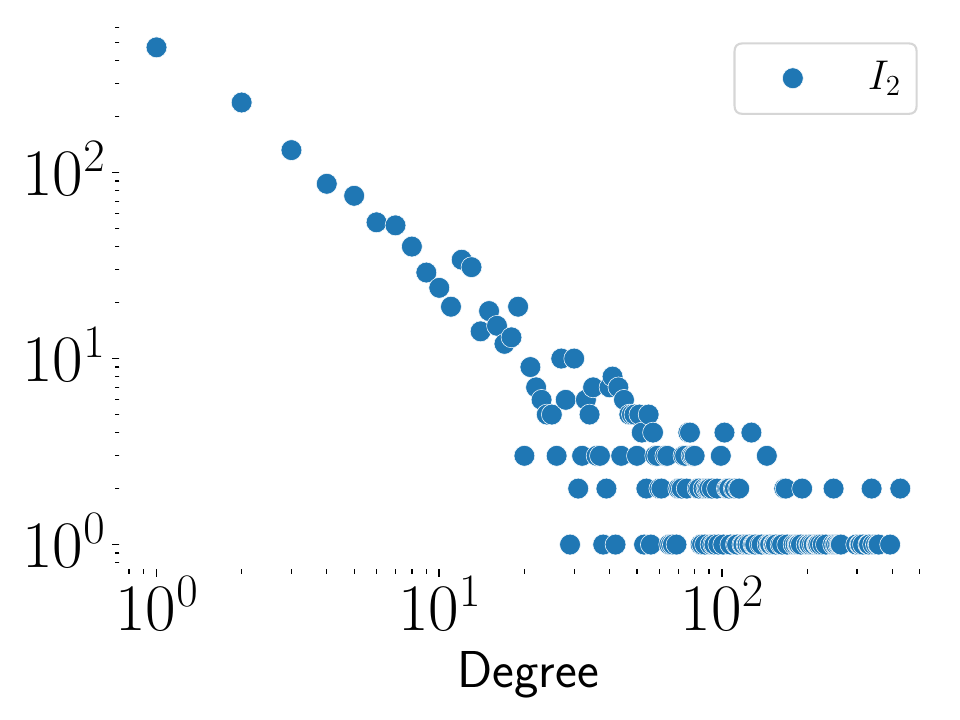}
    \end{tabular}
    \caption{User/item degree distributions for the partitions obtained with $\varepsilon=.01$. The first graph shows the global degree distributions. Graphs 2 and 3 focus on the user communities  $U_1$ and $U_2$, whereas 4 and 5 on item categories $I_1$ and $I_2$.}
    \label{fig:subpopulation_degree_distributions}
\end{figure*}

Finally, Figure~\ref{fig:subpopulation_degree_distributions} presents the degree distributions for users and items in a log-log scale. Specifically, it compares the distributions of the entire dataset with those of the generated communities and categories. Remarkably, the degree distributions of the subpopulations maintain the same characteristics as those of the complete dataset, demonstrating the effectiveness of our method in preserving the overall structure while allowing for distinct community and category formations.

\begin{figure}[!ht]
\centering
   \includegraphics[width=.3\columnwidth]{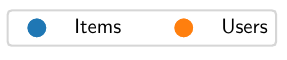}\\
    \centering
    \begin{subfigure}[b]{0.3\columnwidth}
    \centering
    \includegraphics[width=\columnwidth]{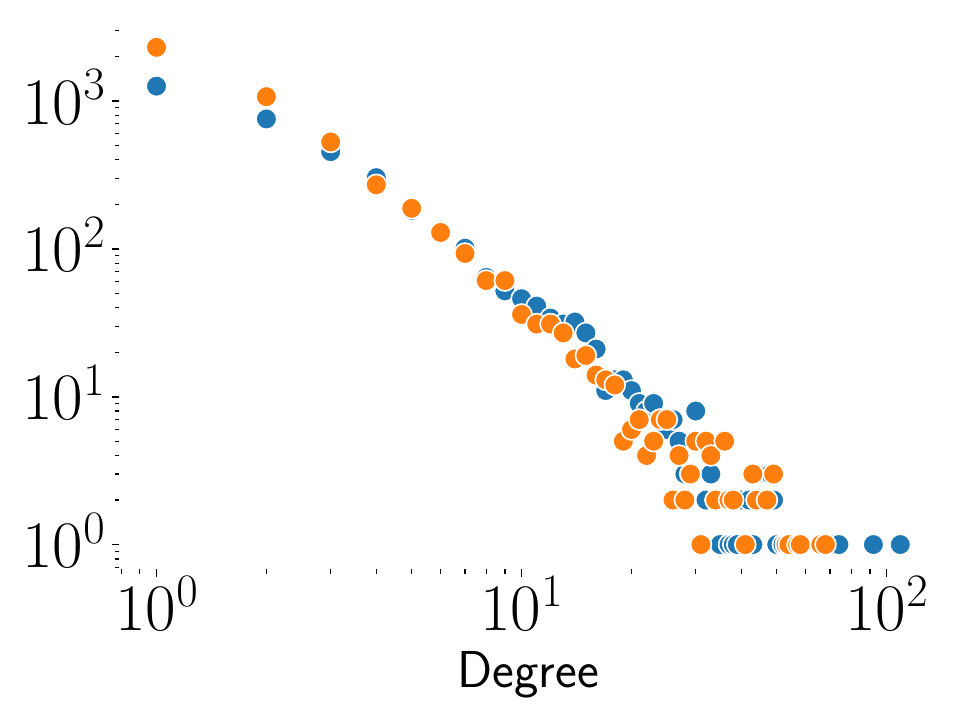}
    \caption{}
    \label{fig:long_tail_distributions_a}
    \end{subfigure}
    \begin{subfigure}[b]{0.3\columnwidth}
    \centering
    \includegraphics[width=\columnwidth]{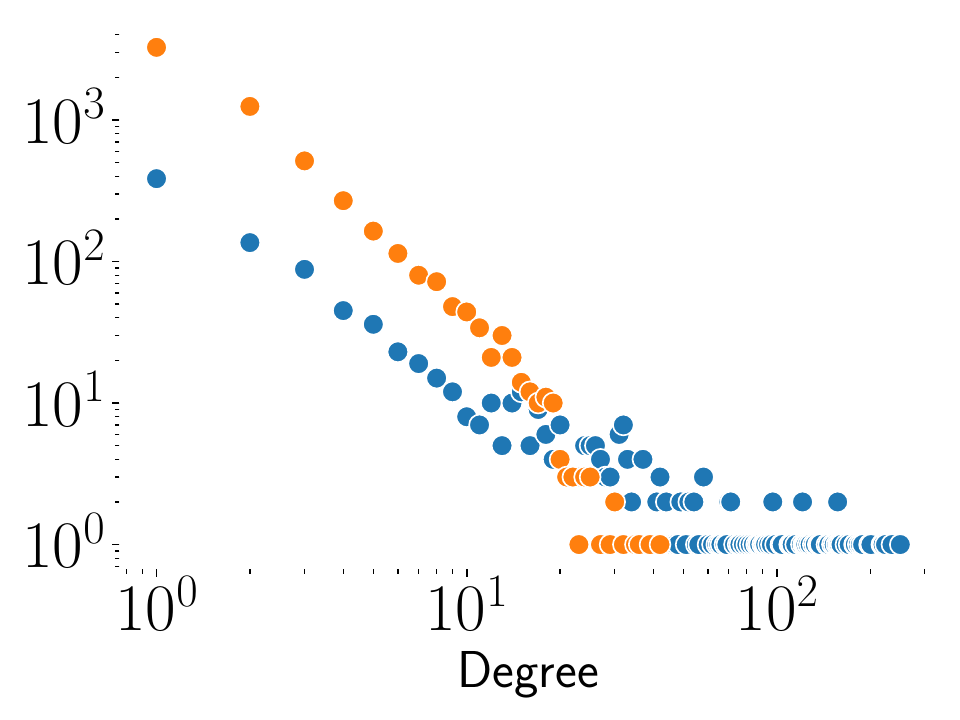}
    \caption{}
    \label{fig:long_tail_distributions_b}
    \end{subfigure}
    
    \begin{subfigure}[b]{0.3\columnwidth}
    \centering
    \includegraphics[width=\columnwidth]{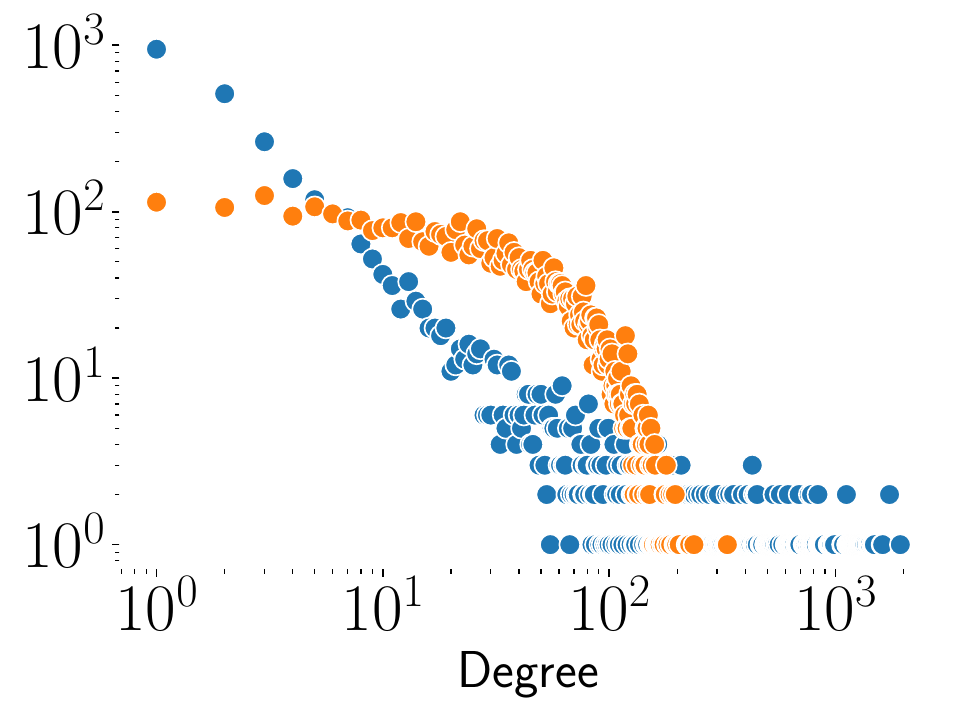}
    \caption{}
    \label{fig:long_tail_distributions_c}
    \end{subfigure}
    \begin{subfigure}[b]{0.3\columnwidth}
    \centering
    \includegraphics[width=\columnwidth]{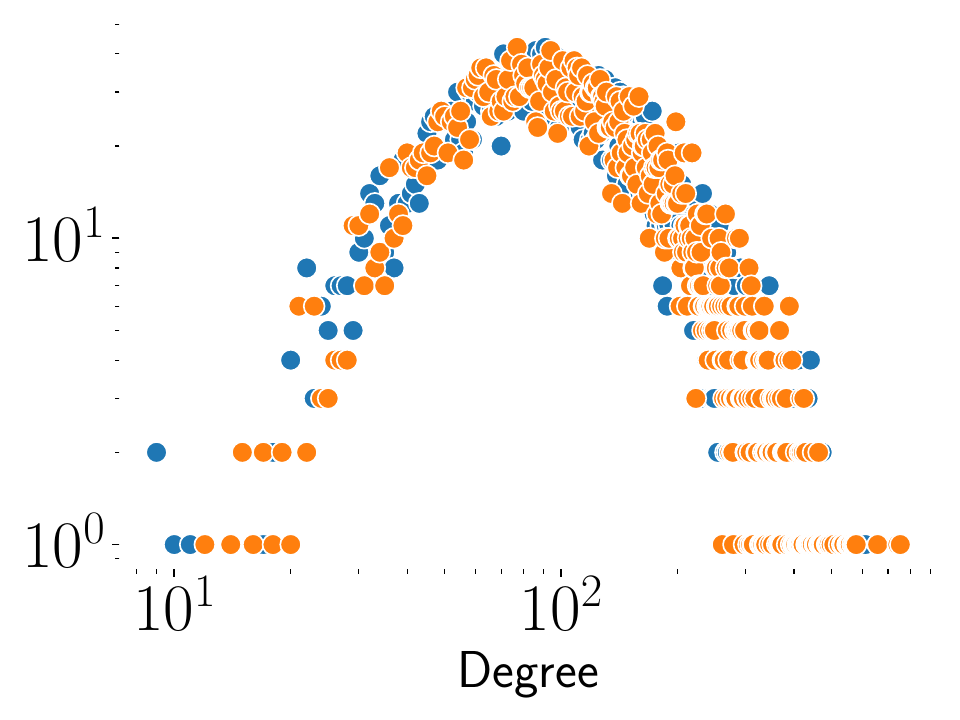}
    \caption{}
    \label{fig:long_tail_distributions_d}
    \end{subfigure}
    \caption{Degree distributions with the following priors: (a) Power-Law with exponential cut-off for users and items; (b) Power-Law with exponential cut-off for users and Power-Law for items; (c) Stretched Exponential for users, Power-Law for items; (d) Log-Normal distribution for users and items.}
    \label{fig:long_tail_distributions}
\end{figure}

\spara{Tweaking Distributions.} 
We validate the flexibility of \ouralgo by demonstrating that the degree distribution functions as a \emph{plug-and-play} component. In Figure~\ref{fig:long_tail_distributions}, we present the results of generation experiments where various combinations of prior distributions were applied. These combinations showcase the model's flexibility by producing a wide range of distribution shapes, emphasizing the precise control possible in the data generation process.

\begin{figure}[!ht]
    \centering
    \begin{subfigure}[b]{0.3\columnwidth}
        \centering
        \includegraphics[width=\columnwidth]{fig/legend.pdf}
    \end{subfigure}
    \\
    \begin{subfigure}[b]{0.3\columnwidth}
    \centering
    \includegraphics[width=\columnwidth]{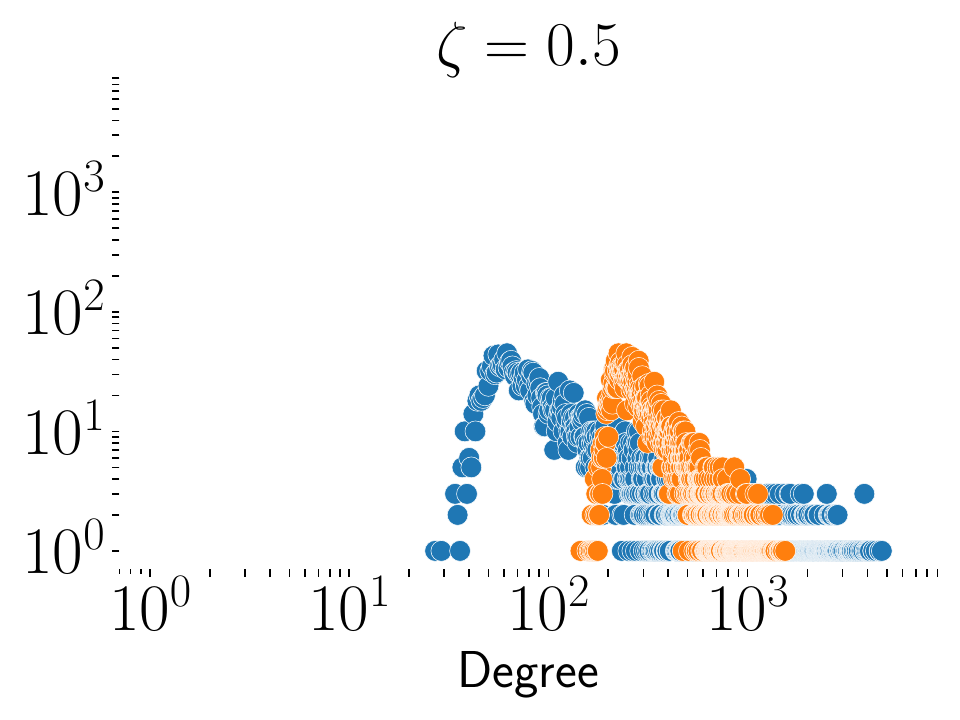}
    \end{subfigure}
    \begin{subfigure}[b]{0.3\columnwidth}
    \centering
    \includegraphics[width=\columnwidth]{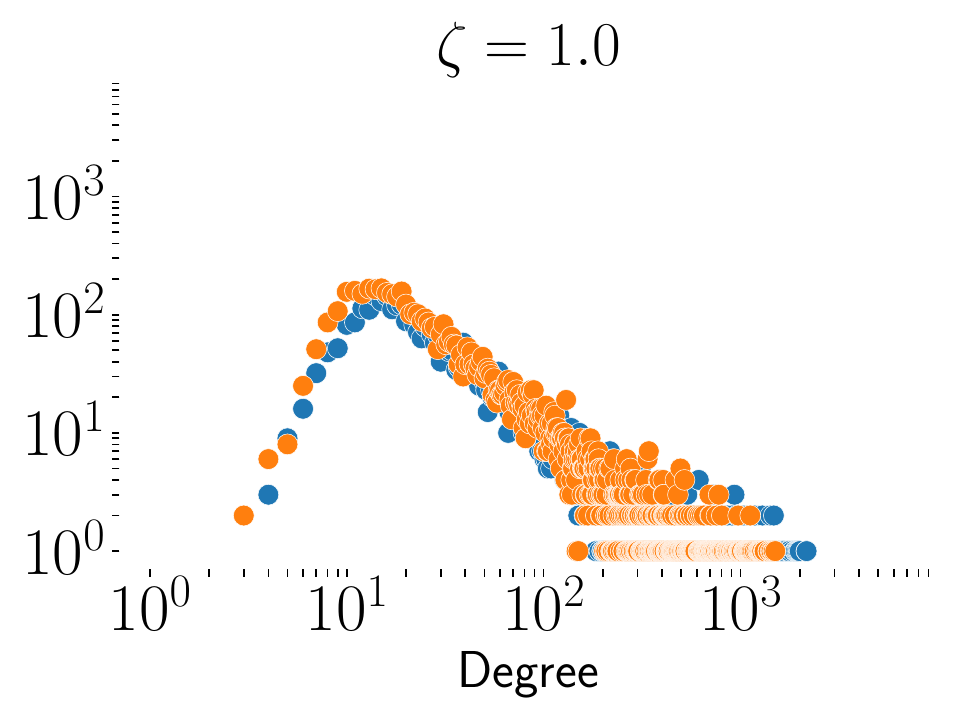}
    \end{subfigure}
    \begin{subfigure}[b]{0.3\columnwidth}
    \centering
    \includegraphics[width=\columnwidth]{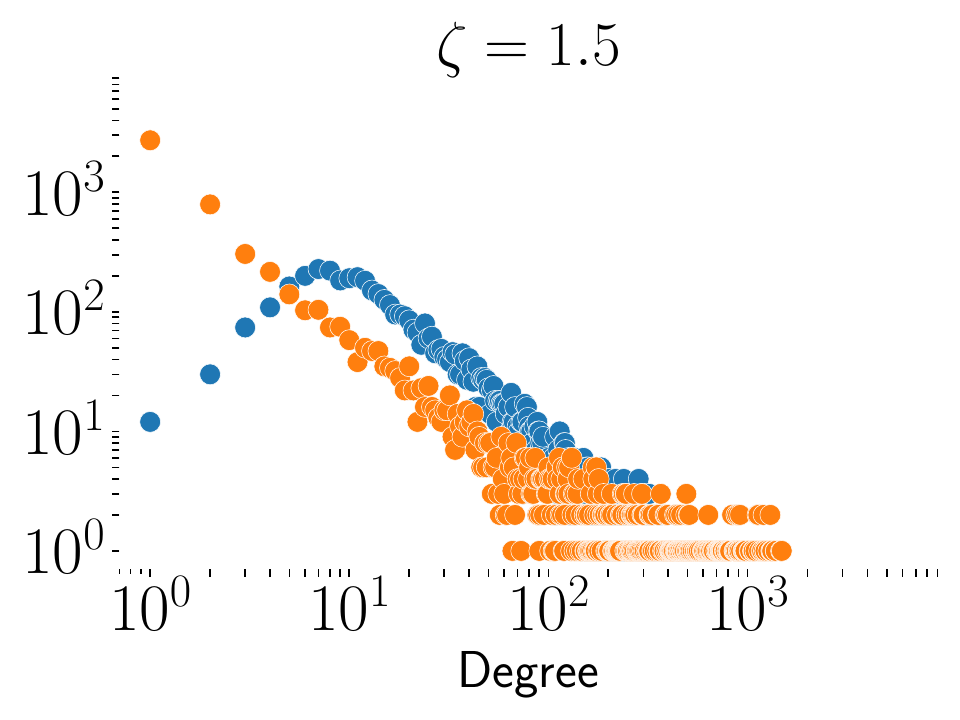}
    \end{subfigure}
    \begin{subfigure}[b]{0.3\columnwidth}
    \centering
    \includegraphics[width=\columnwidth]{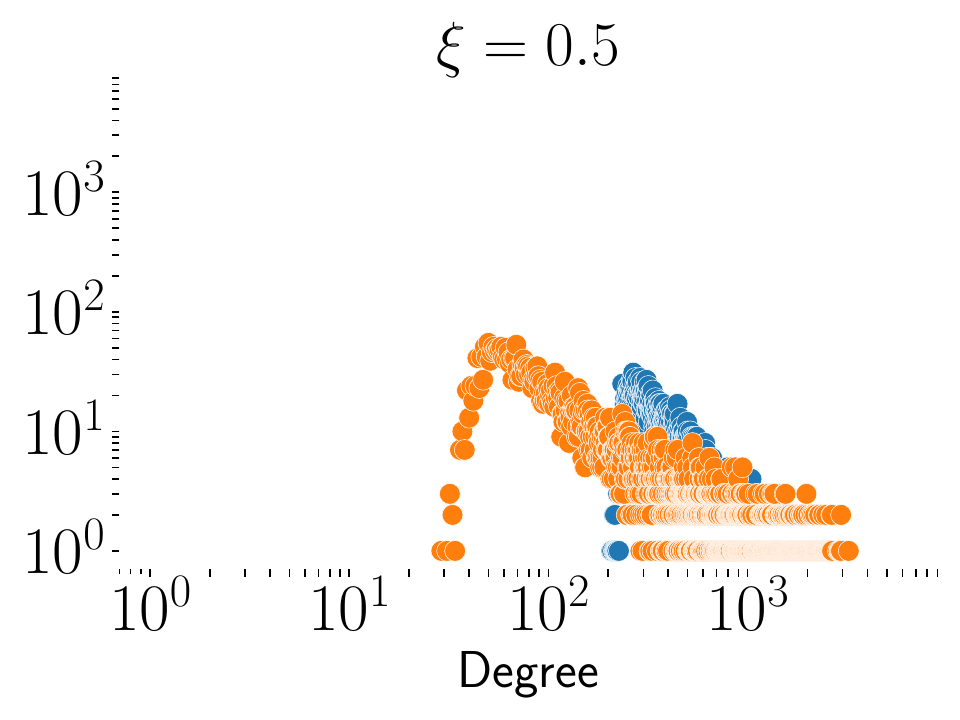}
    \end{subfigure}
    \begin{subfigure}[b]{0.3\columnwidth}
    \centering
    \includegraphics[width=\columnwidth]{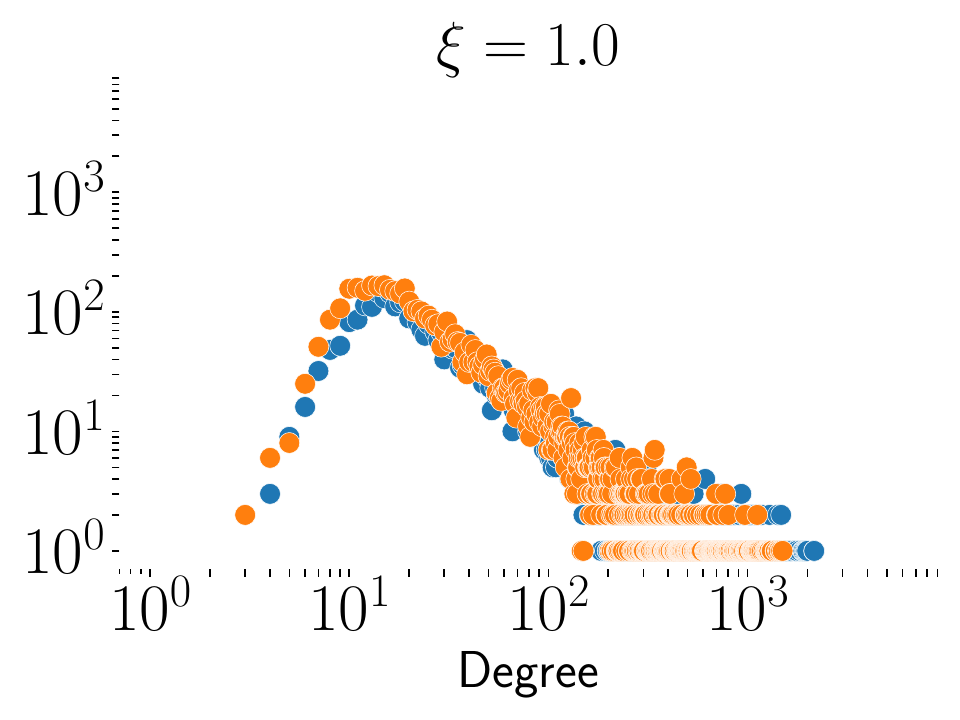}
    \end{subfigure}
    \begin{subfigure}[b]{0.3\columnwidth}
    \centering
    \includegraphics[width=\columnwidth]{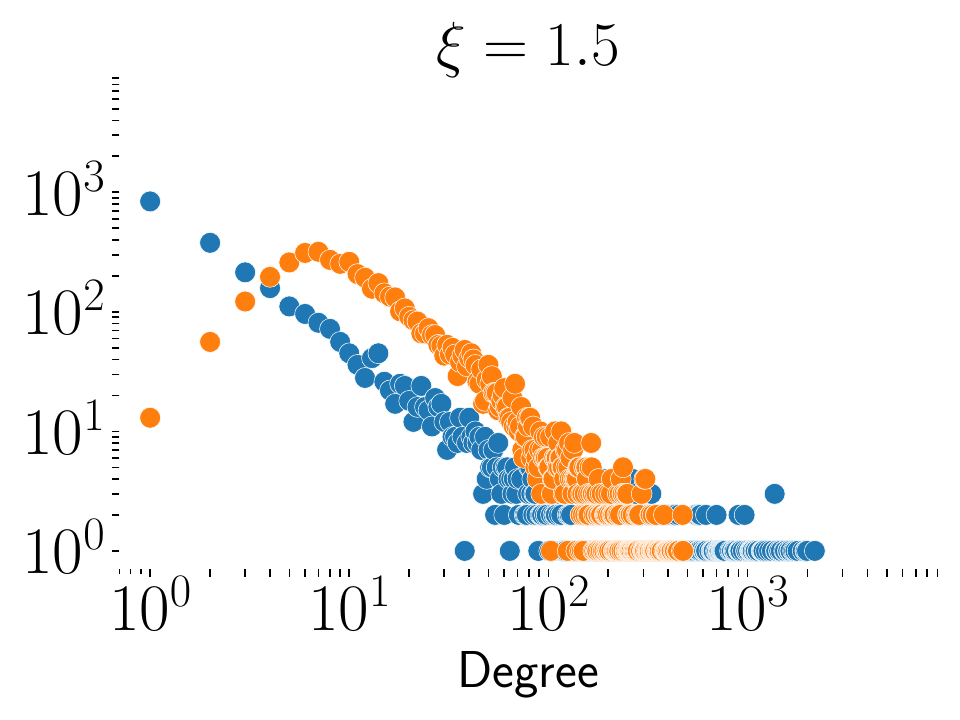}
    \end{subfigure}
    \centering
    \begin{subfigure}[b]{0.3\columnwidth}
    \centering
    \includegraphics[width=\columnwidth]{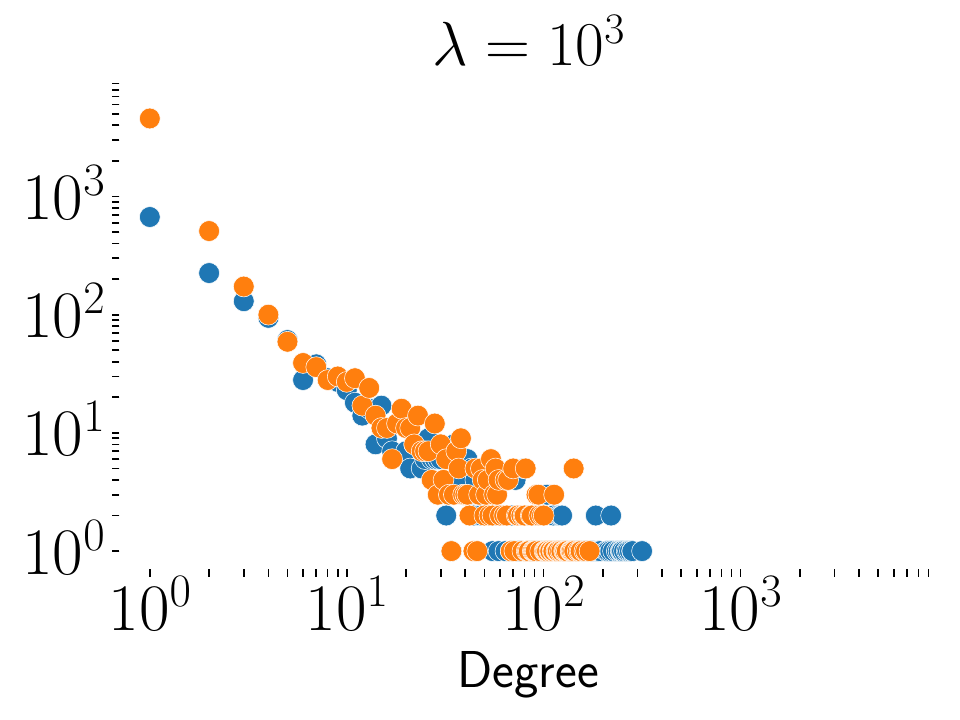}
    \end{subfigure}
    \begin{subfigure}[b]{0.3\columnwidth}
    \centering
    \includegraphics[width=\columnwidth]{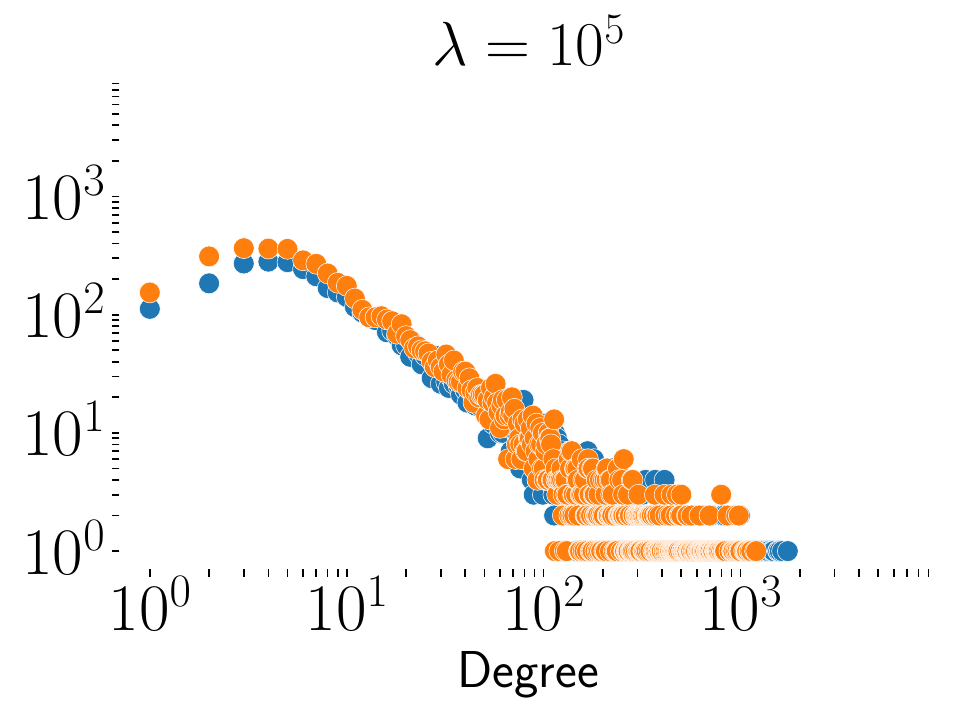}
    \end{subfigure}
    \begin{subfigure}[b]{0.3\columnwidth}
    \centering
    \includegraphics[width=\columnwidth]{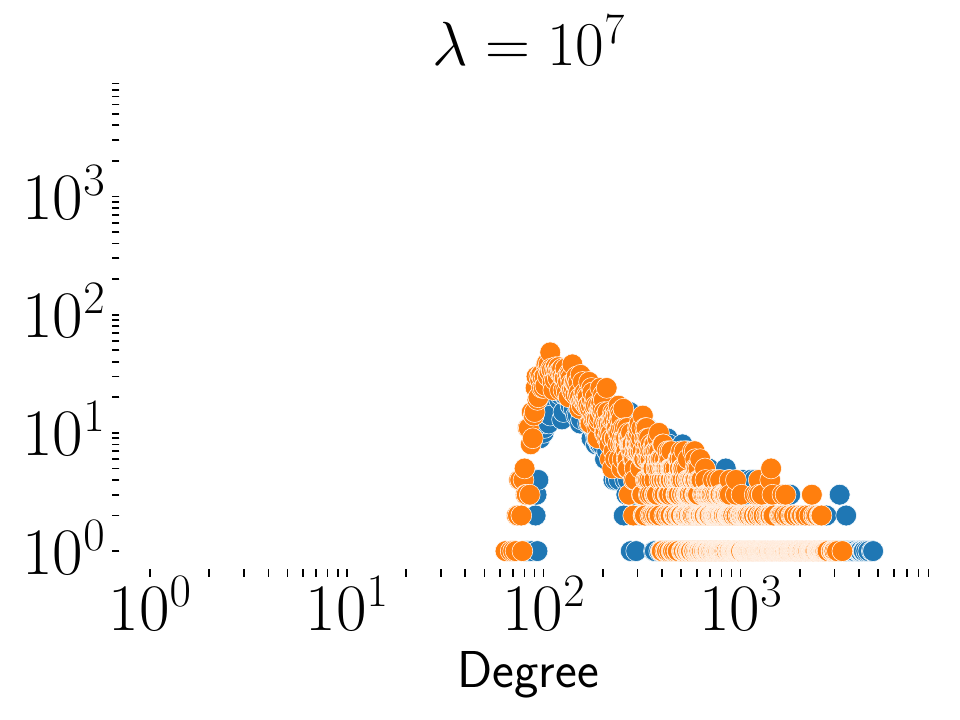}
    \end{subfigure}
    \caption{Effects of the $\zeta, \xi$ and $\lambda$ hyper-parameters on the distributions of the generated data.}
    \label{fig:varying-hypers}
\end{figure}

We also conduct a sensitivity analysis of the weighting hyperparameters $\zeta$, $\xi$, and $\lambda$ to evaluate their impact on the generated preferences and the resulting degree distributions. In this analysis, the frequency distributions are regulated by power-law, although similar behaviors are observed for the other priors listed in Table~\ref{tab:distributions}. Figure~\ref{fig:varying-hypers} illustrates the effects of these hyperparameters. Specifically, as $\zeta$ and $\xi$ approach $0$, the corresponding user engagement and item popularity distributions exhibit a bell-shaped curve, driven by the influence of the \textrm{Beta} distribution on user-item interactions. By contrast, higher values of these parameters sharpen the distribution, producing shapes that range from power-law to more peaked distributions.
Regarding $\lambda$, increasing its value scales the overall sampling probability, thereby influencing the density of both user and item distributions.

Additional experiments exploring the effects of $\zeta$, $\xi$, and $\lambda$ on the degree distributions are provided in the Appendix.

\begin{figure}[!ht]
    \centering
    \begin{subfigure}[b]{0.3\columnwidth}
        \centering
        \includegraphics[width=\columnwidth]{fig/legend.pdf}
    \end{subfigure}

    \begin{subfigure}[b]{0.3\columnwidth}
    \centering
    \includegraphics[width=\columnwidth]{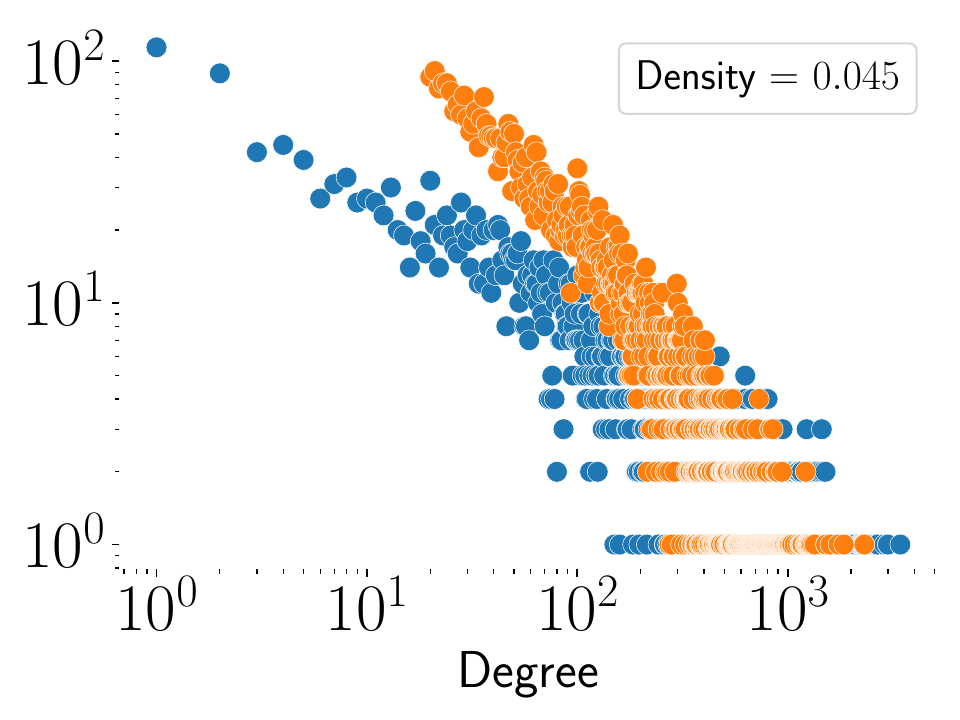}
    \caption{}
    \label{fig:original_movielens}
    \end{subfigure}
    \begin{subfigure}[b]{0.3\columnwidth}
    \centering
    \includegraphics[width=\columnwidth]{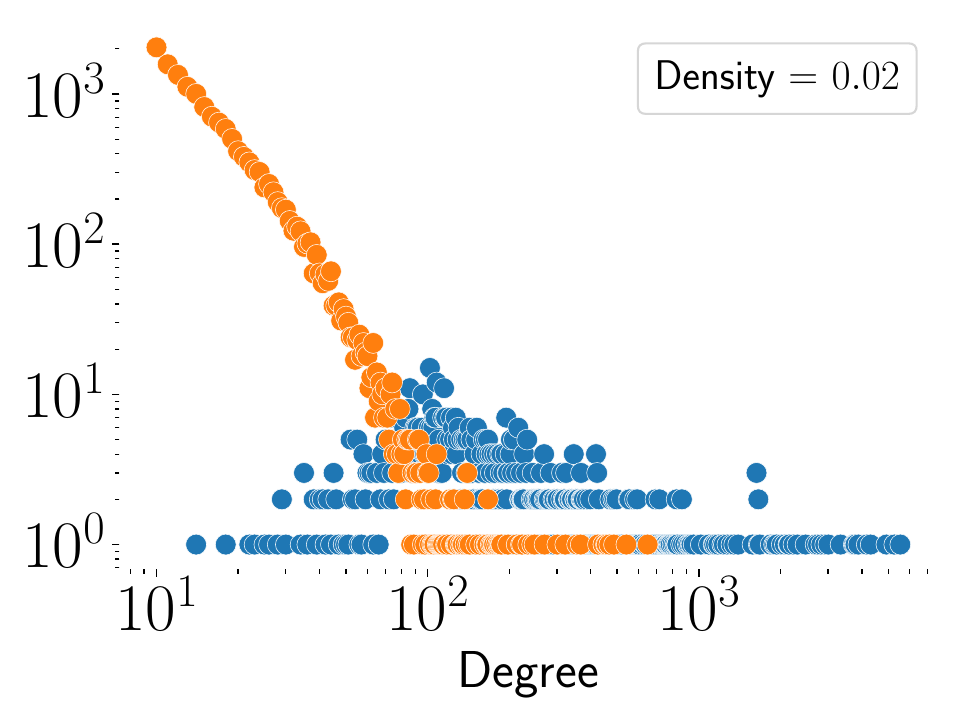}
    \caption{}
    \label{fig:original_yahoo-r3}
    \end{subfigure}
    \begin{subfigure}[b]{0.3\columnwidth}
    \centering
    \includegraphics[width=\columnwidth]{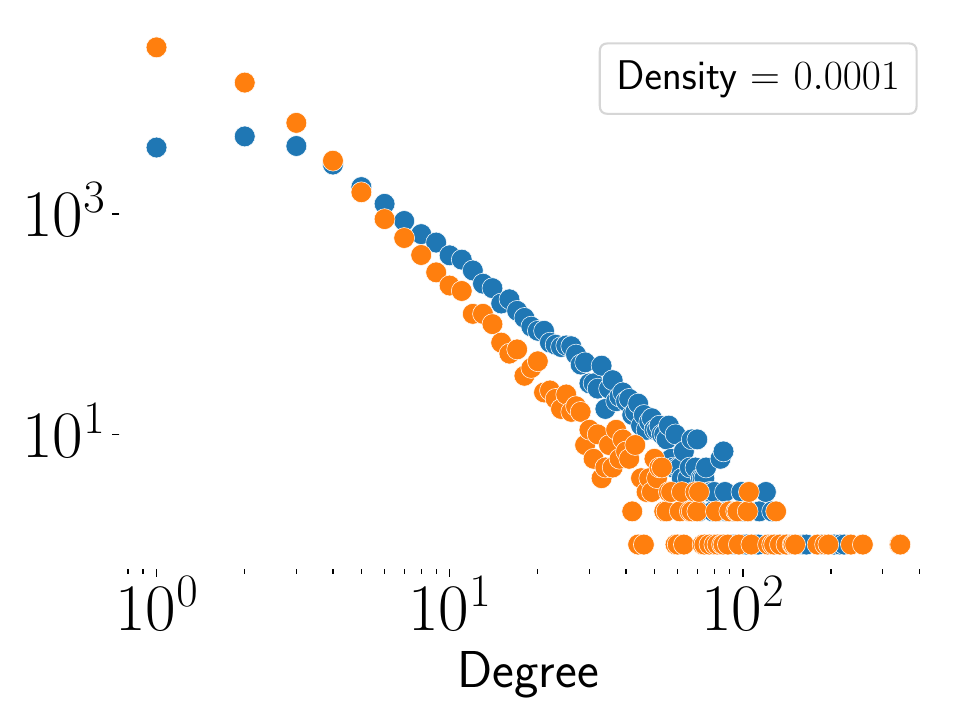}
    \caption{}
    \label{fig:original_amazon}
    \end{subfigure}
    
    \begin{subfigure}[b]{0.3\columnwidth}
    \centering
    \includegraphics[width=\columnwidth]{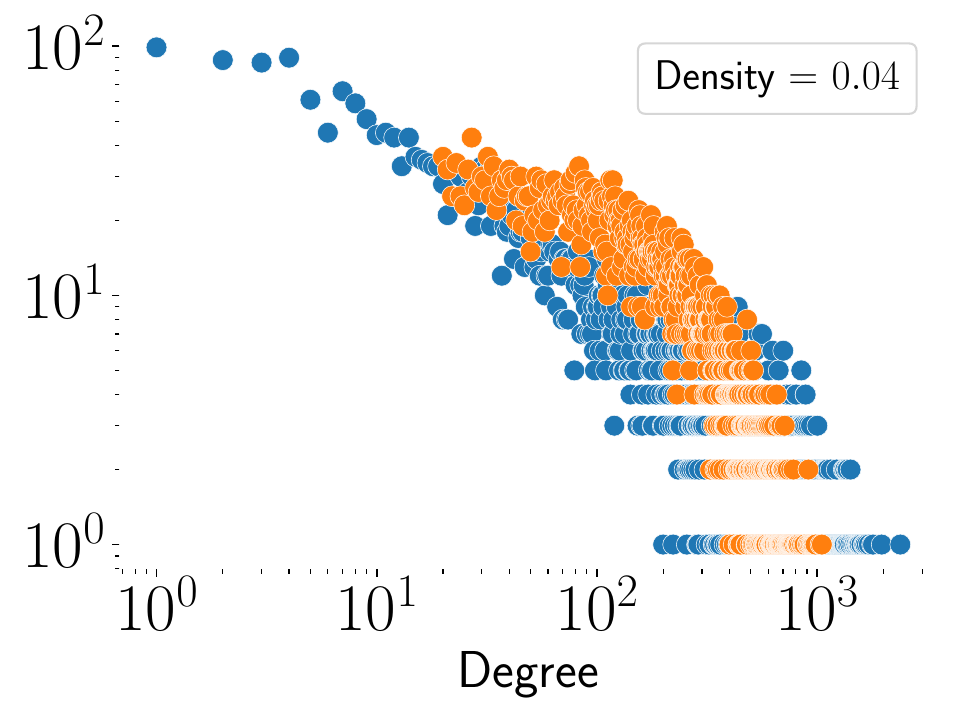}
    \caption{}
    \label{fig:synthetic_movielens}
    \end{subfigure}
    \begin{subfigure}[b]{0.3\columnwidth}
    \centering
    \includegraphics[width=\columnwidth]{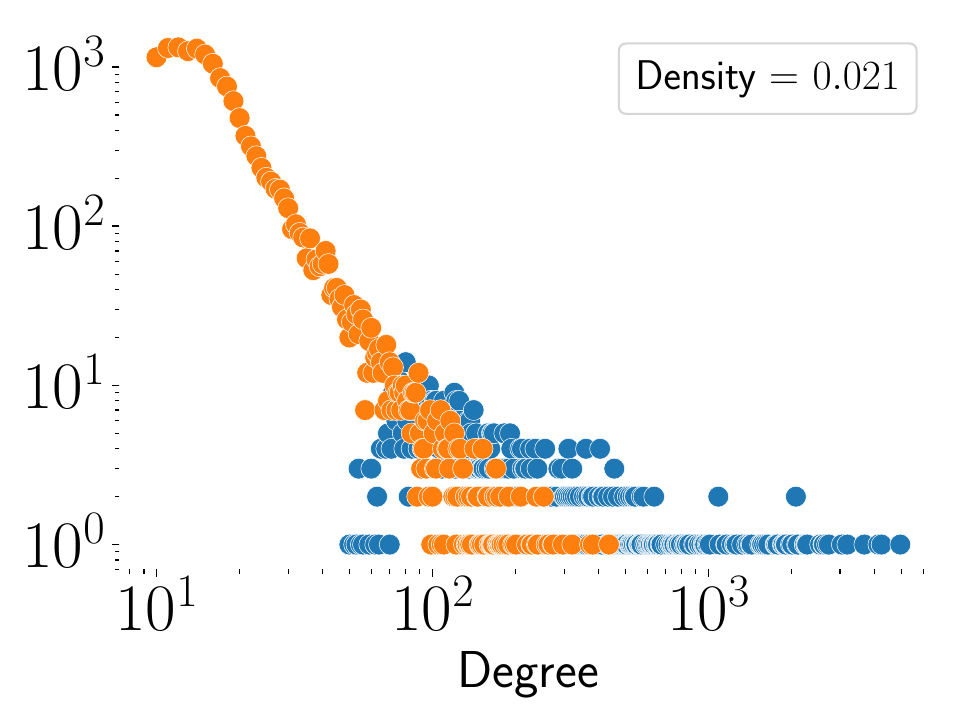}
    \caption{}
    \label{fig:synthetic_yahoo-r3}
    \end{subfigure}
    \begin{subfigure}[b]{0.3\columnwidth}
    \centering
    \includegraphics[width=\columnwidth]{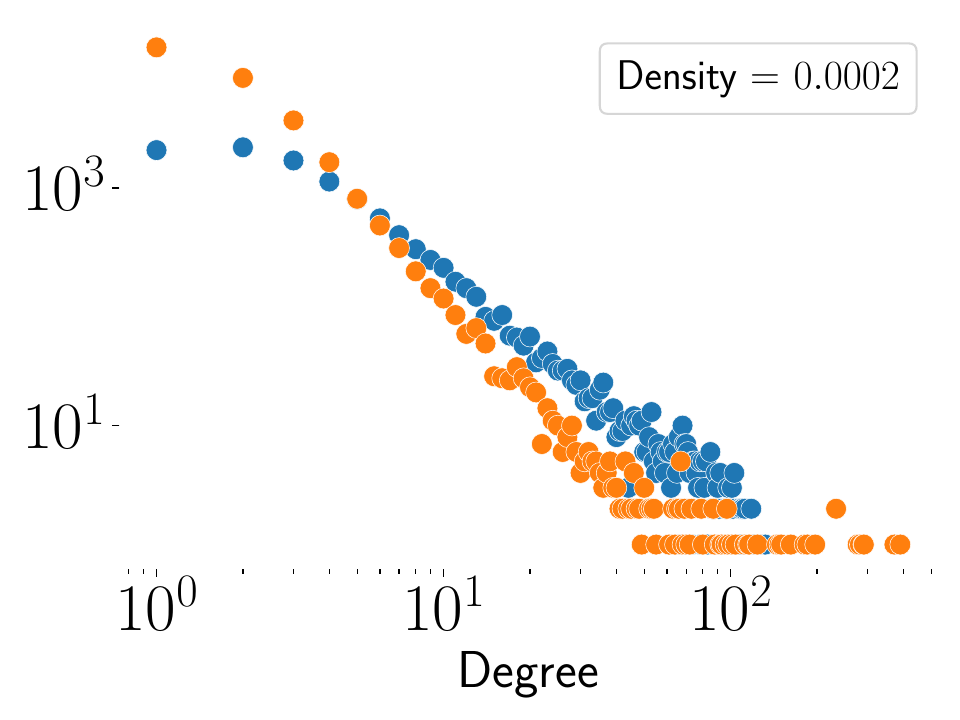}
    \caption{}
    \label{fig:synthetic_amazon}
    \end{subfigure}
    
    \caption{Real and Synthetic degree distributions of users and items. The top row depicts Movielens-1M (a), Yahoo-R3 (b) and Amazon (c), while the bottom row shows the corresponding synthetic samples ((c), (d) and (e)), generated by \ouralgo.
    }
    \label{fig:simulated_yahoo-r3_movielens}
\end{figure}

\vspace{0.5em}
\noindent To answer \textbf{RQ2}, we perform a final set of experiments to test the ability of generating realistic datasets. We consider two aspects. 

\spara{Comparing distributions.} In this study, we utilize existing datasets as reference benchmarks, focusing on Movielens-1M\footnote{\url{https://grouplens.org/datasets/movielens/1m/}}, which pertains to movie ratings, Yahoo! Music\footnote{\url{https://webscope.sandbox.yahoo.com/catalog.php?datatype=r}} (Yahoo-R3), which reflects user preferences collected during normal interactions with the Yahoo Music service, and Amazon Musical Instruments Reviews\footnote{\url{https://www.kaggle.com/datasets/eswarchandt/amazon-music-reviews}}, which contains comments and ratings published by users w.r.t. musical instruments.

The results of the data generation process are illustrated in Figure~\ref{fig:simulated_yahoo-r3_movielens}. The top row shows the actual distributions from the datasets Movielens-1M, Yahoo-R3, and Amazon, while the bottom row presents the synthetic samples generated by \ouralgo. As evident from the comparison, the generated samples closely replicate the structural properties of the original datasets. Details regarding the parameters for the synthetic samples generation can be found in the Appendix.

\spara{Comparing benchmarking capabilities.} To evaluate the quality of the generated data, we compared the performance of several recommendation algorithms on both the real and synthetic datasets. The underlying hypothesis is that if the synthetic data accurately mirrors the structural properties of the real data, the performance trends of recommendation algorithms should be similar across both datasets. We tested a suite of algorithms, including \textit{ItemKNN}~\citep{ItemKNN}, \textit{MultiVAE}~\citep{MULTIVAE}, \textit{BPR}~\citep{BPR}, \textit{Neural Matrix Factorization (NeuMF)}~\citep{NeuMF}, and a popularity-based baseline (\textit{Pop}). Similar recommendation scores and algorithm rankings across the datasets would indicate high similarity in data distributions.

In these experiments, we produced multiple instances of the generated data by varying the coefficient $\delta$, to emulate noisy interactions that typically can occur within real preferences. We trained the recommendation algorithms on both  real and synthetic datasets\footnote{For training, we adopted the Recbole library~\citep{recbole}}, and evaluated their performance in terms of Recall and Hit-Rate. Results with cut-off  $10$ are shown in Figure~\ref{fig:comparison_movielens_yahoo}, with additional results for other cut-offs provided in the Appendix.

The findings indicate a strong alignment between the performance metrics on the real and synthesized datasets, with the exception of cases where excessive noise was introduced. Most importantly, the ranking of algorithms on the synthetic data is consistent with that on the real data, demonstrating that \ouralgo can effectively work as a proxy for real data in benchmarking scenarios.

    
\begin{figure}[ht!]
    \centering
    \includegraphics[width=\linewidth]{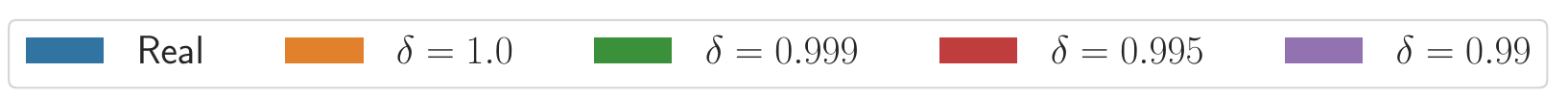}\\
    \includegraphics[width=.4\linewidth]{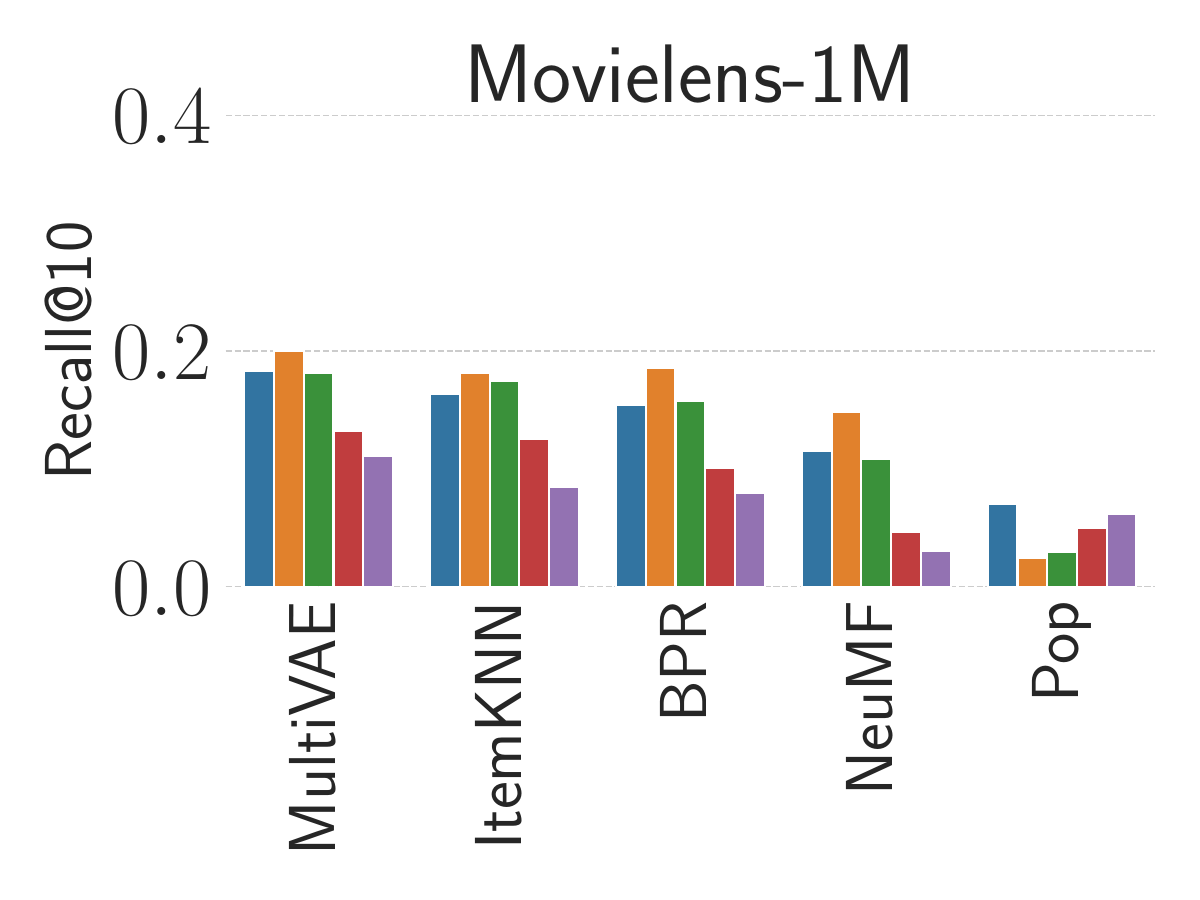}
    \includegraphics[width=.4\linewidth]{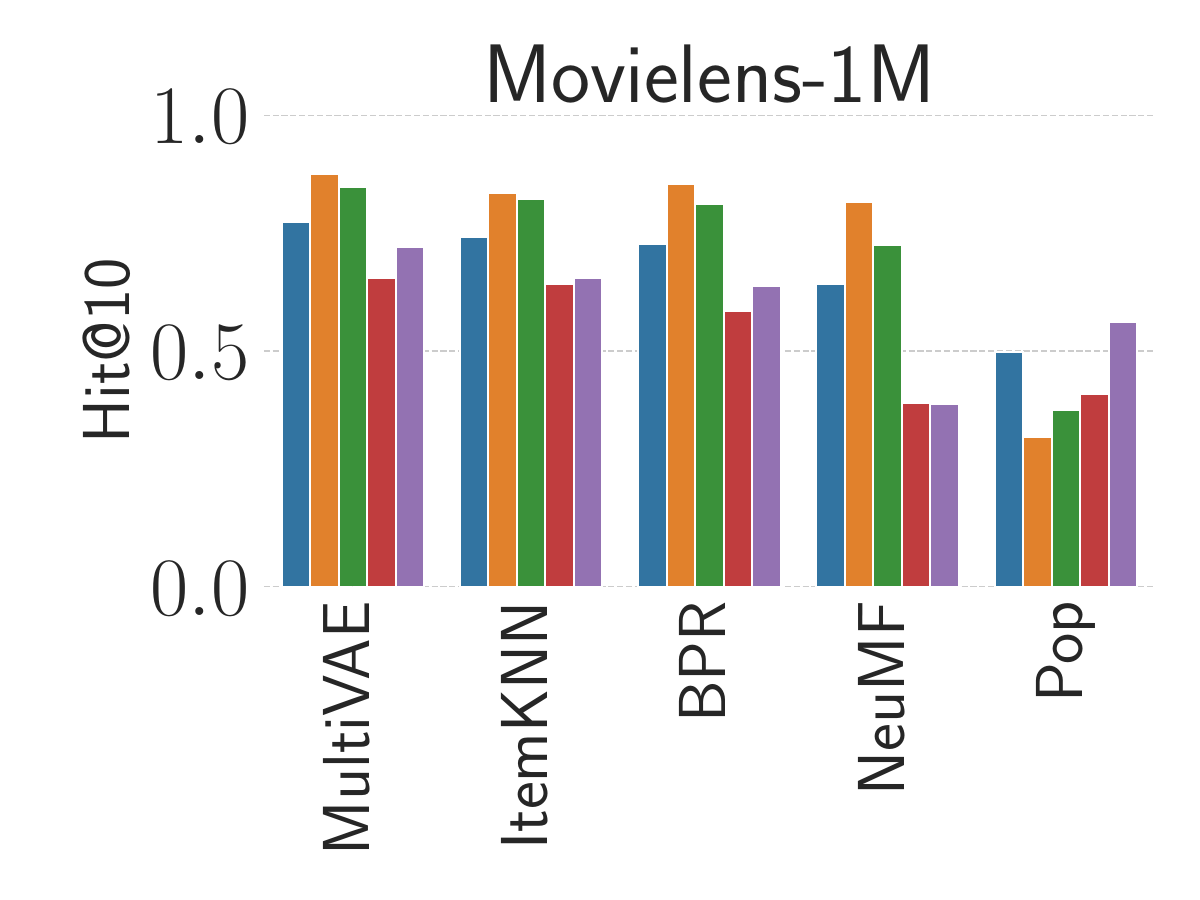}
    
    \includegraphics[width=.4\linewidth]{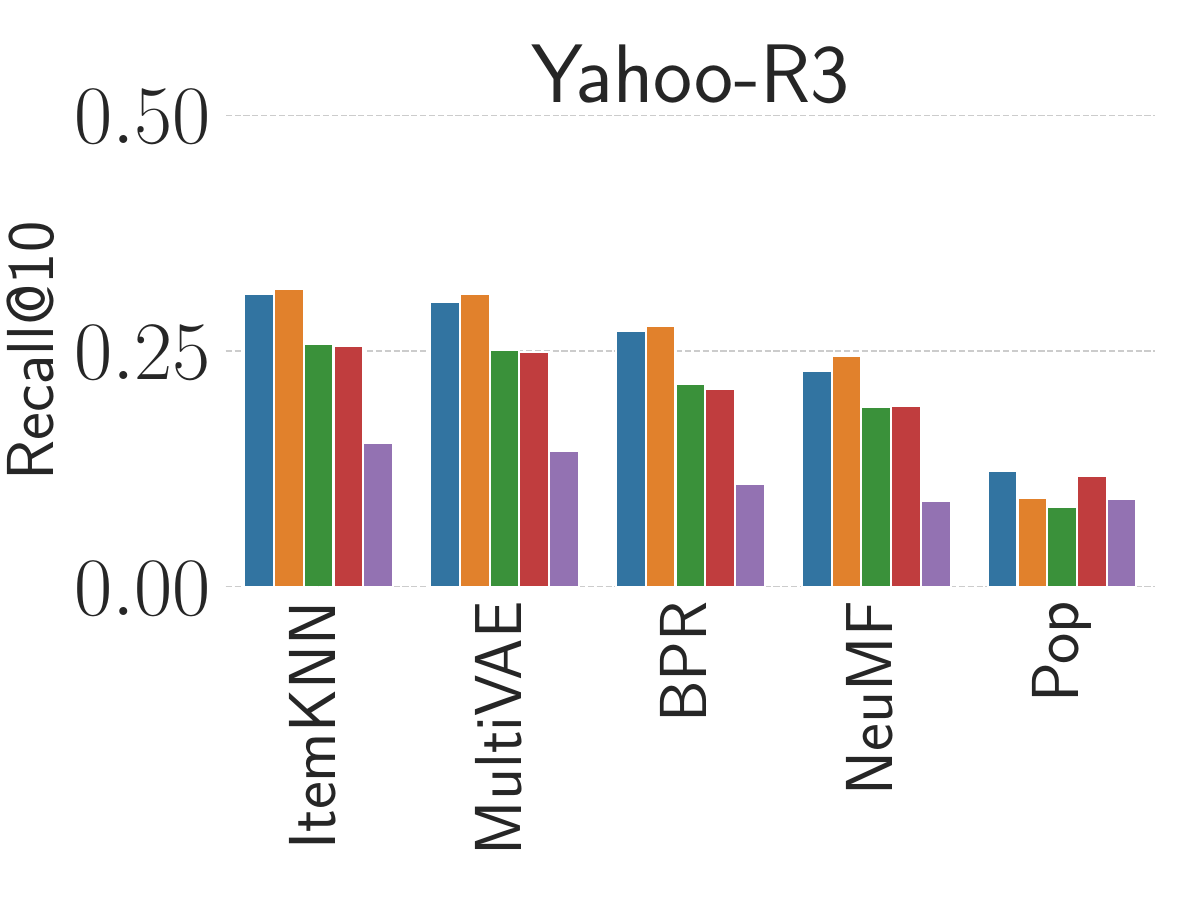}
    \includegraphics[width=.4\linewidth]{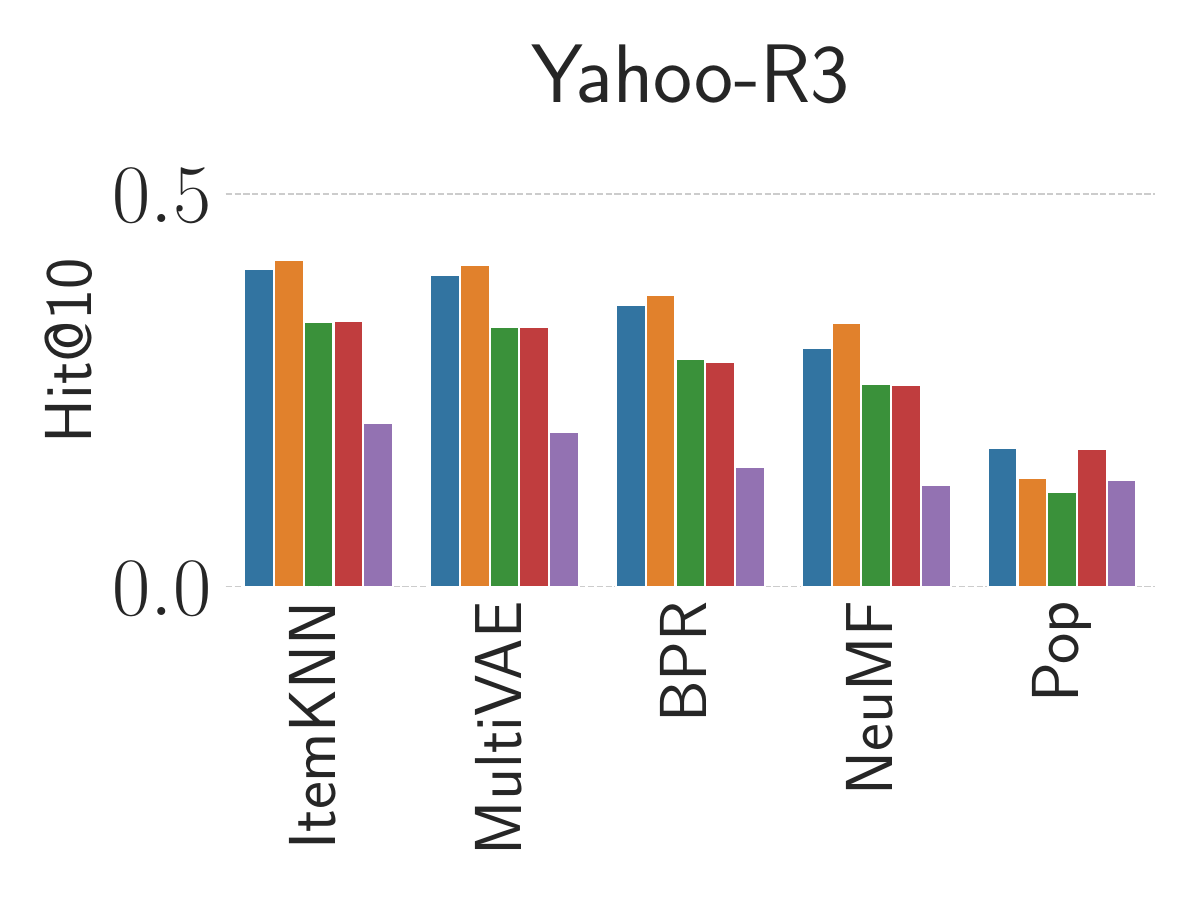}
    
    \includegraphics[width=.4\linewidth]{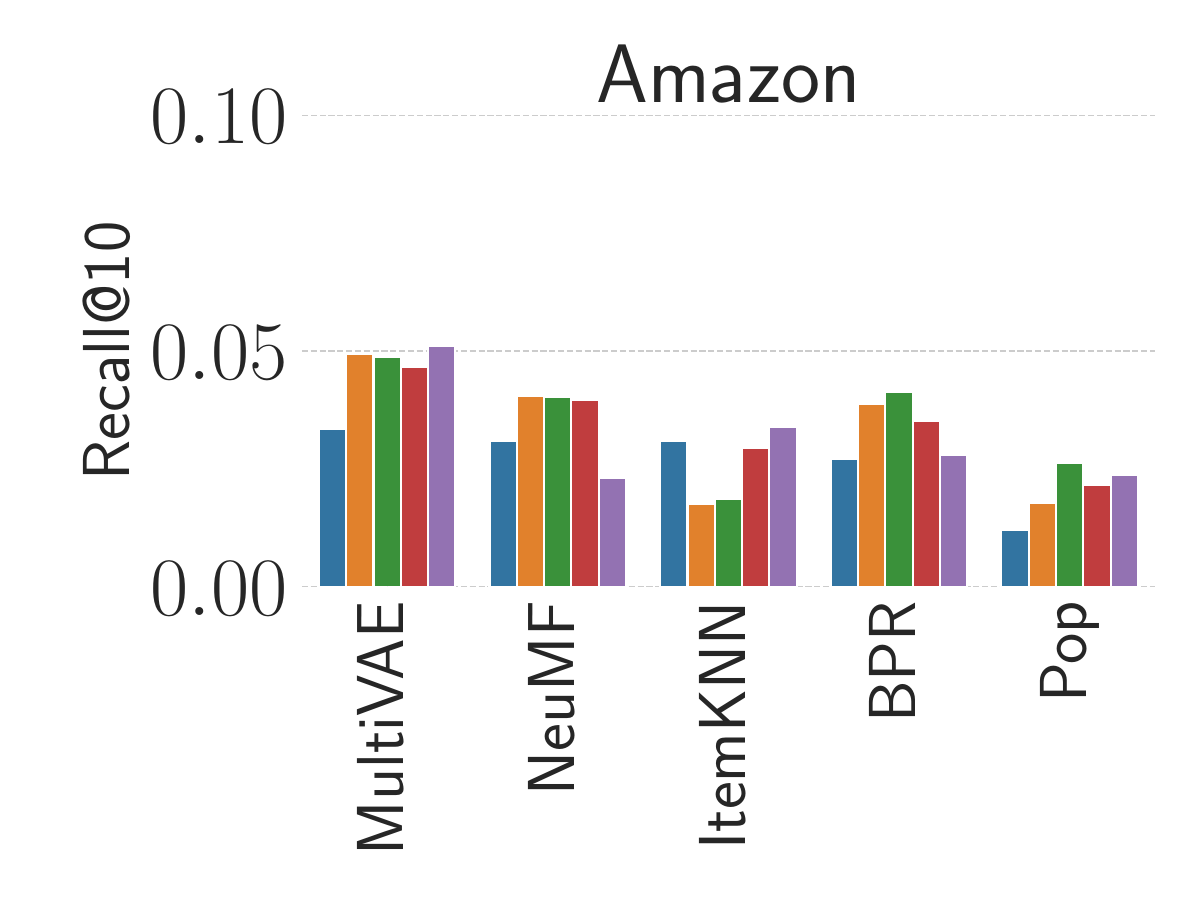}
    \includegraphics[width=.4\linewidth]{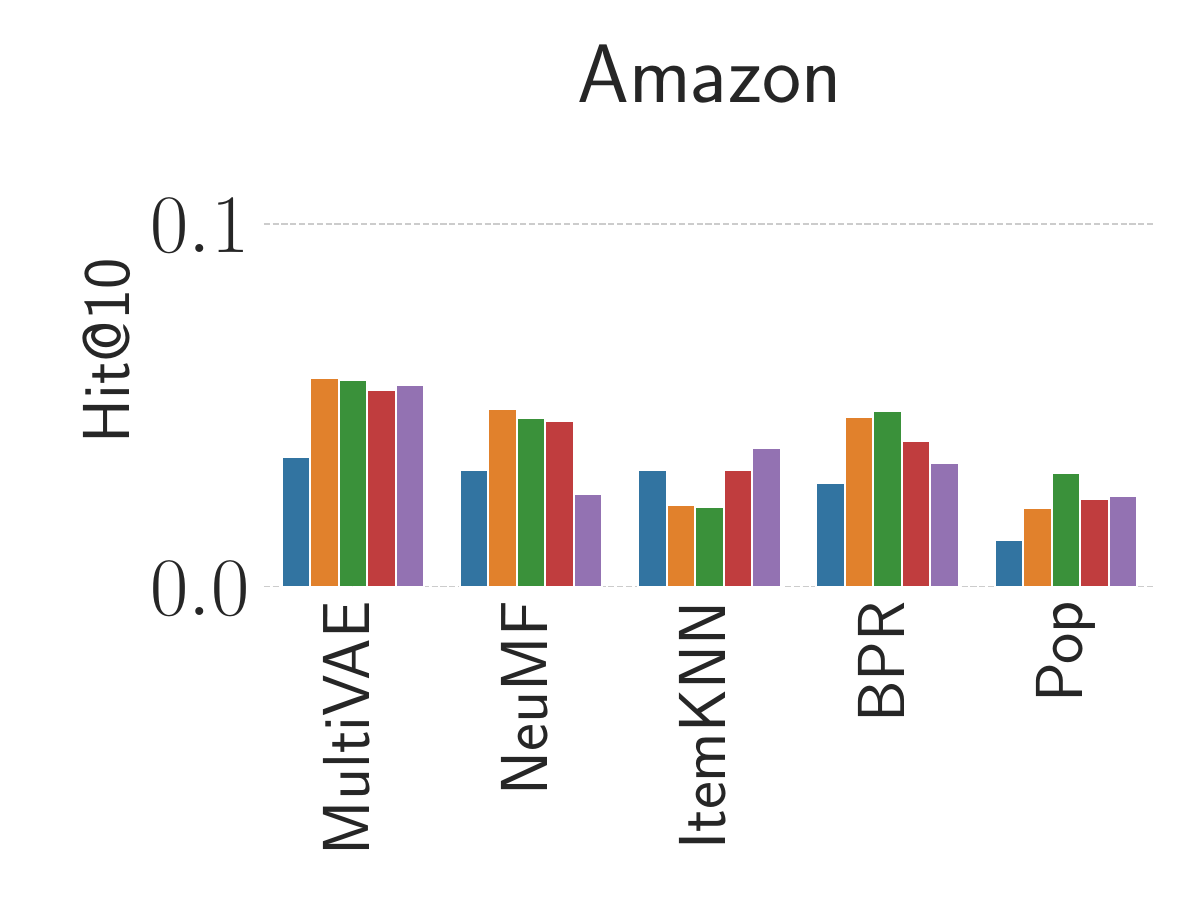}
    \vspace{-0.15cm}
    \caption{Performance comparison on Movielens-1M, Yahoo-R3, and Amazon, in terms of Recall@10 (left-column) and Hit-Rate@10 (right-column). Colors represent results on the real dataset and generated samples with varying values of $\delta$.}
    \label{fig:comparison_movielens_yahoo}
\end{figure}





\section{Conclusions and Future Work}
\label{sec:conclusions}

The paper has proposed \ouralgo, a customizable generative probabilistic framework for user-item interactions.
Our main intuition consists in modeling the generation process as a combination of three main factors: \textit{User-Item Matching}, \textit{User Engagement Level}, and \textit{Item Popularity}, thus allowing control of each individual component.
We implemented the theoretical formulation by endowing the framework with several control parameters that enable substantial control of the data generation process, under the aforementioned perspectives.
We conducted extensive experimentation with \ouralgo, showing its effectiveness concerning these tasks. 

The proposed framework is also amenable for further extensions.  In particular, the model can be adapted to cope with more complex features, such as explicit ratings, or textual content associated with preferences (e.g., reviews~\citep{peng2024reviewllmharnessinglargelanguage}). Additionally, the framework can be further extended to include user-user interactions or item-item relationships.

\clearpage

\section*{Acknowledgements}
This work has been partially funded by MUR on D.M. 351/2022, PNRR Ricerca, CUP H23C22000440007, and on D.M. 352/2022, PNRR Ricerca, CUP H23C22000550005. It is also supported by project SERICS (PE00000014) under the MUR National Recovery and Resilience Plan funded by the European Union -- NextGenerationEU.





\appendix
\begin{figure}[!ht]
    \centering
    \begin{subfigure}[b]{0.3\linewidth}
    \centering
    \includegraphics[width=\linewidth]{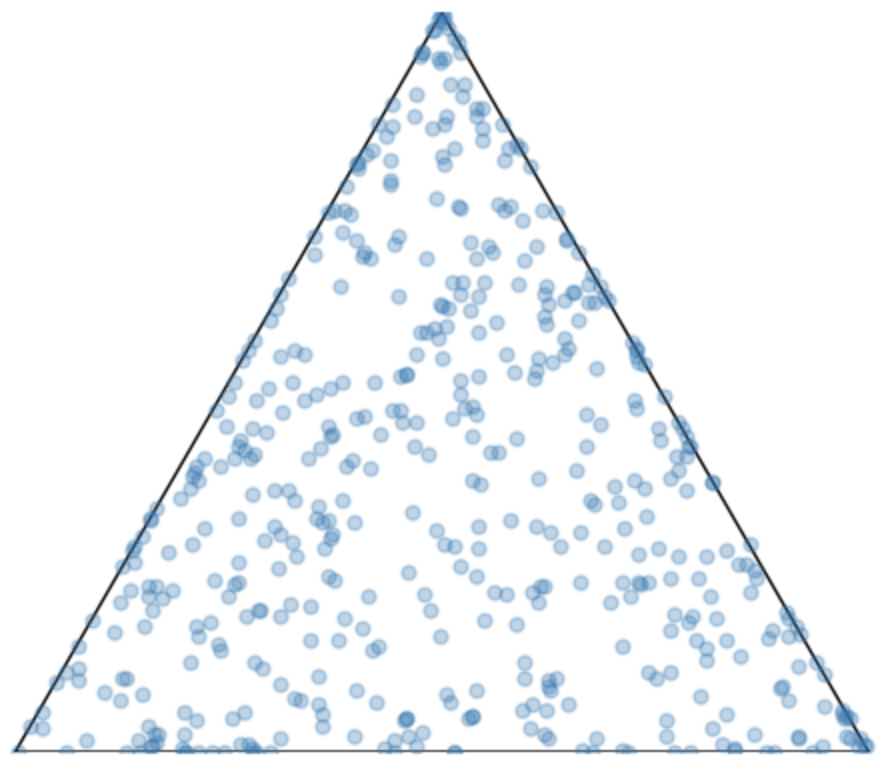}
    \label{fig:multi dir gen}
    \end{subfigure}
    \qquad
    \begin{subfigure}[b]{0.3\linewidth}
    \centering
    \includegraphics[width=\linewidth]{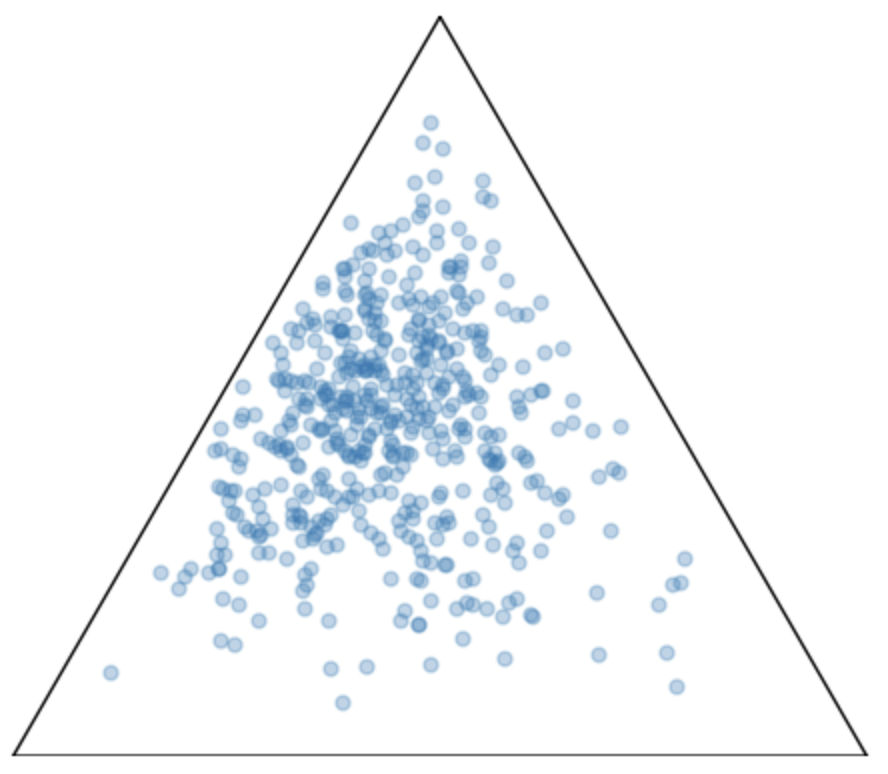}
    \label{fig:single dir gen}
    \end{subfigure}
    \caption{User profiles generated (in a three-dimensional feature space) through multiple prior sampling (left) as opposed to single prior sampling (right). Each point represents a sample in the hierarchical sampling process. By enabling a specific prior $\mu^{\rho}_u$ for each user, the resulting samples are evenly distributed over the whole latent space. By contrast, a single prior $\mu^{\rho}$ results in a concentration within specific regions of the space.}
    \label{fig:multi vs single dir}
\end{figure}

\begin{figure}[!ht]
    \centering
    \begin{subfigure}[b]{0.3\linewidth}
    \centering
    \includegraphics[width=\linewidth]{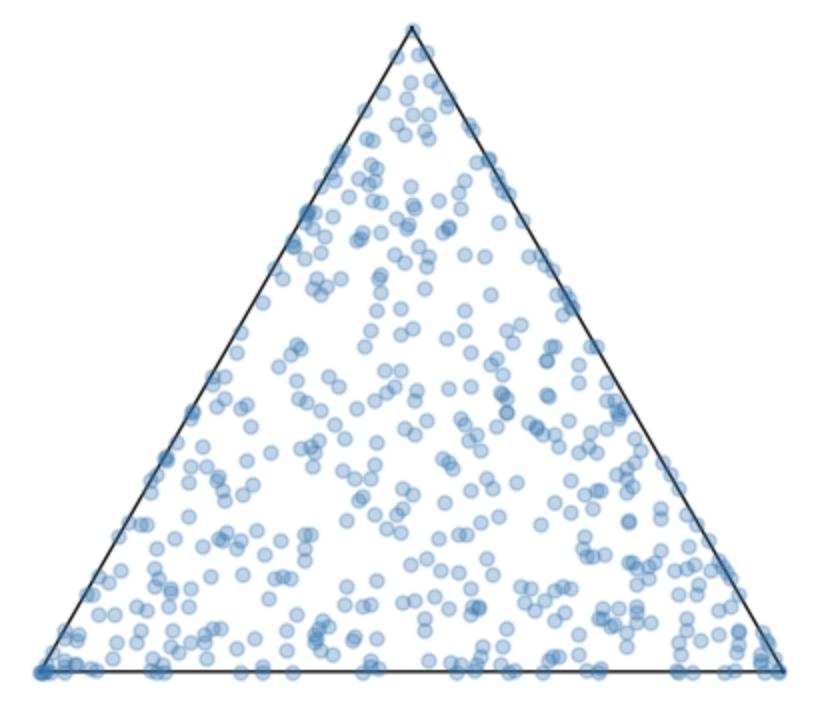}
    \end{subfigure}
    \qquad
    \begin{subfigure}[b]{0.3\linewidth}
    \centering
    \includegraphics[width=\linewidth]{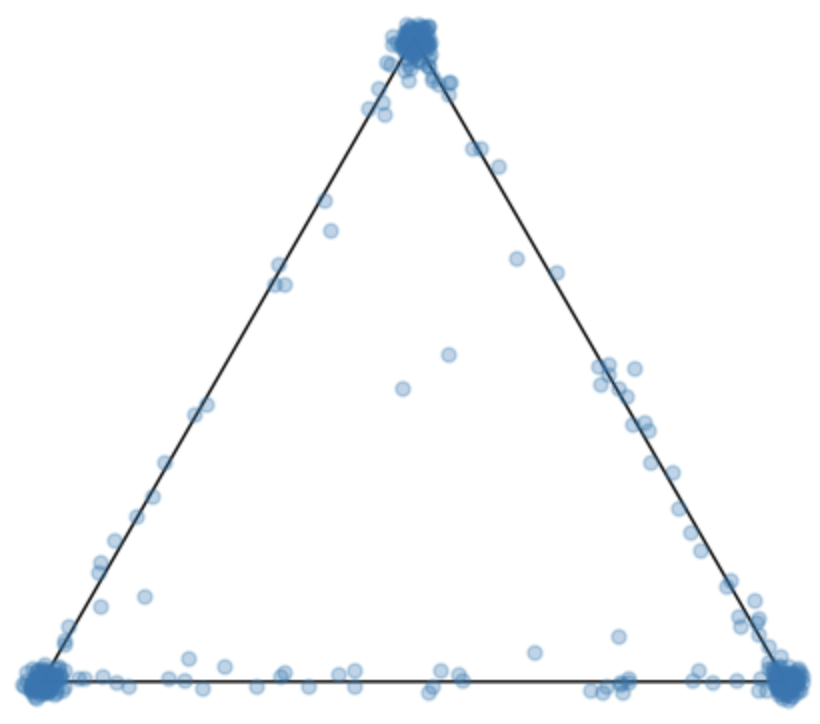}
    \end{subfigure}
    \caption{User (left) and item (right, with jitter) representations within a three-dimensional feature space, according to the proposed User-Item Matching model. Users are evenly distributed within the latent space. By contrast, item samples tend to concentrate along specific subsets of the latent features, representing edges/corners in the feature space.}
    \label{fig:user item dir}
\end{figure}

\begin{figure}[!ht]
    \centering
    \begin{subfigure}[b]{0.23\textwidth}
    \centering
    \includegraphics[width=\textwidth]{fig/generated_datasets/multiple_populations_multiple_categories/two_pop_two_cat/eta_0.01/intersections.pdf}
    \end{subfigure}
    \begin{subfigure}[b]{0.23\textwidth}
    \centering
    \includegraphics[width=\textwidth]{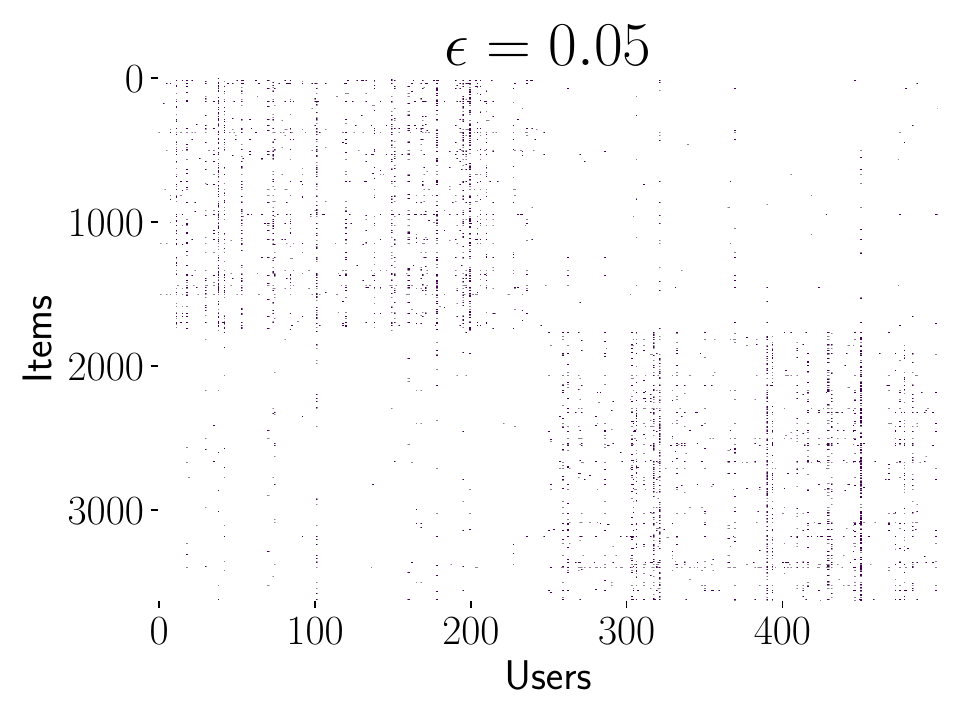}
    \end{subfigure}
    
    \centering
    \begin{subfigure}[b]{0.23\textwidth}
    \centering
    \includegraphics[width=\textwidth]{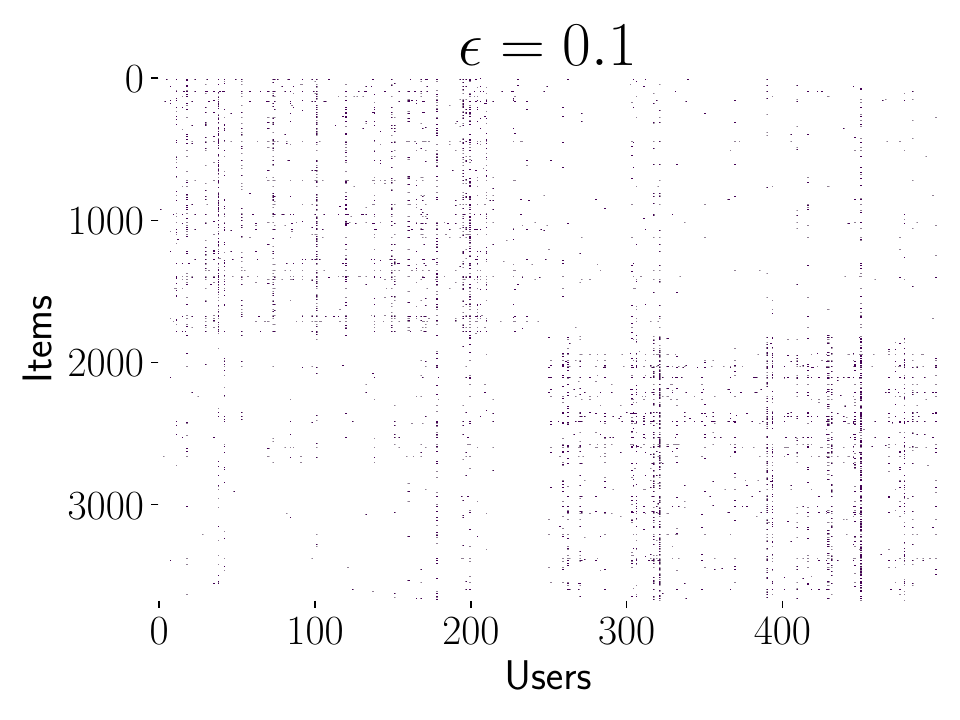}
    \end{subfigure}
    \begin{subfigure}[b]{0.23\textwidth}
    \centering
    \includegraphics[width=\textwidth]{fig/generated_datasets/multiple_populations_multiple_categories/two_pop_two_cat/eta_0.5/intersections.pdf}
    \end{subfigure}
    \caption{Visualization of the user-item interaction matrix by varying the $\varepsilon$ parameter. The X-axis reports the users, while the Y-axis represents the items. A dot in the position ($u, i$) indicates the user $u$ interacted with item $i$.}
    \label{fig:interactions_matrix_appendix}
\end{figure}

\section{User-Item Interaction}\label{sec:appendix_user_item}

Figures~\ref{fig:multi vs single dir} and~\ref{fig:user item dir} show how users and items latent factors are distributed by employing the hierarchical sampling process described in Section~\ref{sec:modeling}.

Figures~\ref{fig:interactions_matrix_appendix} depicts the impact of the $\varepsilon$ parameter over the user-item interaction matrix. Specifically, as $\varepsilon$ increases, the interaction matrix becomes gradually denser, eventually reaching full homogeneity. 
\begin{figure}[ht!]
    \centering
    \includegraphics[width=.2\textwidth]{fig/legend_u1_u2.pdf}\\
    \includegraphics[width=.23\textwidth]{fig/generated_datasets/multiple_populations_multiple_categories/two_pop_two_cat/eta_0.01/category_percentage_distribution.pdf}
    \includegraphics[width=.23\textwidth]{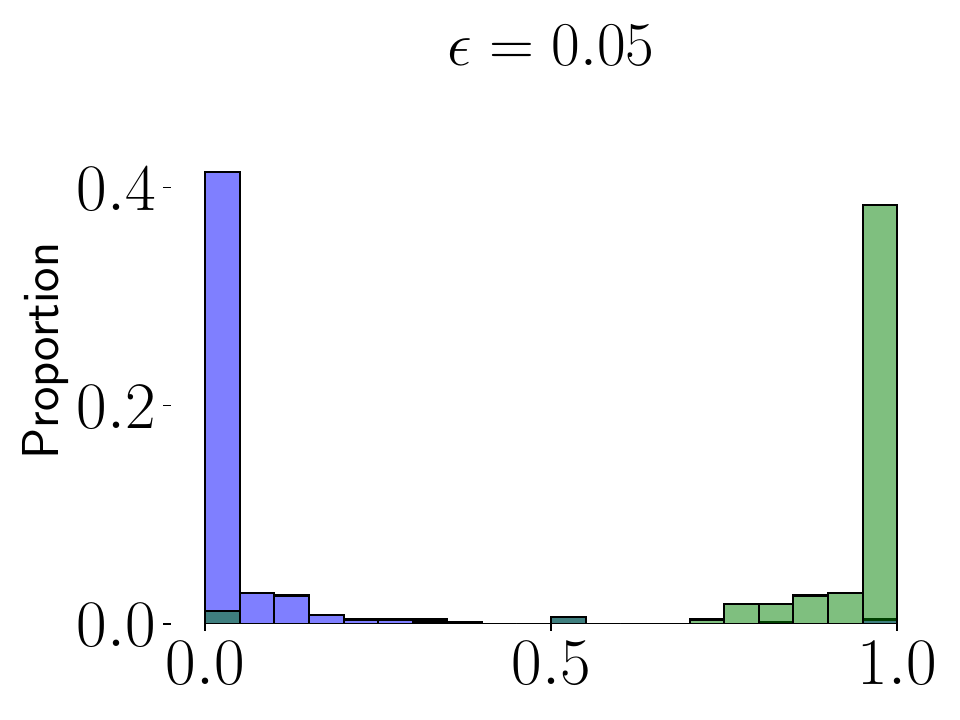}
    \includegraphics[width=.23\textwidth]{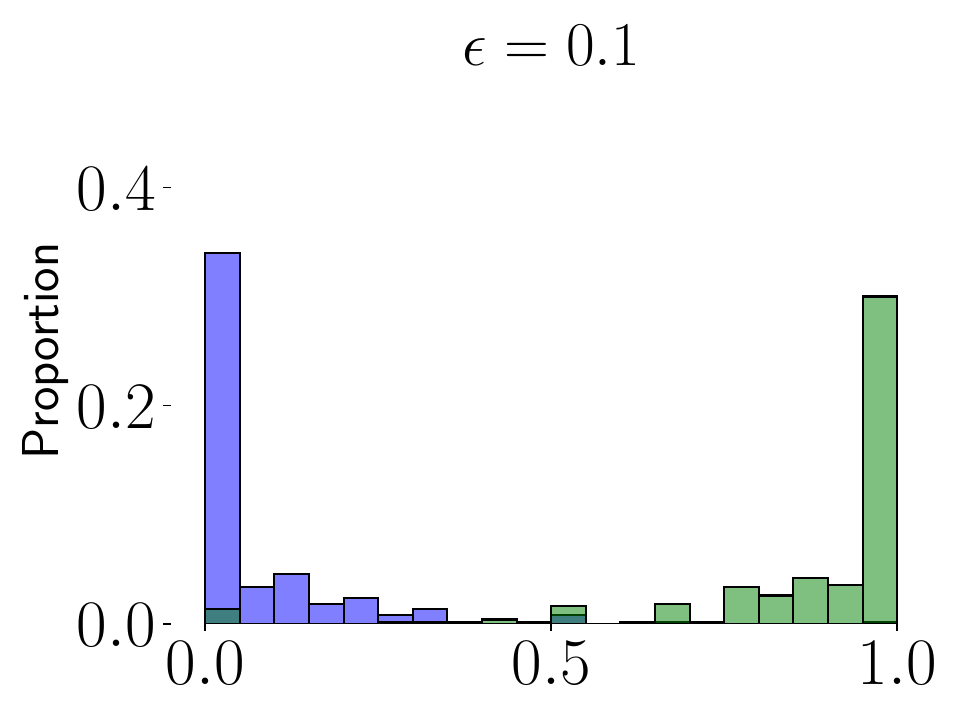}
    \includegraphics[width=.23\textwidth]{fig/generated_datasets/multiple_populations_multiple_categories/two_pop_two_cat/eta_0.5/category_percentage_distribution.pdf}
    \caption{Histograms of user interactions with a specific topic of interest. The X-axis represents the percentage of items in $I_1$ within the users history. The Y-axis shows the proportion of users having that percentage.} \label{fig:category_percentage_distributions_appendix}
\end{figure}

As a complementary analysis, Figure~\ref{fig:category_percentage_distributions_appendix} presents the proportion of items in $I_1$ in the interaction history of each communities ($U_1$ and $U_2$). We show that, when $\varepsilon$ approaches 0, the histograms are totally clustered; vice-versa, when it is close to 1, the interactions overlap.

\section{Hyper-parameter tuning}
\begin{figure}[H]
    \centering
    \begin{subfigure}[b]{0.4\columnwidth}
        \centering
        \includegraphics[width=\columnwidth]{fig/legend.pdf}
    \end{subfigure}
    \\
    \begin{subfigure}[b]{0.32\columnwidth}
    \centering
    \includegraphics[width=\columnwidth]{fig/generated_datasets/varying_zeta_parameter/degree_distributions_0.5.pdf}
    \label{fig:varying_zeta_0.5}
    \end{subfigure}
    \centering
    \begin{subfigure}[b]{0.32\columnwidth}
    \centering
    \includegraphics[width=\columnwidth]{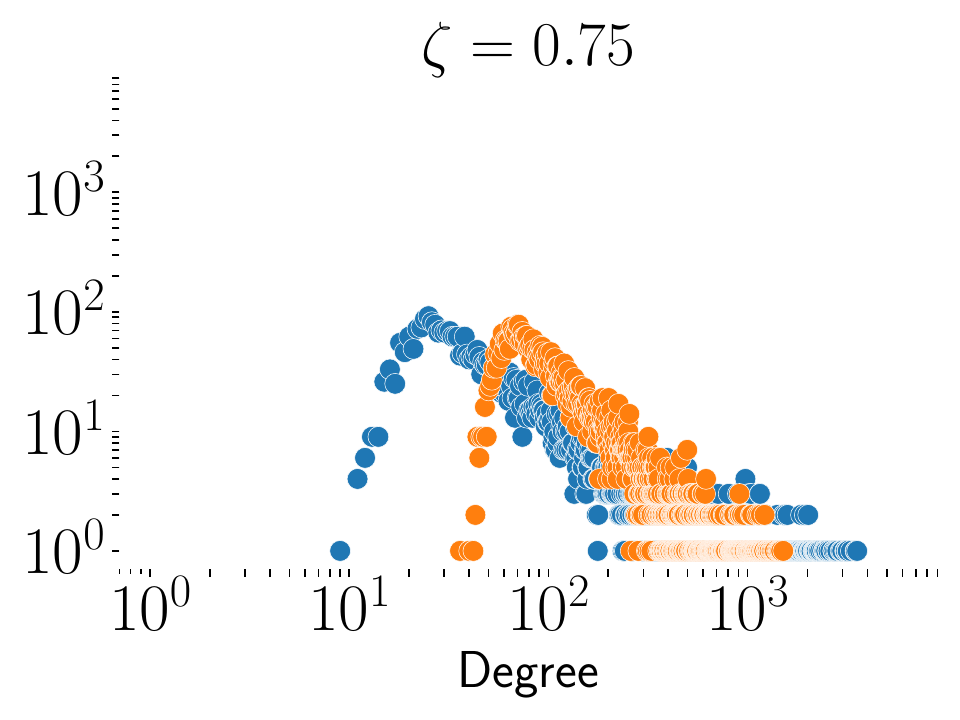}
    \label{fig:varying_zeta_0.75}
    \end{subfigure}
    \begin{subfigure}[b]{0.32\columnwidth}
    \centering
    \includegraphics[width=\columnwidth]{fig/generated_datasets/varying_zeta_parameter/degree_distributions_1.0.pdf}
    \label{fig:varying_zeta_1.0}
    \end{subfigure}
    
    \begin{subfigure}[b]{0.32\columnwidth}
    \centering
    \includegraphics[width=\columnwidth]{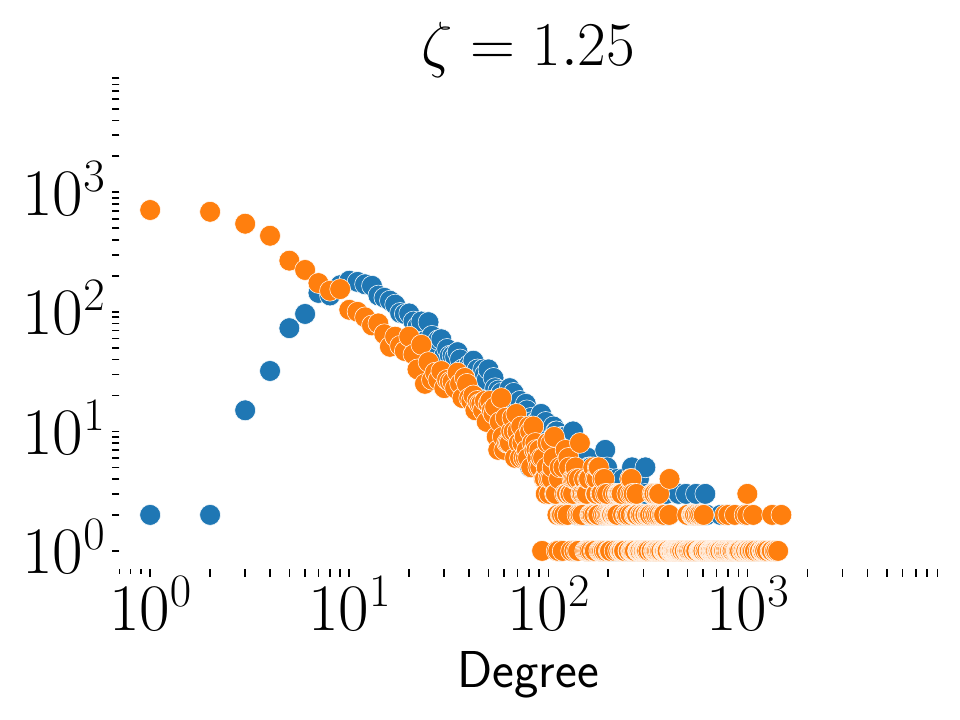}
    \label{fig:varying_zeta_1.25}
    \end{subfigure}
    \begin{subfigure}[b]{0.32\columnwidth}
    \centering
    \includegraphics[width=\columnwidth]{fig/generated_datasets/varying_zeta_parameter/degree_distributions_1.5.pdf}
    \label{fig:varying_zeta_1.5}
    \end{subfigure}
    \begin{subfigure}[b]{0.32\columnwidth}
    \centering
    \includegraphics[width=\columnwidth]{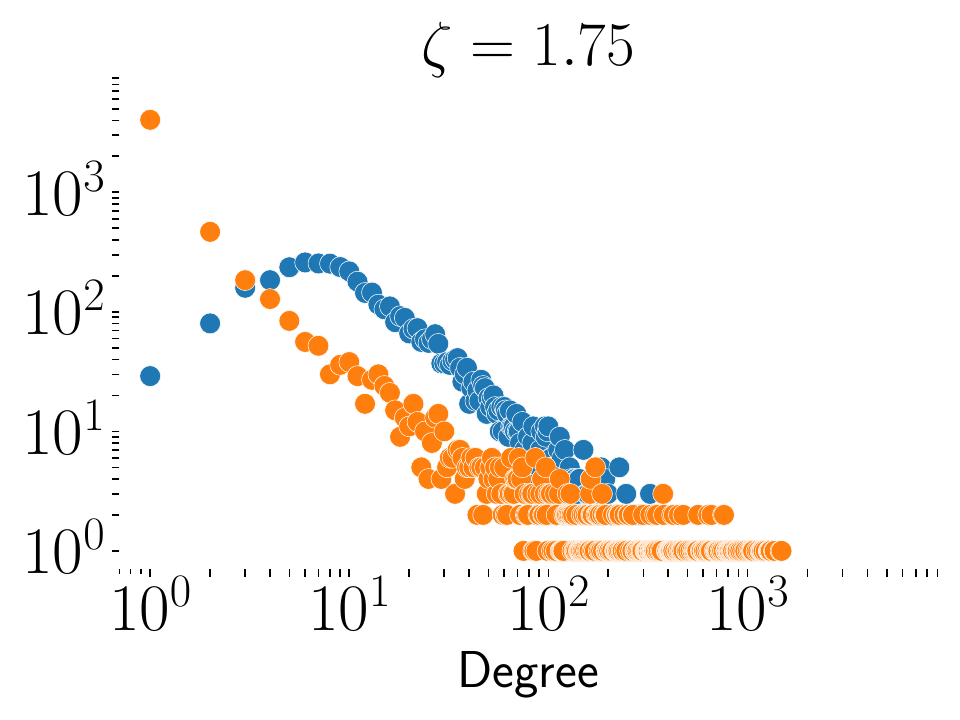}
    \label{fig:varying_zeta_1.75}
    \end{subfigure}
    \caption{Effects of $\zeta$ on the generated data distributions.}
    \label{fig:varying_zeta}
\end{figure}

\begin{figure}[H]
\begin{subfigure}[b]{0.4\columnwidth}
        \centering
        \includegraphics[width=\columnwidth]{fig/legend.pdf}
    \end{subfigure}
    \\
    \begin{subfigure}[b]{0.32\columnwidth}
    \centering
    \includegraphics[width=\columnwidth]{fig/generated_datasets/varying_xi_parameter/degree_distributions_0.5.pdf}
    \end{subfigure}
    \centering
    \begin{subfigure}[b]{0.32\columnwidth}
    \centering
    \includegraphics[width=\columnwidth]{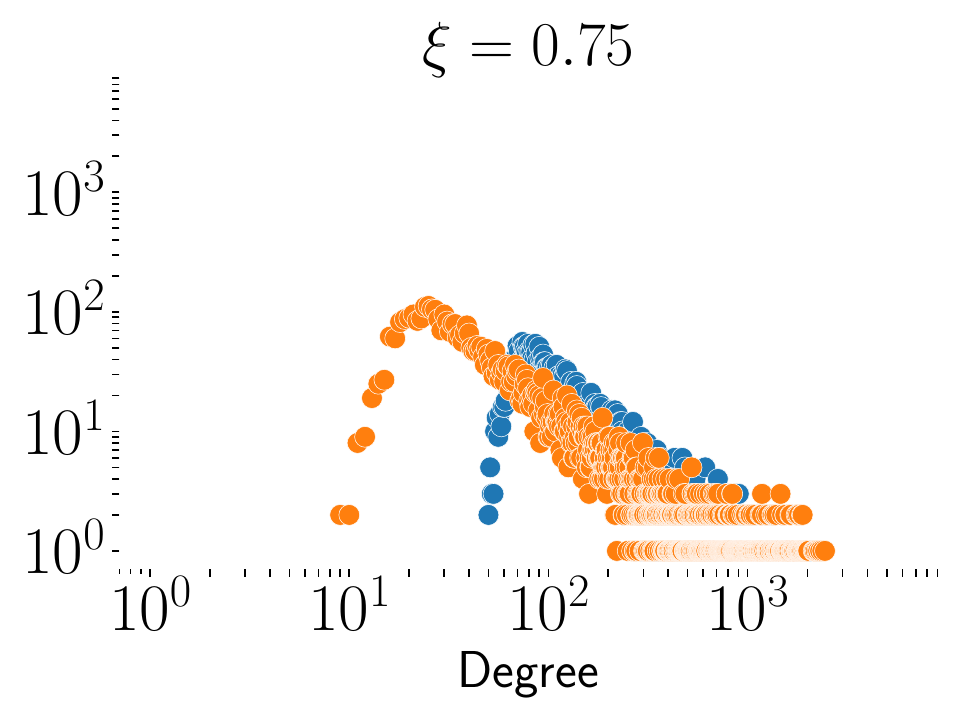}
    \end{subfigure}
    \begin{subfigure}[b]{0.32\columnwidth}
    \centering
    \includegraphics[width=\columnwidth]{fig/generated_datasets/varying_xi_parameter/degree_distributions_1.0.pdf}
    \end{subfigure}
    
    \begin{subfigure}[b]{0.32\columnwidth}
    \centering
    \includegraphics[width=\columnwidth]{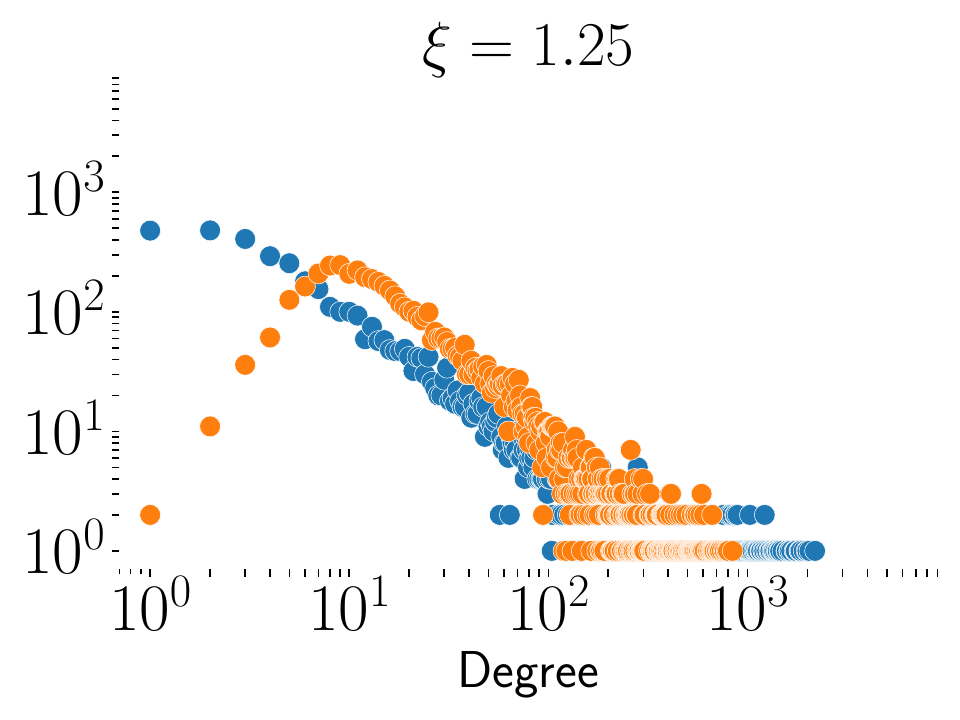}
    \end{subfigure}
    \begin{subfigure}[b]{0.32\columnwidth}
    \centering
    \includegraphics[width=\columnwidth]{fig/generated_datasets/varying_xi_parameter/degree_distributions_1.5.pdf}
    \end{subfigure}
    \begin{subfigure}[b]{0.32\columnwidth}
    \centering
    \includegraphics[width=\columnwidth]{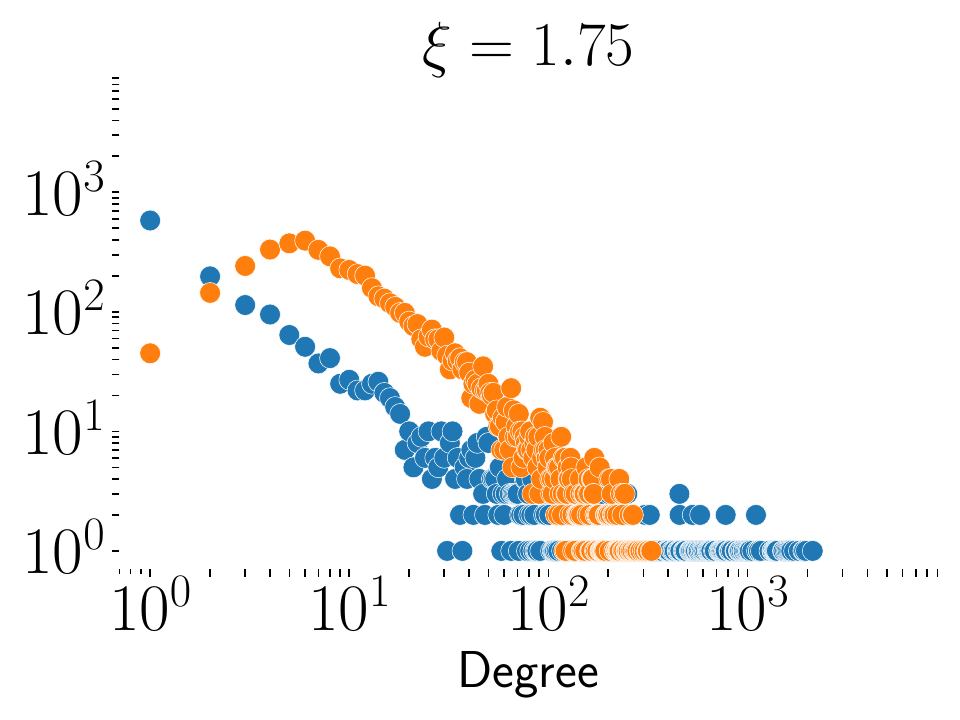}
    \end{subfigure}
    
    \caption{Effects of $\xi$ on the generated data distributions.}
    \label{fig:varying_xi}
\end{figure}

\begin{figure}[H]
\begin{subfigure}[b]{0.4\columnwidth}
        \centering
        \includegraphics[width=\columnwidth]{fig/legend.pdf}
    \end{subfigure}
\\
    \begin{subfigure}[b]{0.32\columnwidth}
    \centering
    \includegraphics[width=\columnwidth]{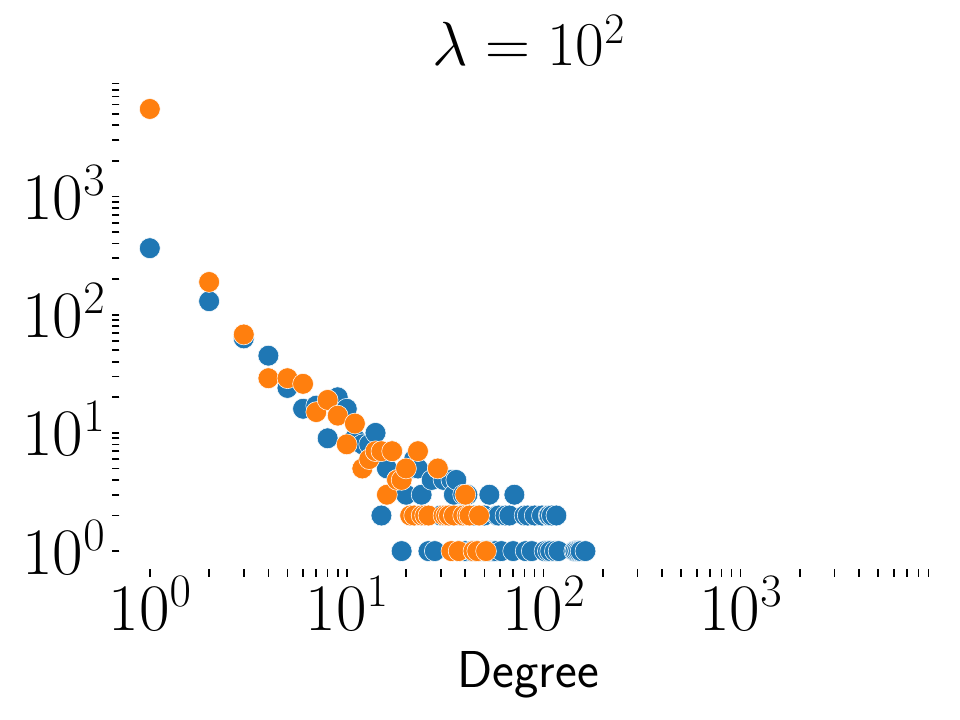}
    \end{subfigure}
    \centering
    \begin{subfigure}[b]{0.32\columnwidth}
    \centering
    \includegraphics[width=\columnwidth]{fig/generated_datasets/varying_lambda_parameter/degree_distributions_10_3.pdf}
    \end{subfigure}
    \begin{subfigure}[b]{0.32\columnwidth}
    \centering
    \includegraphics[width=\columnwidth]{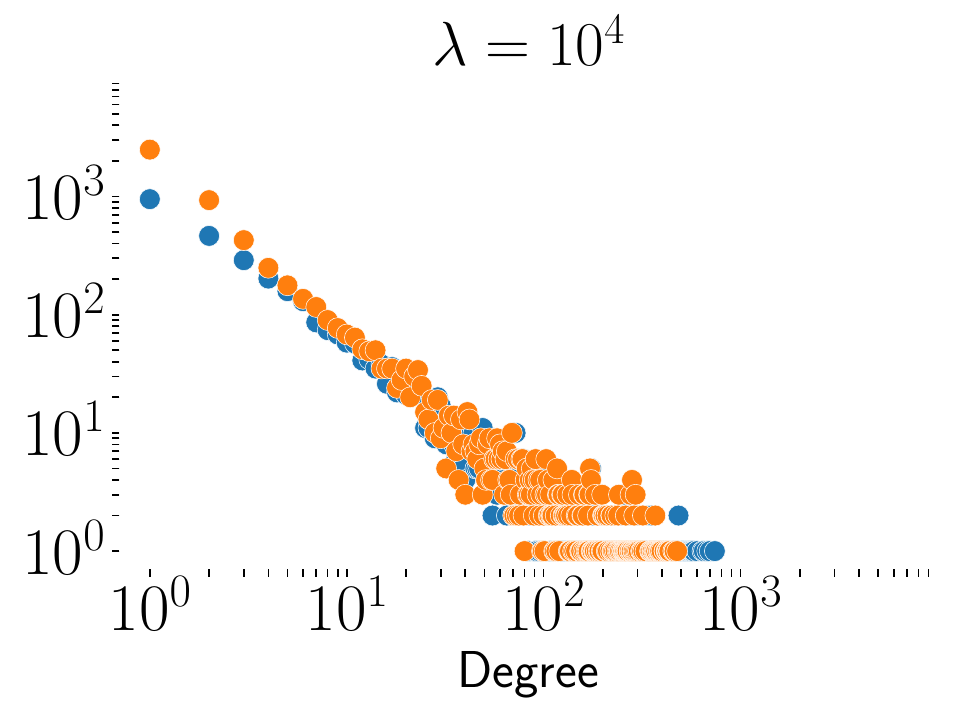}
    \end{subfigure}
    
    \begin{subfigure}[b]{0.32\columnwidth}
    \centering
    \includegraphics[width=\columnwidth]{fig/generated_datasets/varying_lambda_parameter/degree_distributions_10_5.pdf}
    \end{subfigure}
    \begin{subfigure}[b]{0.32\columnwidth}
    \centering
    \includegraphics[width=\columnwidth]{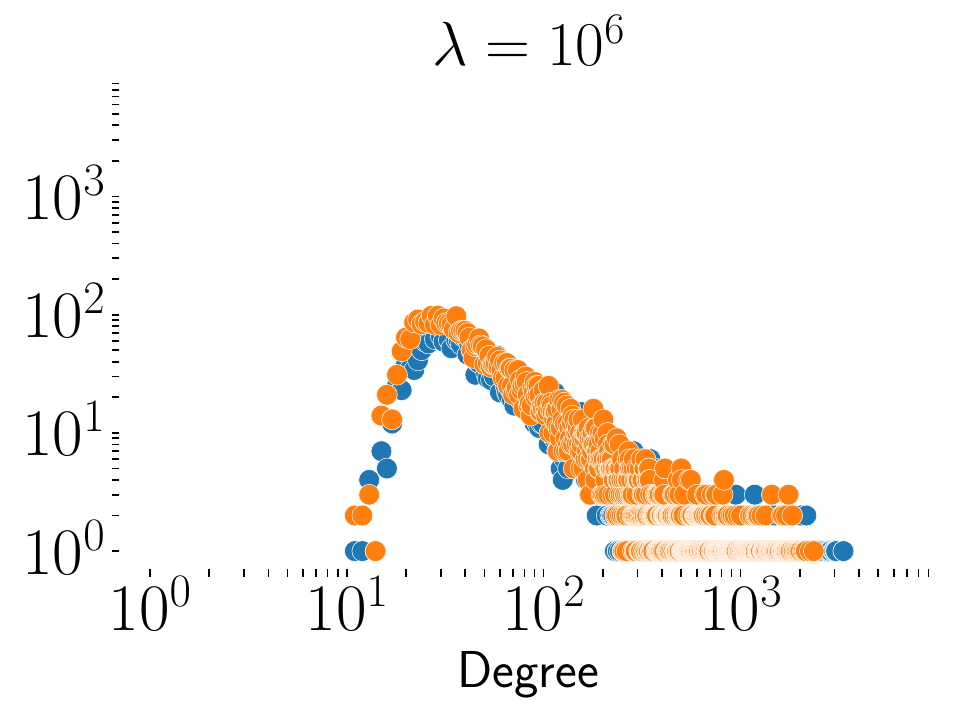}
    \end{subfigure}
    \begin{subfigure}[b]{0.32\columnwidth}
    \centering
    \includegraphics[width=\columnwidth]{fig/generated_datasets/varying_lambda_parameter/degree_distributions_10_7.pdf}
    \end{subfigure}
    
    \caption{Effects of $\lambda$ on the generated data distributions.}
    \label{fig:varying_lambda}
\end{figure}

We here report some additional experiments concerning the impact of the weighting factors $\zeta$, $\xi$, and $\lambda$ over the generated degree distributions, as shown in Figures~\ref{fig:varying_zeta},~\ref{fig:varying_xi}, and~\ref{fig:varying_lambda}, respectively. As we can see from the plots, fixed $\xi$ (resp. $\zeta$) to 1, when $\zeta$ (resp., $\xi$) increases, the distributions of users (resp., items) become more sharpened, conversely following a Normal pattern when the control value is low, due to the Bernoullian sampling process.
Regarding $\lambda$, the higher the value, the higher the minimum degree of users and items, hence the density of the generated dataset. 

\section{Performance comparison}

\begin{figure}[!ht]
    \centering
    \begin{subfigure}[b]{\columnwidth}
        \centering
        \includegraphics[width=\columnwidth]{fig/comparison_real_generated/barplot_legend_2.pdf}
    \end{subfigure}
    \begin{subfigure}[b]{0.48\columnwidth}
    \centering
    \includegraphics[width=\columnwidth]{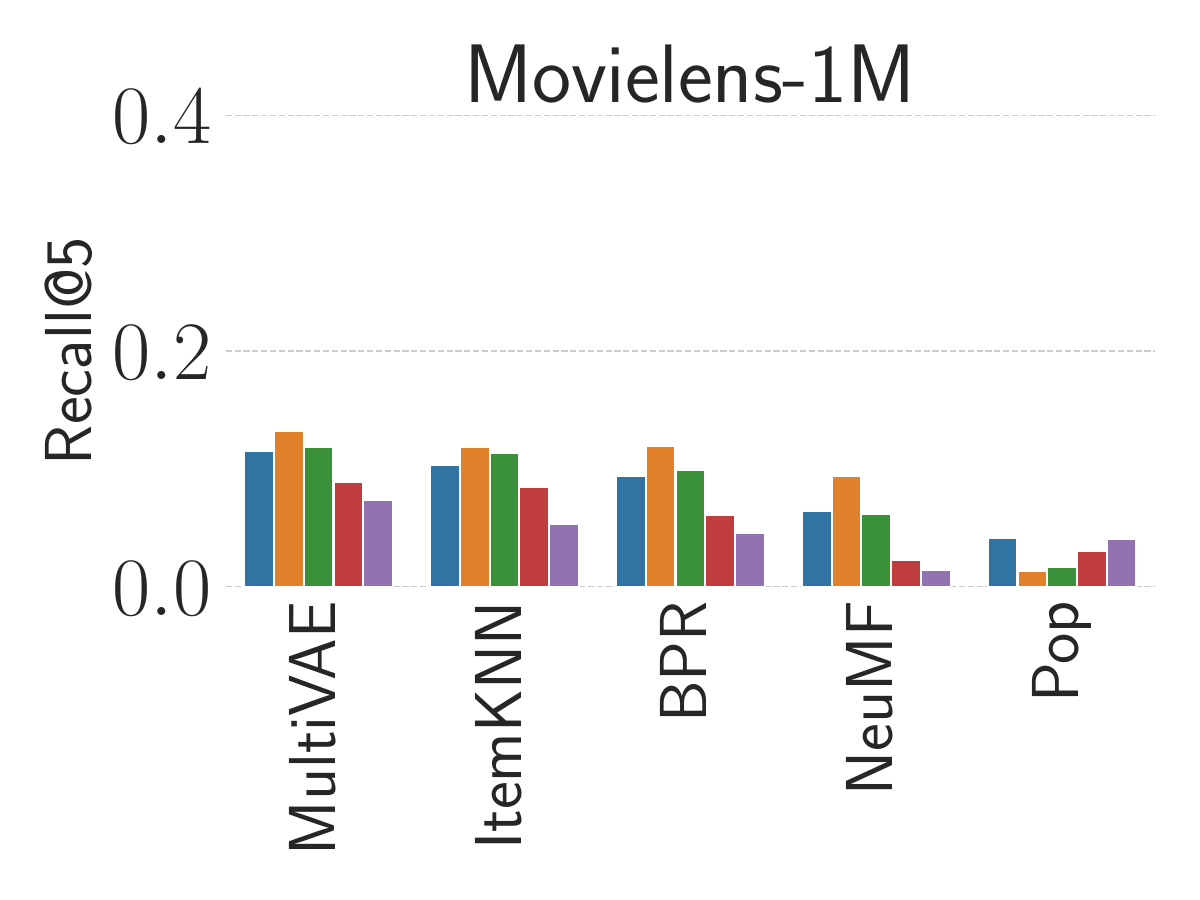}
    \end{subfigure}
    \begin{subfigure}[b]{0.48\columnwidth}
    \centering
    \includegraphics[width=\columnwidth]{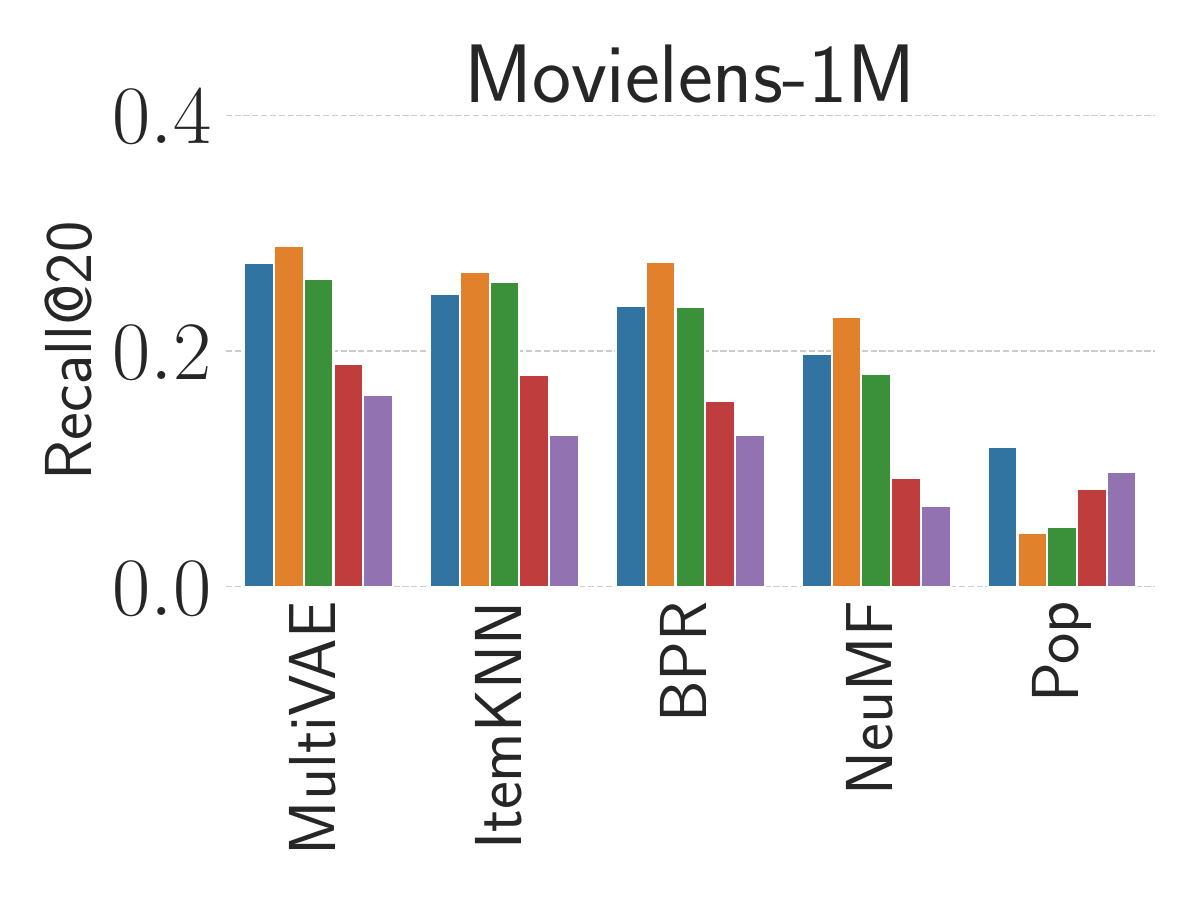}
    \end{subfigure}
    
    \begin{subfigure}[b]{0.48\columnwidth}
    \centering
    \includegraphics[width=\columnwidth]{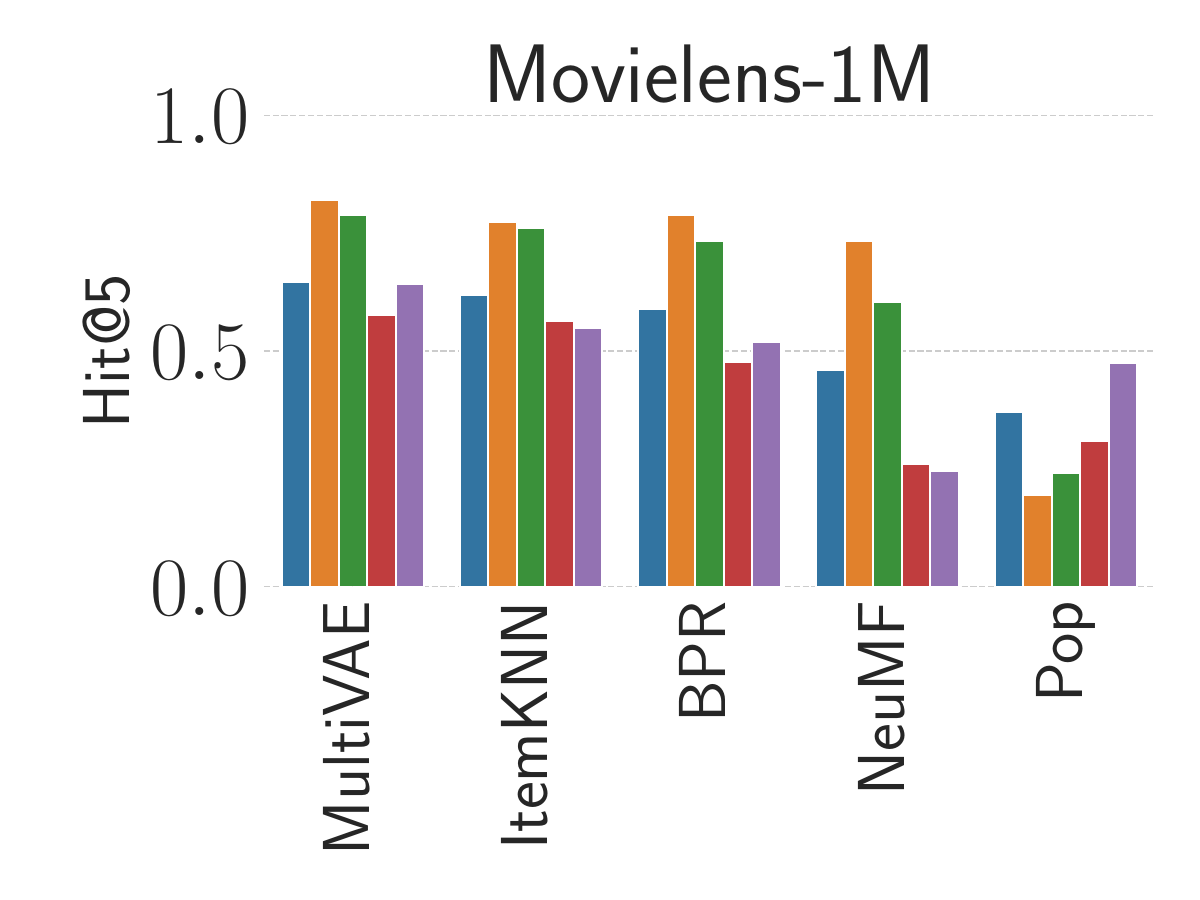}
    \end{subfigure}
    \begin{subfigure}[b]{0.48\columnwidth}
    \centering
    \includegraphics[width=\columnwidth]{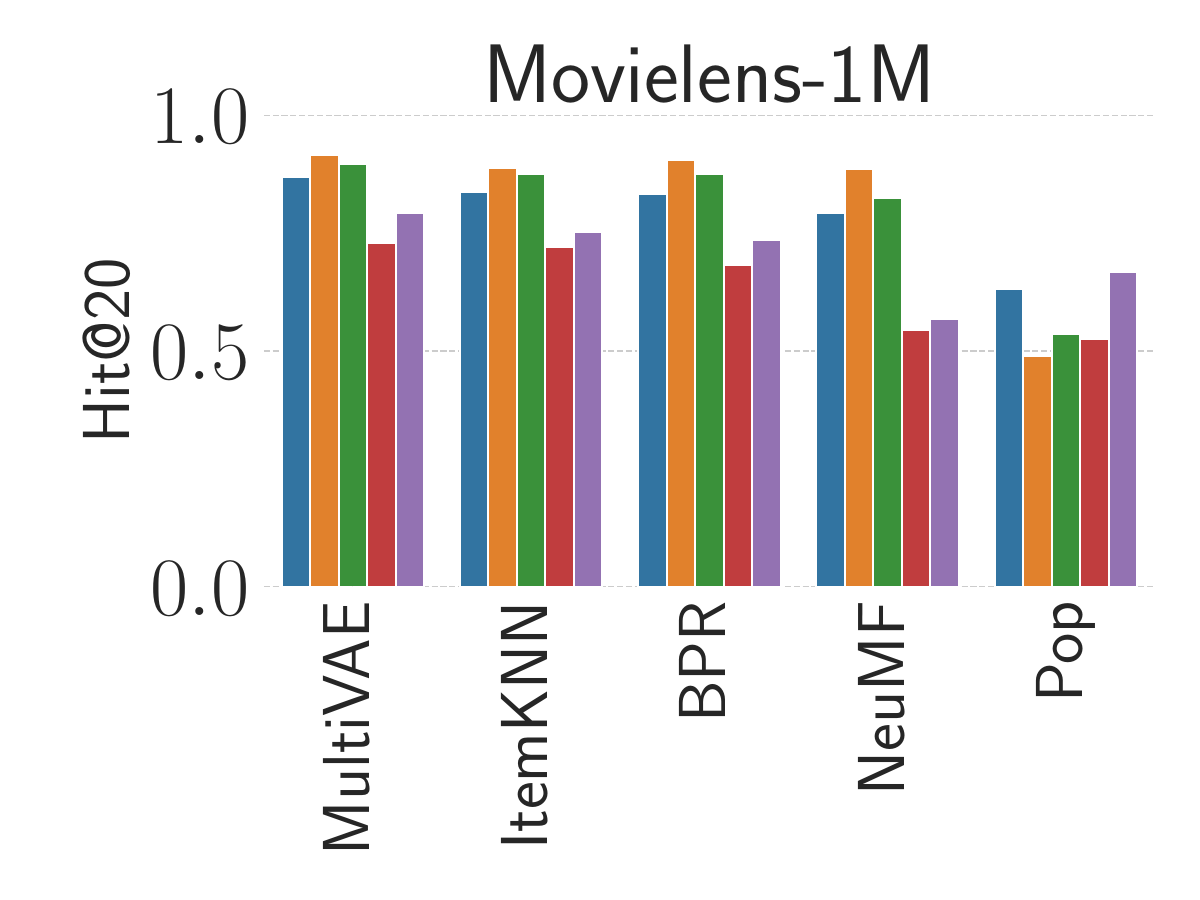}
    \end{subfigure}
    \caption{Performance comparison on Movielens-1M at cut-offs equal 5 and 20. Colors indicate the results over the real dataset and on the generated sample at varying of $\delta$.}
    \label{fig:comparison_movielens-5-20}
\end{figure}

\begin{figure}[!ht]
    \centering
    \begin{subfigure}[b]{\columnwidth}
        \centering
        \includegraphics[width=\columnwidth]{fig/comparison_real_generated/barplot_legend_2.pdf}
    \end{subfigure}
    
    \begin{subfigure}[b]{0.48\columnwidth}
    \centering
    \includegraphics[width=\columnwidth]{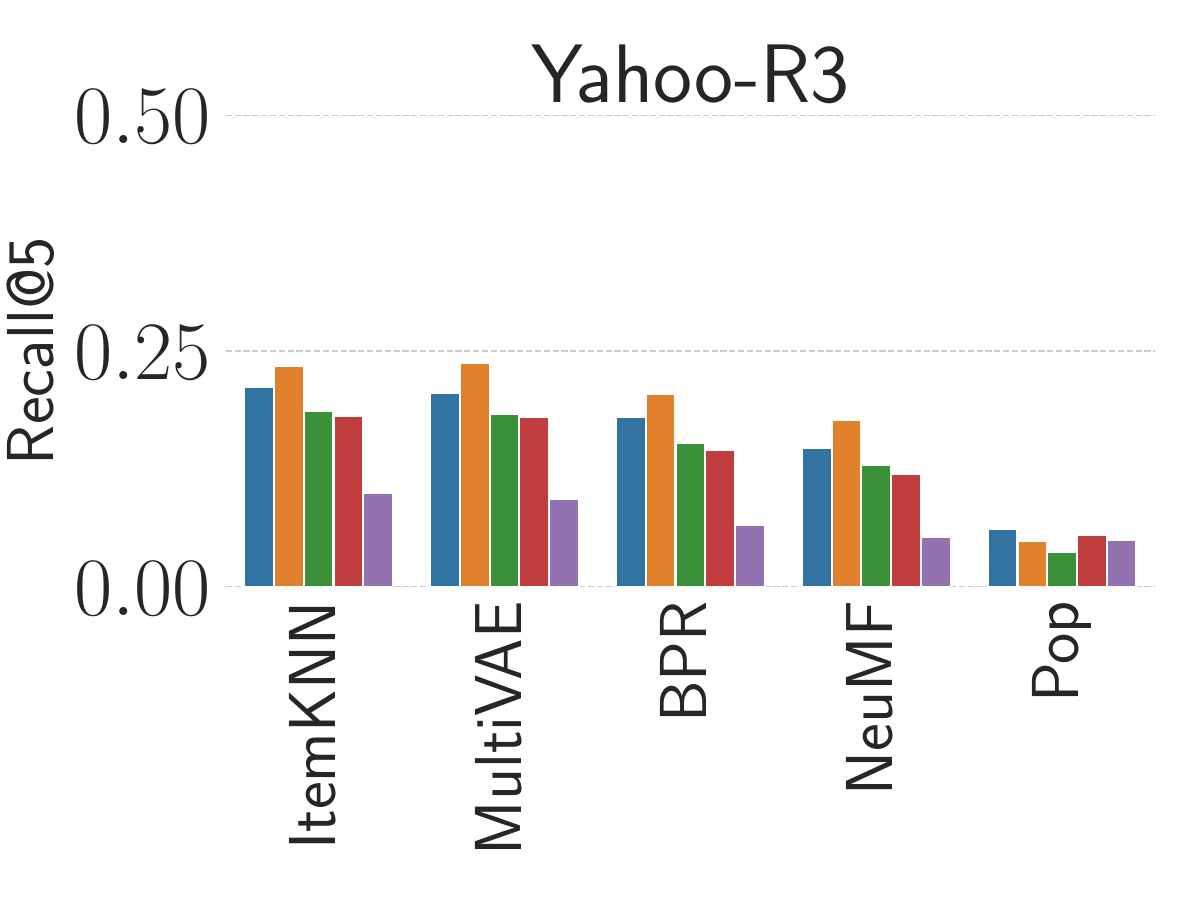}
    \end{subfigure}
    \begin{subfigure}[b]{0.48\columnwidth}
    \centering
    \includegraphics[width=\columnwidth]{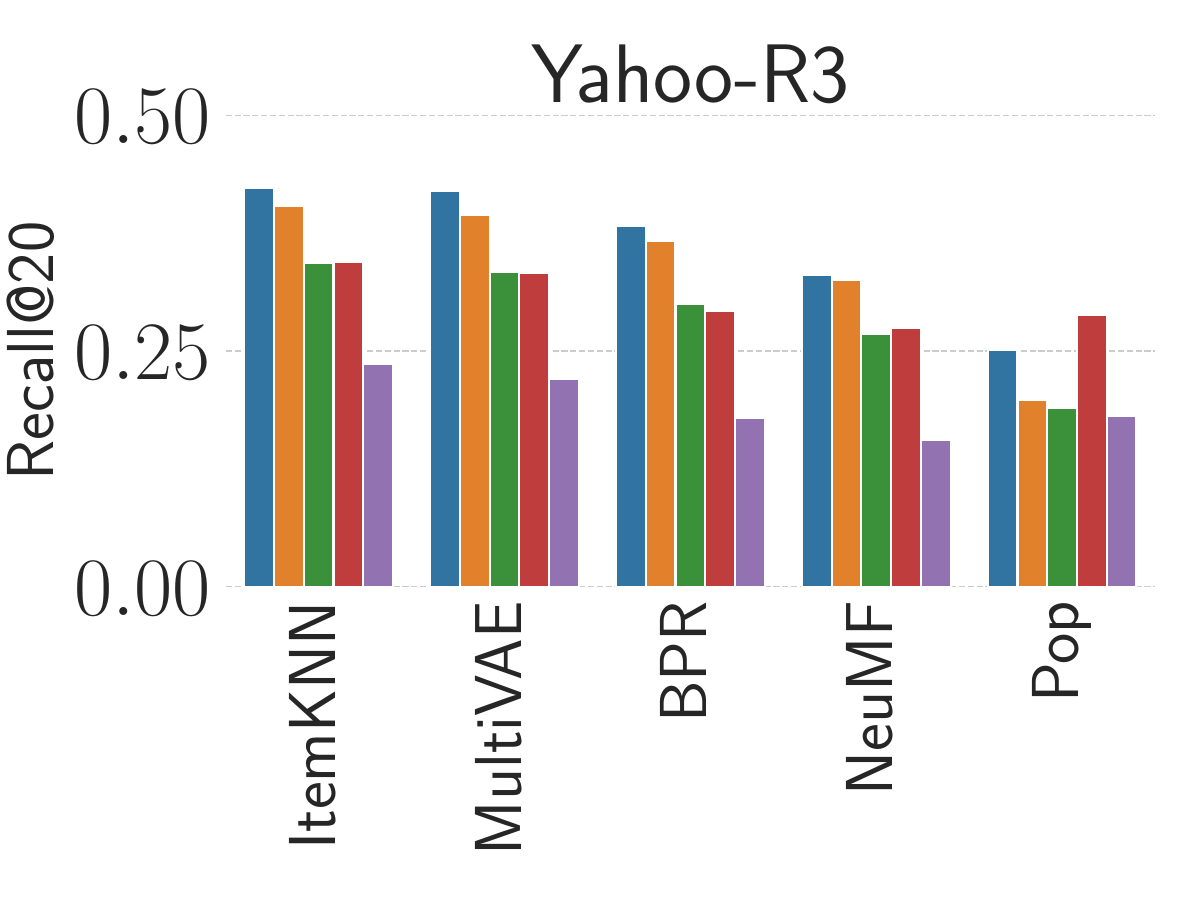}
    \end{subfigure}
    
    \begin{subfigure}[b]{0.48\columnwidth}
    \centering
    \includegraphics[width=\columnwidth]{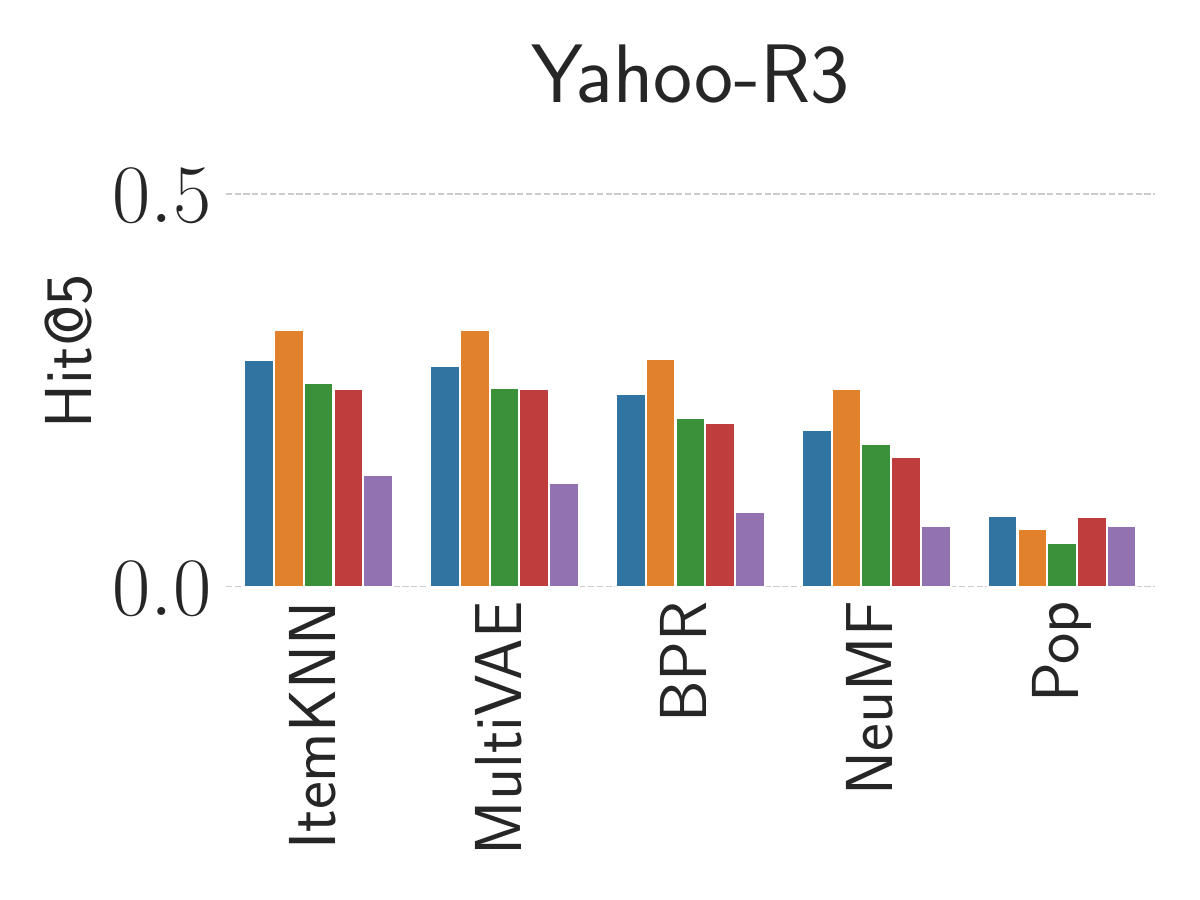}
    \end{subfigure}
    \begin{subfigure}[b]{0.48\columnwidth}
    \centering
    \includegraphics[width=\columnwidth]{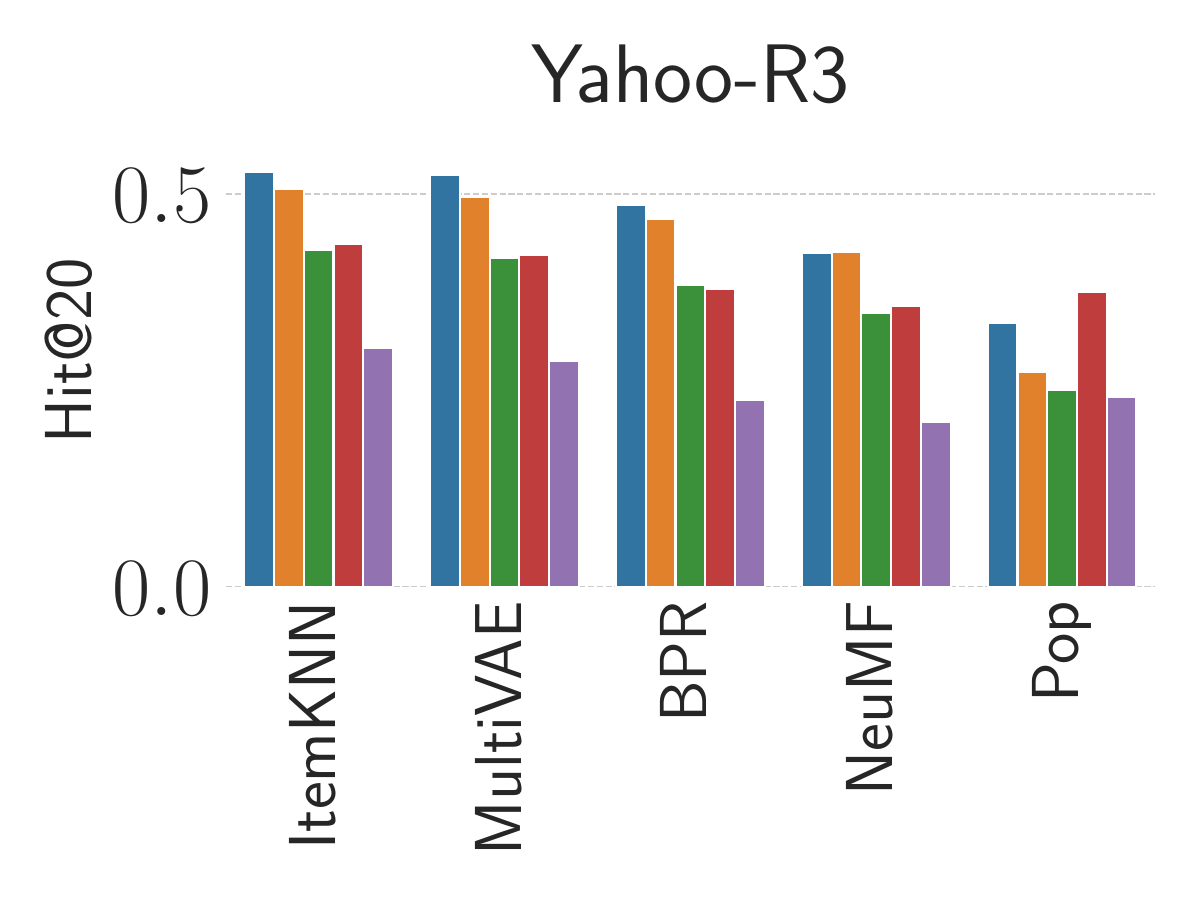}
    \end{subfigure}
    \caption{Performances comparison on Yahoo-R3 at cut-offs equal 5 and 20. The colors indicate the results over the real dataset and on the generated sample at varying of $\delta$.}
    \label{fig:comparison_yahoo-r3-5-20}
\end{figure}

\begin{figure}[!ht]
    \centering
    \begin{subfigure}[b]{\columnwidth}
        \centering
        \includegraphics[width=\columnwidth]{fig/comparison_real_generated/barplot_legend_2.pdf}
    \end{subfigure}
    
    \begin{subfigure}[b]{0.48\columnwidth}
    \centering
    \includegraphics[width=\columnwidth]{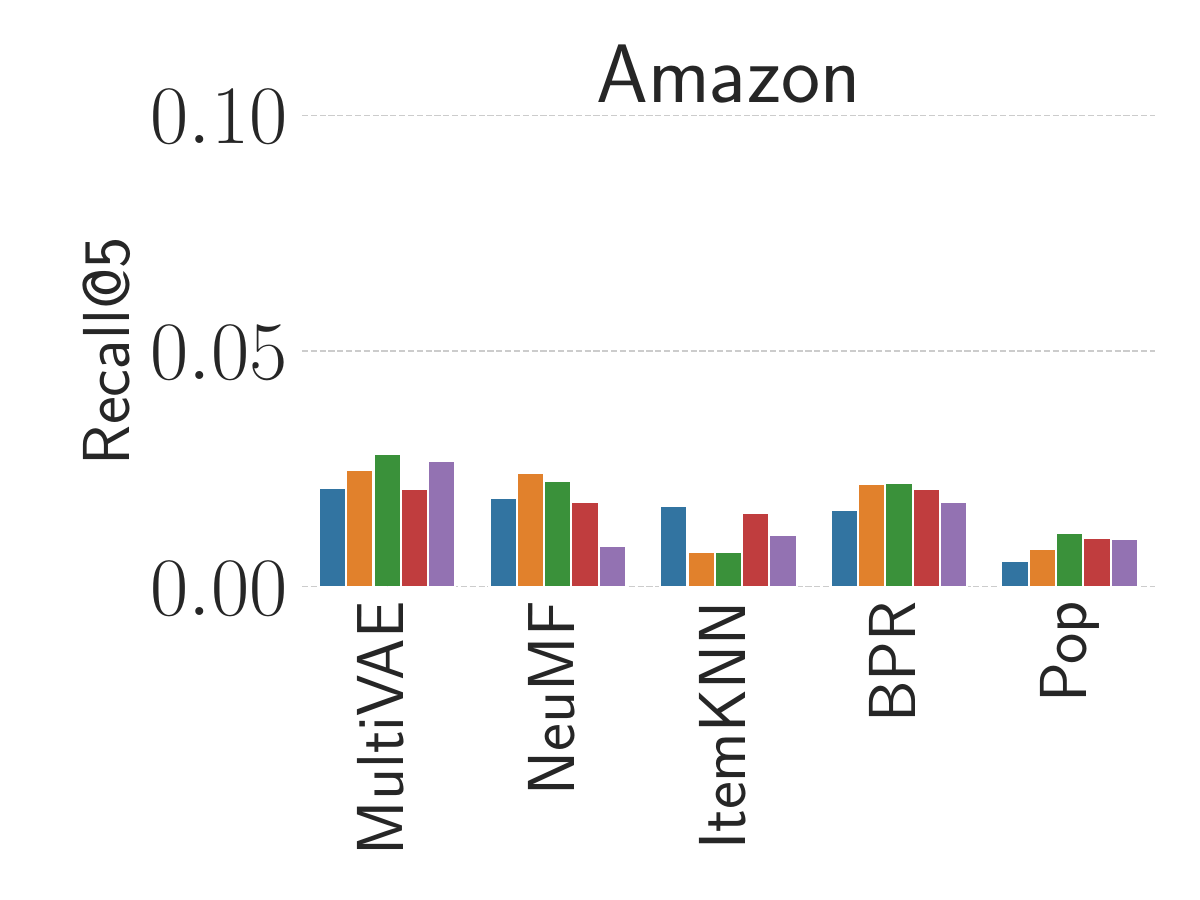}
    \end{subfigure}
    \begin{subfigure}[b]{0.48\columnwidth}
    \centering
    \includegraphics[width=\columnwidth]{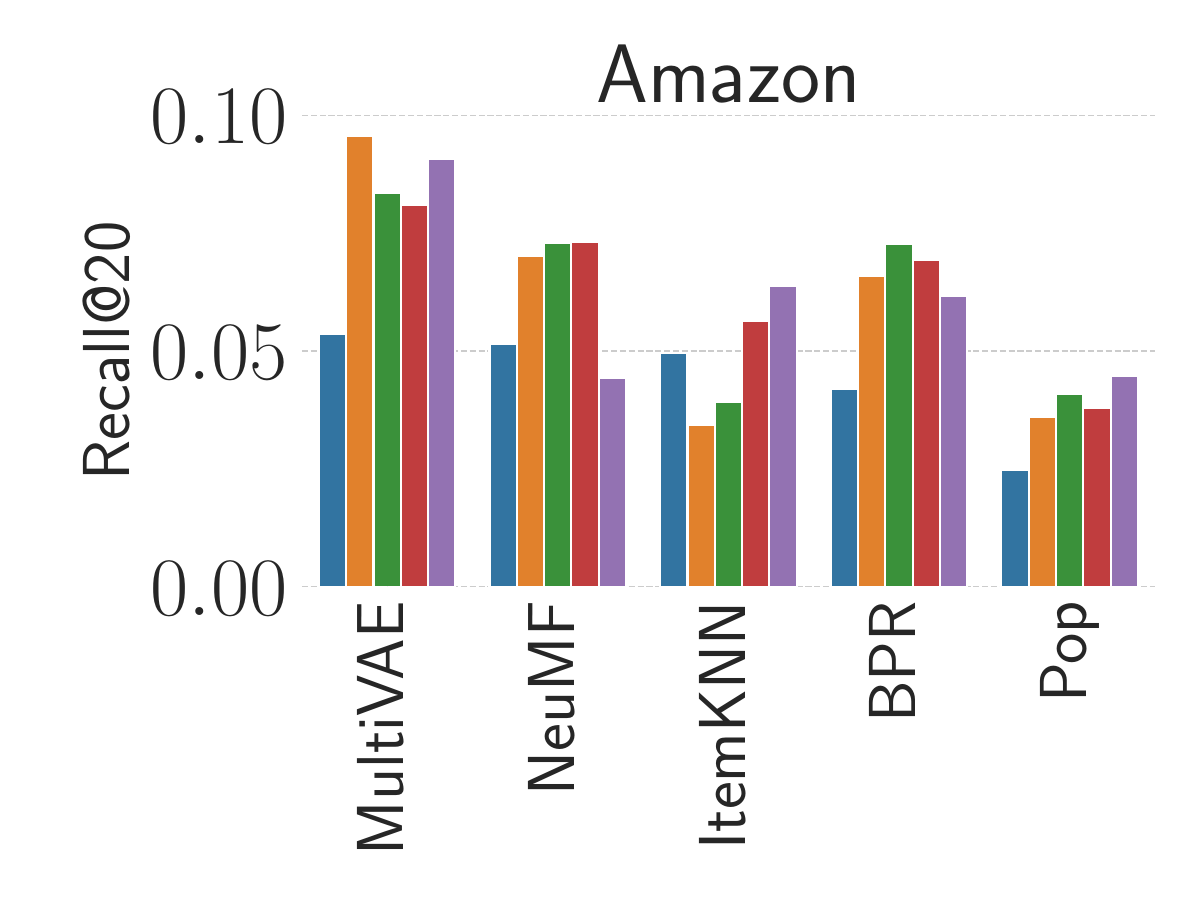}
    \end{subfigure}
    \begin{subfigure}[b]{0.48\columnwidth}
    \centering
    \includegraphics[width=\columnwidth]{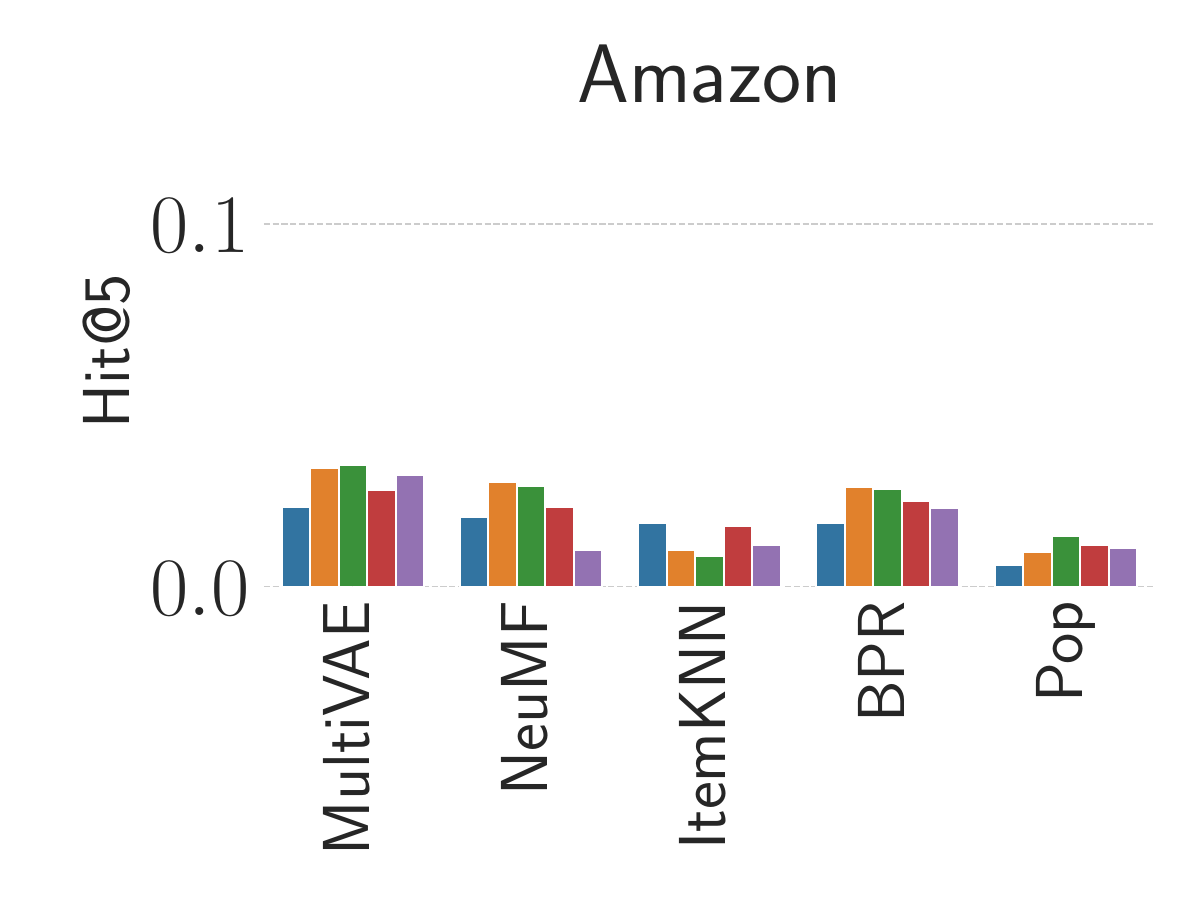}
    \end{subfigure}
    \begin{subfigure}[b]{0.48\columnwidth}
    \centering
    \includegraphics[width=\columnwidth]{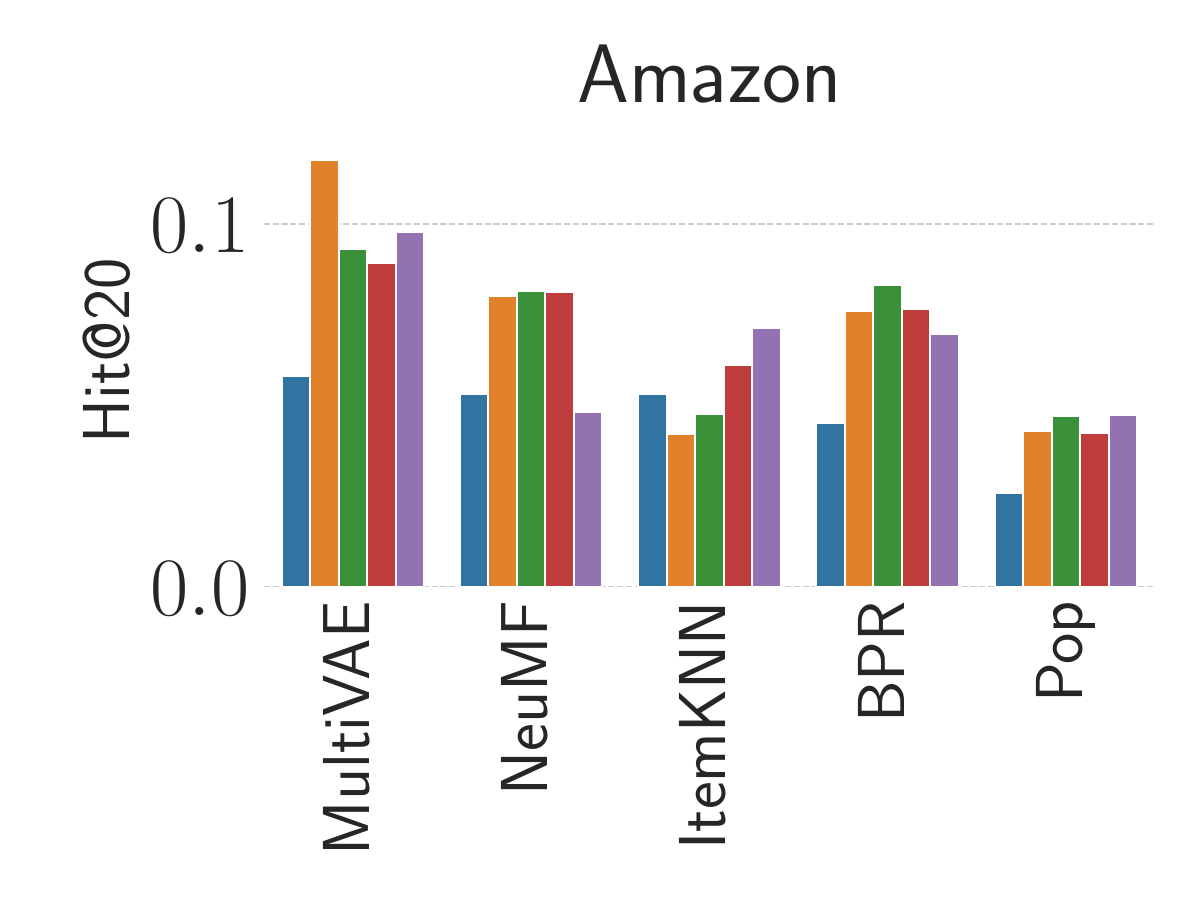}
    \end{subfigure}
    \caption{Performances comparison on Amazon at cut-offs equal 5 and 20. The colors indicate the results over the real dataset and on the generated sample at varying of $\delta$.}
    \label{fig:comparison_amazon-5-20}
\end{figure}

Figures~\ref{fig:comparison_movielens-5-20},~\ref{fig:comparison_yahoo-r3-5-20} and~\ref{fig:comparison_amazon-5-20} depict additional performance comparisons between the real benchmark datasets Movielens-1M, Yahoo-R3, and Amazon, respectively, and the synthetic samples generated by \ouralgo, in terms of Recall and Hit-Rate with cut-offs equal 5 and 20. The colors indicate the performance on the original dataset (in blue) and on the synthetic sample with different levels on noise injected (parameterized by the $\delta$ factor).
On Amazon, due to the extreme sparsity of the dataset, we perform a pre-processing step where we removed the users and items with having degree lower than 5. As we can see, the performance over the generated samples are comparable to that obtained over the original dataset, thus proving the efficacy of \ouralgo in reproducing existing benchmarks.


\section{Synthetic samples parameters}
\begin{table*}[hbt!]
    \centering
    \resizebox{\linewidth}{!}{
    \begin{tabular}{cl}
         \toprule
         \textbf{Symbol} & \textbf{Meaning}  \\
        \midrule         
        
         $U=\{1,\ldots,n\}$ & Set of users \\
         $I=\{1,\ldots, m\}$ & Set of items \\
         $U = \{U_1, U_2, \ldots, U_c \}$ and $I = \{I_1, I_2, \ldots, I_g\}$, & Populations of users and items that exhibit strong internal homophily \\
         $\dataset$ & Dataset of user-item interactions \\
         $r \in \{0, 1\}$ & interaction occurrence (relative to $(u,i)$: if $(u,i)$ exists then $r = 1$)\\
         $\nu$ & Parameters governing the preferences \\
         
         $z_u = |\{i: (u,i)\in \dataset\}|, \varrho_u$ & Engagement level and relative likelihood for user $u$\\
         $y_i = |\{u: (u,i)\in \dataset\}|, \varphi_i$ & Popularity frequency and likelihood for item $i$ \\
         
         $\pi_k, k \in \{1, .., K\}, P_{\theta_k}(z_u)$ & User engagement distribution and probability governed by the parameters of $\pi$ \\
         $\psi_h, h \in \{1, .., H\}, P_{\vartheta_h}(y_i)$ & Item popularity distribution and probability governed by the parameters of $\psi$\\
         $\rho_u$ & User latent factors \\
         $\mu_u^\rho$ & User parameters for sampling $\rho_u$\\
         $\alpha_i$ & Item latent factors \\
         $\mu_i^\alpha$ & Item parameters for sampling $\alpha_i$\\
         $\hat{\mathcal{D}}= \{\dataset, \mathbf{z}, \mathbf{y}, \nu\}$ & Estimated dataset \\
         $Q(h,k|\hat{\mathcal{D}}, \Theta)$ & Proposal probability distribution \\
        \bottomrule
    \end{tabular}
    }
    \caption{Notation table.}
    \label{tab:notation}
\end{table*}
For the original user/item distributions, we determined that the best fits are provided by Log-Normal/Stretched Exponential distributions for Movielens, Power-Law/Log-Normal distributions for Yahoo-R3, and Power-Law/Power-Law for Amazon. Using these estimated parameters, we regenerated the datasets artificially:
\begin{itemize}[leftmargin=2em]
    \item Movielens-1M: $x_u \sim \text{Log-Normal}(\mu = 4.61, \sigma = 1.2)$, $y_i \sim \text{StretchedExponential}(\lambda = 0.025, \beta = 0.691)$, $\zeta = 1.0$, $\xi = 1.4$ and $\lambda = 55k$.
    \item Yahoo-R3: $x_u \sim \text{Power-Law}(\alpha = 3.8)$, $y_i \sim \text{Log-Normal}(\mu = -0.98, \sigma = 2.44)$, $\zeta = 1.9$, $\xi = 1.0$ and $\lambda = 650k$.
    \item Amazon: $x_u \sim \text{Power-Law}(\alpha = 2.6)$, $y_i \sim \text{Power-Law}(\alpha = 1.9)$, $\zeta = 1.1$, $\xi = 0.95$ and $\lambda = 20k$.
\end{itemize}

\clearpage

\end{document}